\documentclass[acmsmall]{acmart}

\usepackage{soul}
\usepackage{wrapfig}
\usepackage{enumitem}
\usepackage[ruled,linesnumbered]{algorithm2e}
\usepackage{minibox}
\usepackage{float}
\usepackage{multirow}
\usepackage{listings}
\usepackage{subcaption}
\usepackage[most]{tcolorbox}
\usepackage{pifont}

\newcommand{\act}[0]{\textsc{ACT}}
\newcommand{\idl}[0]{TAIDL}
\newcommand{\pii}[0]{\textsf{pii}}
\newcommand{\hlo}[0]{\textsc{XLA-HLO}}
\newcommand{\tmul}[0]{\textsf{tdpbusd}}

\definecolor{myred}{HTML}{e06666}
\definecolor{myblue}{HTML}{388edc}
\definecolor{mygreen}{HTML}{6aa84f}
\definecolor{myorange}{HTML}{f6b26b}

\newcommand{\cmark}[0]{{\color{mygreen}\ding{51}}}
\newcommand{\xmark}[0]{{\color{red}\ding{55}}}

\newcommand{\hbm}[1]{{\color{myred} #1}}
\newcommand{\spad}[1]{{\color{myblue} #1}}
\newcommand{\ibuf}[1]{{\color{mygreen} #1}}

\newcommand{\slice}[0]{\mathsf{slice}}
\newcommand{\upslice}[0]{\mathsf{upslice}}
\newcommand{\copyop}[0]{\mathsf{copy}}
\newcommand{\concat}[0]{\mathsf{concat}}
\newcommand{\reshape}[0]{\mathsf{reshape}}
\newcommand{\bitcvt}[0]{\mathsf{bitcvt}}
\newcommand{\broadcast}[0]{\mathsf{broadcast}}
\newcommand{\reduce}[0]{\mathsf{reduce}}
\newcommand{\transpose}[0]{\mathsf{transpose}}
\newcommand{\tdot}[0]{\mathsf{dot}}
\newcommand{\texp}[0]{\mathsf{exp}}
\newcommand{\tdiv}[0]{\mathsf{div}}
\newcommand{\tcopy}[0]{\mathsf{copy}}

\newcommand{\add}[0]{\mathsf{add}}
\newcommand{\sub}[0]{\mathsf{sub}}
\newcommand{\negative}[0]{\mathsf{neg}}

\newcommand{\tmin}[0]{\mathsf{min}}
\newcommand{\tmax}[0]{\mathsf{max}}
\newcommand{\tclamp}[0]{\mathsf{clamp}}
\newcommand{\treverse}[0]{\mathsf{reverse}}

\newcommand{\loadrm}[0]{\mathsf{load\_rm}}
\newcommand{\loadcm}[0]{\mathsf{load\_cm}}
\newcommand{\storerm}[0]{\mathsf{store\_rm}}
\newcommand{\storecm}[0]{\mathsf{store\_cm}}
\newcommand{\gemm}[0]{\mathsf{gemm}}
\newcommand{\softmax}[0]{\mathsf{softmax}}
\newcommand{\mov}[0]{\mathsf{mov}}
\newcommand{\piig}[0]{\mathsf{PG}}
\newcommand{\phase}[0]{\mathsf{Phase}}

\newcommand{\bflat}[0]{\mathsf{bflat}}
\newcommand{\asm}[0]{\mathsf{ASM}}
\newcommand{\dm}[0]{\mathsf{D}}
\newcommand{\isa}[0]{\mathsf{ISA}}
\newcommand{\compiler}[0]{\mathsf{Compiler}}
\newcommand{\compgen}[0]{\mathsf{CompGen}}

\newcommand{\fail}[0]{\mathsf{FAIL}}
\newcommand{\nterm}[0]{\perp}

\newcommand{\type}[0]{\mathsf{type}}
\newcommand{\mydim}[0]{\mathsf{dim}}
\newcommand{\mem}[0]{\mathsf{mem}}
\newcommand{\esun}[0]{\mathsf{esun}}
\newcommand{\csp}[0]{\mathsf{CSP}}
\newcommand{\nrows}[0]{\mathsf{n}}
\newcommand{\addin}[0]{\mathsf{addr_{in}}}
\newcommand{\addout}[0]{\mathsf{addr_{out}}}

\theoremstyle{acmdefinition}
\newtheorem{definition}{Def}[section]
\theoremstyle{acmplain}
\newtheorem{theorem}{Theorem}[section]
\newtheorem{lemma}[theorem]{Lemma}

\newtcolorbox{definitionboxinner}[1]{%
  enhanced,
  breakable,
  colback=gray!5,
  colframe=gray!80!black,
  fonttitle=\bfseries,
  title={#1},
}
\newenvironment{definitionbox}[2][]{%
  \refstepcounter{definition}%
  \if\relax\detokenize{#1}\relax\else\label{#1}\fi
  \begin{definitionboxinner}{Def~\thedefinition: #2}%
}{%
  \end{definitionboxinner}%
}

\definecolor{codered}{rgb}{0.6,0,0}
\definecolor{codegreen}{rgb}{0,0.6,0}
\definecolor{codeblue}{rgb}{0,0,0.6}
\definecolor{codegold}{rgb}{0.6,0.6,0}
\definecolor{codepurple}{rgb}{0.58,0,0.82}
\definecolor{codebg}{rgb}{0.97,0.97,0.97}

\lstdefinestyle{backend-snippet-a}{
  language=C++,
  morekeywords={calls, fusion, fused\_subgraph, scf}
}

\lstdefinestyle{backend-snippet-b}{
  language=C++,
  basicstyle=\ttfamily\linespread{0.85}\footnotesize\color{codegreen},
  emphstyle=\bfseries\ttfamily,
  emph={llvm, inline\_asm, fused\_subgraph, tiled\_kernel},
}

\lstdefinestyle{qkv-isa}{
  emph={set\_hbm, add\_data\_model, add\_instr, set\_inputs, set\_output, set\_constraints, set\_abstract\_computation, extract},
  morekeywords={name, alpha, beta, start, S0, S1, E}
}

\lstdefinestyle{csp1}{
  language=C++,
  emph={INSTR, get\_h0, get\_e},
  morekeywords={std, vector}
}

\lstdefinestyle{template-theta}{
  language=C++,
  emph={INSTR, get\_h0},
  morekeywords={std, vector, pair, ortools, IntExpr, str, override, int64}
}

\lstdefinestyle{template-loadrm}{
  language=C++,
  emph={INSTR, get\_h0},
  morekeywords={std, vector, pair, ortools, IntExpr, str, override, int64}
}

\lstdefinestyle{outer-algo}{
  emph={cost\_model},
}

\lstdefinestyle{act-algo}{
  morekeywords={yield},
}

\lstdefinestyle{loadrm-precondition}{
  language=C++,
  morekeywords={pub, fn, let}
}

\usepackage{csquotes}
\usepackage{placeins}   

\lstdefinestyle{hlo}{
  language={},
  morekeywords={ENTRY, ROOT, parameter, constant},
  keywordstyle=\bfseries\ttfamily\color{codeblue},
  emph={convert, transpose, reshape, slice, dot, add, multiply, maximum,
        minimum, clamp, broadcast, tuple, relu},
  emphstyle=\bfseries\ttfamily\color{codered},
}

\lstdefinestyle{shell}{
  language={},
  numbers=left,
  basicstyle=\ttfamily\linespread{0.85}\footnotesize,
  keywordstyle=\ttfamily,
}

\lstdefinestyle{amx-asm}{}

\newif\ifactapxtoc
\newcommand{\actapxtocstart}{\actapxtoctrue}

\setcopyright{cc}
\setcctype{by}
\acmDOI{10.1145/3839457}
\copyrightyear{2026}
\acmYear{2026}
\acmJournal{PACMPL}
\acmVolume{10}
\acmNumber{OOPSLA2}
\acmArticle{325}
\acmMonth{10}
\received{2026-03-17}
\received[accepted]{2026-08-06}

\ccsdesc[500]{Software and its engineering~Compilers}
\ccsdesc[500]{Software and its engineering~Retargetable compilers}

\keywords{ML Compilers, AI Accelerators, Automated Compiler Construction}

\author{Devansh Jain}
\affiliation{%
  \institution{University of Illinois Urbana-Champaign}
  \country{USA}
}
\email{devansh9@illinois.edu}
\orcid{0009-0006-1442-1502}

\author{Akash Pardeshi}
\affiliation{%
  \institution{University of Illinois Urbana-Champaign}
  \country{USA}
}
\email{pardesh2@illinois.edu}
\orcid{0009-0006-2792-589X}

\author{Marco Frigo}
\affiliation{%
  \institution{University of Illinois Urbana-Champaign}
  \country{USA}
}
\email{mfrigo3@illinois.edu}
\orcid{0009-0008-2320-9989}

\author{Kaustubh Khulbe}
\affiliation{%
  \institution{University of Illinois Urbana-Champaign}
  \country{USA}
}
\email{kkhulbe2@illinois.edu}
\orcid{0009-0006-2971-4260}

\author{Krut Patel}
\authornote{Work done while at University of Illinois Urbana-Champaign.}
\affiliation{%
  \institution{NVIDIA}
  \country{USA}
}
\email{krutp@nvidia.com}
\orcid{0009-0001-2802-2354}

\author{Saatvik Lochan}
\affiliation{%
  \institution{University of Illinois Urbana-Champaign}
  \country{USA}
}
\email{slochan2@illinois.edu}
\orcid{0009-0003-7966-5653}

\author{Jai Arora}
\affiliation{%
  \institution{University of Illinois Urbana-Champaign}
  \country{USA}
}
\email{jaia3@illinois.edu}
\orcid{0009-0004-7759-481X}

\author{Charith Mendis}
\affiliation{%
  \institution{University of Illinois Urbana-Champaign}
  \country{USA}
}
\email{charithm@illinois.edu}
\orcid{0000-0002-8140-2321}

\renewcommand{\shortauthors}{Jain et al.}

\title{Automatically Generating ML Compiler Backends from Tensor Accelerator ISA Descriptions}

\begin{document}

\begin{abstract}
  Machine learning (ML) compilers play a key role in enabling high-performance implementations of ML workloads.
  These compilers use existing CPU and GPU backends to generate device-specific code.
  In recent years, many tensor accelerators (or AI accelerators) have been designed to further accelerate these workloads, with commercial products like AWS Trainium publicly available.
  However, compared to commodity hardware, a majority of tensor accelerators do not have mature ML compiler backends with robust code generation support.
  Moreover, tensor accelerator designs are subject to fast iteration cycles, making it difficult to manually develop and maintain ML compiler backends.
  Therefore, to enable faster integration of novel tensor accelerator designs in ML infrastructure, we need to make the compiler backend construction process more agile.

  In this paper, we introduce \act{}, a compiler backend generator that automatically generates compiler backends for tensor accelerators, given just the instruction set architecture (ISA) descriptions.
  These backends are integrated with XLA, a production ML compiler.
  \act{} uses a novel ISA-parameterized compilation algorithm to generate a compiler backend with an equality-saturation-based instruction selection phase and a constraint- programming-based memory allocation phase.
  We generated compiler backends for 6 accelerator platforms from industry (e.g., AWS Trainium, Intel AMX) and academia (e.g., Gemmini).
  We showed that these generated backends match or outperform commercial compiler backends and expert-written kernel libraries, while maintaining low compilation overheads.
  Notably, \act{}-generated backend for AWS NKI ISA improved the code generation coverage for AWS Trainium by 2.3x compared with AWS's production compiler, \texttt{neuronx-cc}.

  \act{} is part of a larger open-source ecosystem, built around our ISA description language \idl{}, that automatically generates essential software tools, such as test oracles and compiler backends, from ISA descriptions of tensor accelerators.
  Our tooling has been adopted by multiple academic and industry teams designing novel tensor accelerators.
  The ecosystem is available at \url{https://github.com/act-compiler/act}.
\end{abstract}

\maketitle


\section{Introduction}
\label{sec:introduction}

In recent years, machine learning (ML) workloads have gained popularity, specifically deep learning models.
ML practitioners use ML frameworks such as Tensorflow~\cite{tensorflow}, PyTorch~\cite{pytorch}, and JAX~\cite{jax} to express their computations, and ML compilers such as XLA~\cite{xla}, TorchInductor~\cite{pytorch}, and TVM~\cite{tvm} to lower these computations into highly performant executables targeting CPUs and GPUs.

\begin{figure}[!h]
  \centering
  \includegraphics[width=0.95\textwidth]{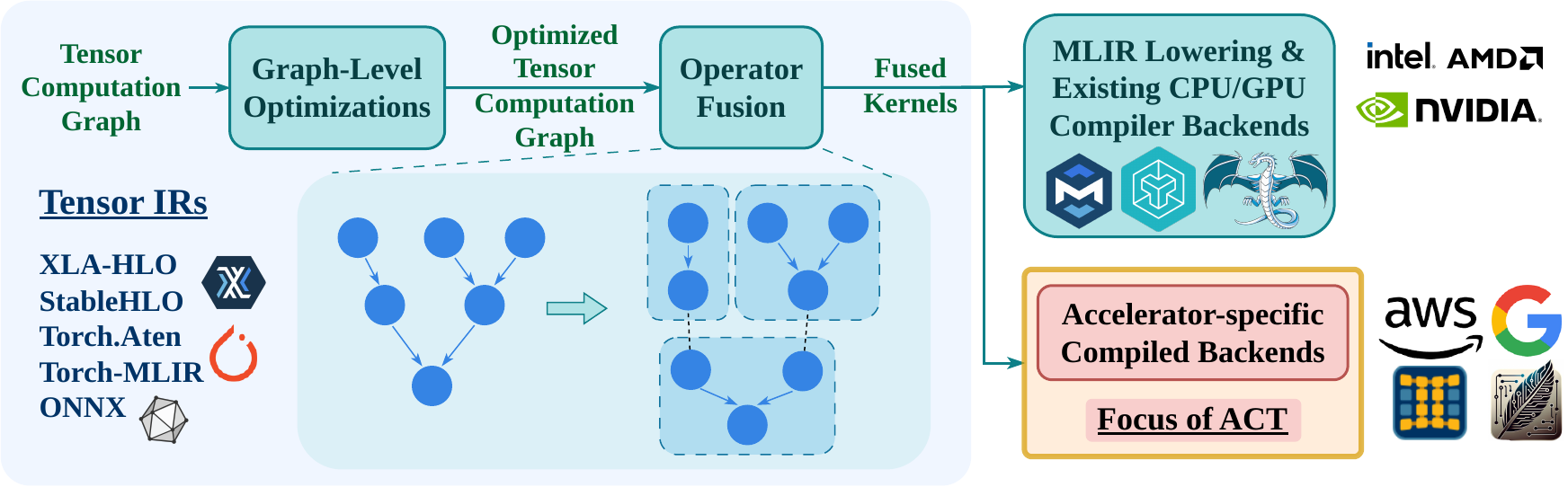}
  \vspace{-1em}
  \caption{Typical hierarchical compilation pipeline present in ML compilers. Our solution, \act{}, focuses on compiler backend construction (shaded red) for tensor accelerators like AWS Trainium~\cite{trainium} and Gemmini~\cite{gemmini}.
    We use XLA compiler~\cite{xla} as our reference ML compiler, which uses XLA-HLO IR~\cite{xla-hlo} and StableHLO IR~\cite{stablehlo}.
  }
  \Description{Typical hierarchical compilation pipeline present in ML compilers. Our solution, \act{}, focuses on compiler backend construction (shaded red) for tensor accelerators like AWS Trainium~\cite{trainium} and Gemmini~\cite{gemmini}.}
  \label{fig:intro-tensor-compilers}
  \vspace{-1.2em}
\end{figure}

Akin to general-purpose compilers, ML compilers also use a hierarchical compilation pipeline (Fig.~\ref{fig:intro-tensor-compilers}).
The frontend lowers programs written using ML frameworks to a tensor-operator-based intermediate representation (IR) known as tensor computation graphs.
These graphs are then optimized by several graph-level optimization passes~\cite{xla-autotune} such as algebraic simplification and layout selection.
The scale of these optimized graphs~\cite{tpugraphs} can range up to 1000s of nodes.
This is followed by operator fusion and tiling passes, which emit 100s of \textit{tiled kernels} (sub-graphs with typically 10--50 nodes).
These tiled kernels are compiled to device-specific assembly code individually using \emph{existing} compiler backends.
For example, XLA~\cite{xla}, a popular ML compiler, uses the existing lowering pipeline in MLIR~\cite{mlir} and LLVM~\cite{llvm} to compile these tiled kernels in XLA-HLO IR~\cite{xla-hlo}, XLA's tensor IR, into assembly code for commodity hardware (CPUs and GPUs).

To get the best performance out of ML workloads, many domain-specific tensor accelerators (a.k.a. AI accelerators) have been designed.
They have shown superior runtime performance, and the fast proliferation of such accelerator designs in both academia and industry is a testament to their importance and popularity.
For example, consider commercial tensor accelerators such as Google TPUs~\cite{tpuv1, tpuv3}, AWS Trainium~\cite{inferentia, trainium} and Intel AMX~\cite{intel-amx}, and a large body of academic designs such as Gemmini~\cite{gemmini}, MAERI~\cite{maeri}, FEATHER~\cite{feather}, EVA~\cite{eva}.
These accelerators are diverse, with different software-managed memory hierarchies and instruction sets (compute capabilities).

\vspace{-0.5em}
\subsection{Need for Generation of Accelerator-specific Compiler Backends}
\label{subsec:introduction-motivation}

Widespread use of a hardware architecture requires compiler backends that generate device-specific code, as evident for CPUs and GPUs --- companies like Google and Amazon~\cite{amazon-article} have invested heavily in compiler toolchains for their tensor accelerators~\cite{tpuv3, tpuv4, trainium,inferentia}.
In contrast, a majority of tensor accelerators proposed in the literature do not have mature backends and are often evaluated only on synthetic workloads~\cite{3la}, which \emph{inhibits their integration} with popular ML compilers.

Tensor accelerators without mature compiler backends instead rely on custom hand-written kernel libraries, which support only a limited set of kernels and miss out on fusion opportunities outside this set~\cite{flashattention, korch, dnnfusion, smartmem}.
For the Gemmini~\cite{gemmini} accelerator, we observe that operator fusion can lead to over 1.7x better performance and a 50\% reduction in data movement (\S\ref{subsec:evaluation-perf-gemmini}).
Therefore, compiler backends that support code generation for \emph{arbitrary tiled kernels} are needed.

Manually constructing such backends for every accelerator is too tedious and too slow to adapt to the fast pace of accelerator research (\S\ref{subsec:background-accelerator}).
Therefore, we propose an \emph{agile} methodology --- \emph{automated generation} of compiler backends for the multitude of tensor accelerator designs.

\vspace{-0.5em}
\subsection{Challenges \& Goals}
\label{subsec:introduction-challenges}

Automatically generating compiler backends that support code generation for arbitrary tiled kernels for many diverse tensor accelerators comes with new challenges not present in commodity hardware (e.g., prior works like vectorizer generators for vector ISAs~\cite{vegen, isaria, diospyros}).

\vspace{0.2em}
\textbf{Challenge 1: Code generation and optimizations for software-managed scratchpads.}
Most tensor accelerators have software-managed scratchpads with different memory hierarchies.
The data access patterns of tensor accelerators vary significantly in shape and size with multi-dimensional addressing (e.g., AWS Trainium~\cite{nki}) and multi-dimensional base elements (e.g., $16 {\times} 64$ matrices in Intel AMX~\cite{intel-amx}).
An ideal compiler backend needs to \emph{automatically} manage data allocation and reuse for these scratchpads.
Works such as Exo~\cite{exo} model scratchpads themselves, but leave the data allocation to end-users, whereas 3LA~\cite{3la} does not model scratchpad reuse.

\vspace{0.2em}
\textbf{Challenge 2: Handling parameterized ISA instructions.}
Tensor accelerators have instructions with parameters that control execution counts.
For example, the Google TPU~\cite{tpuv1} and AWS Trainium~\cite{nki} instructions are parameterized by input shapes (e.g., $B {\times} 128$).
Methodologies adopted by existing vectorizer generators~\cite{isaria, diospyros, vegen} can only handle fixed-size vector instructions.

\vspace{0.2em}
\textbf{Challenge 3: Handling complex ISA descriptions.}
Tensor accelerator ISA descriptions often include complex data layout transformations~\cite{intel-amx, feather, trainium}, that require accelerator-specific legalization passes to support such instructions.
Existing kernel libraries~\cite{exo,intel-amx} often rely on long and tedious pattern-matching rules manually written by experts for each new accelerator.

Moreover, the compiler backend generator should ideally achieve three goals.

\vspace{0.2em}
\textbf{Goal 1: Increased compilation coverage.}
Tensor accelerator designs often require expensive host fallback to execute certain operators (or sub-graphs) due to their limited compute capabilities.
The compiler backend should have a large compilation coverage, i.e., correctly compile for all tiled kernels supported by the accelerator --- in other words, a \emph{sound and complete compiler backend}.

\vspace{0.2em}
\textbf{Goal 2: Full automation.}
To minimize the engineering effort, the compiler backend generator should \emph{fully automate} the backend generation process with just the ISA descriptions as input.

\vspace{0.2em}
\textbf{Goal 3: Interoperability with existing ML compilers.}
The generated compiler backends should be \emph{compatible with existing ML compilers} to facilitate seamless integration and deployment.

\vspace{-0.5em}
\subsection{Our Solution}
\label{subsec:introduction-solution}

In this paper, we introduce \textbf{\act{} (Accelerator Compiler Toolkit)}, the first compiler backend generator that \emph{automatically} generates \emph{sound and complete} compiler backends for tensor accelerators from just their ISA descriptions.
\act{} consumes ISA descriptions written in \idl{}~\cite{taidl}, a popular ISA description language that supports diverse accelerator designs~\cite{intel-amx, tpuv1, nki, gemmini, feather, eva}.
\act{} achieves automatic generation (goal 2) by \emph{parameterizing} the compilation algorithm with \idl{}~\cite{taidl} constructs, which are concretized for each accelerator given their ISA descriptions.
\act{}-generated backends are compatible with XLA~\cite{xla}, a popular ML compiler (goal 3).
These backends consume tiled kernels in XLA-HLO IR as input and produce accelerator-specific assembly code.

We introduce a novel intermediary, \emph{partially instantiated instructions} (\textbf{\pii{}}), to handle instructions parameterized by input matrix shapes (challenge 2), prevalent in tensor accelerator ISAs.
We term such instruction parameters/attributes as \emph{computational} ($\boldsymbol{\alpha}$), with the rest termed as \emph{addressing} ($\boldsymbol{\beta}$).

\act{}'s \emph{ISA-parameterized} compilation algorithm consists of two main phases---one for solving each attribute set.
The first phase introduces \emph{parameterized} instruction selection based on equality saturation~\cite{eqsat} and uses multiple types of rewrites (e.g., IR-to-IR, IR-to-ISA).
\act{} generates the IR-to-ISA rewrites from the \idl{} descriptions.
These generated rewrites, combined with the XLA's IR-to-IR rewrites, perform IR legalization (challenge 3) and resolve the computational attributes ($\alpha$).
The second phase introduces \emph{parameterized} memory allocation for multi-dimensional buffer accesses (challenge 1) based on constraint programming on the addressing attributes ($\beta$).
We introduce intra- and inter-phase fallback paths to ensure completeness of the algorithm (goal 1).

\vspace{0.2em}
In summary, the paper makes the following contributions.

\begin{itemize}[noitemsep, nolistsep]
  \item We introduce \act{}, the first compiler backend generator that \emph{automatically generates} compiler backends for tensor accelerators, given just their ISA descriptions in \idl{}.
        As a core component of \act{}, we present a novel \emph{ISA-parameterized} compilation algorithm.
        (\S\ref{sec:overview}--\S\ref{sec:csp})

  \item We augment \act{}-generated backends with cost models to generate performant code.
        (\S\ref{sec:cost-model})

  \item We use \act{} to generate compiler backends integrated with the XLA compiler for \textbf{six} popular accelerator platforms from industry (AWS Trainium, Intel AMX, Google TPUv1) and academia (Gemmini, FEATHER, EVA), demonstrating the generality of \act{}.
        (\S\ref{subsec:evaluation-support})

  \item We show that the \act{}-generated compiler backends match or outperform existing accelerator-specific backends and expert-written kernel libraries (up to \textbf{6.88x} speedup).
        We also improve compilation coverage for AWS NKI by \textbf{2.3x} over AWS's compiler \texttt{neuronx-cc}.
        (\S\ref{subsec:evaluation-perf}--\S\ref{subsec:evaluation-perf-gemmini})

  \item We show that compilation times of the \act{}-generated compiler backends are less than a second for a wide range of kernel sizes, even for large kernels (\textbf{529 ms} for 390 nodes).
        (\S\ref{subsec:evaluation-time})
\end{itemize}

\vspace{0.2em}
\noindent
\act{} is part of a larger open-source ecosystem built around our ISA description language TAIDL~\cite{taidl}.
The ecosystem and its associated tutorials are accessible at \url{https://github.com/act-compiler/act}.

\vspace{-0.5em}
\section{Background}
\label{sec:background}

We first briefly discuss XLA-HLO IR, tensor accelerators, and equality saturation workflow, providing the background needed for the problem formulation (\S\ref{sec:problem}) and our solution (\S\ref{sec:overview}--\S\ref{sec:cost-model}).

\vspace{-0.5em}
\subsection{XLA-HLO IR: XLA Compiler's Tensor IR}
\label{subsec:background-tensor}

XLA-HLO (High Level Operations) IR~\cite{xla-hlo} is a tensor IR used in the XLA compiler~\cite{xla}, the default ML compiler for TensorFlow~\cite{tensorflow} and JAX~\cite{jax}.
Here, we describe the key concepts of XLA-HLO IR relevant to this paper.
These concepts are also applicable to other tensor IRs such as Torch.Aten~\cite{pytorch}.

\vspace{0.2em}
\textbf{Tensors} are $n$-dimensional objects (generalization of 0-D scalars, 1-D vectors, 2-D matrices) that are usually represented as multi-dimensional arrays.
A tensor of \emph{tensor-type} $\mathbb{E}[S]$ has a shape $S$, a tuple of integers denoting the dimension sizes, with elements of scalar basetype $\mathbb{E}$ such as \textsf{uint8} (\textsf{u8}), \textsf{int32} (\textsf{i32}), \textsf{fp32} (\textsf{f32}), \textsf{bfloat16} (\textsf{bf16}).
A \emph{tensor value} has a fixed tensor-type with fixed data for each of its elements.
A \emph{tensor variable} has a fixed tensor-type but can hold variable data at runtime.

\vspace{0.2em}
\textbf{Tensor Operators.}
The XLA compiler manipulates tensors as first-class objects using tensor operators, which are usually tensor-type agnostic and parameterized by various parameters, collectively called \emph{operator attributes}.
For example, the operator $\slice(x,s,e,p)$ extracts a sub-tensor from an input tensor $x$ of any tensor-type, from indices $s$ to $e$ with a stride of $p$.
Here, $s,e,p$ and the tensor-type of $x$ are operator attributes of $\slice$, with validity conditions such as $0 \leq s < e \leq \text{size}(x)$ and $p > 0$.
We call tensor operators parameterized by these attributes \emph{abstract tensor operators}.
Table~\ref{tab:imp-ops} briefly describes the XLA-HLO tensor operators used in this paper.
Appendix~\ref{app:viz} visualizes these, with their operational and denotational semantics documented in \cite{xla-hlo, stablehlo} and \cite{tensorright}, respectively.

During the compilation of a given model, all operator attributes are instantiated.
We call these instantiated tensor operators \emph{concrete tensor operators}, which take tensor variables as input and produce a tensor variable as output.
Consider an abstract tensor operator $f$ instantiated with attributes $A$, producing a concrete tensor operator $f_A$.
It takes $n$ tensor variables $X = (x_i \vert_{i=1}^{n})$ as input and computes a tensor variable $y$ as output, i.e., $y = f_A(x_1, \dots x_{n})$, or in short, $f_A(X)$.

\begin{table}[!ht]
  \vspace{0.5em}
  \centering
  \begin{tabular}{|| c | c | c ||}
    \hline
    Tensor Operator                    & Operator Attributes           & Description                                     \\
    \hline\hline
    $y = \slice_{[s{:}e]}(x)$          & $\type(x), s, e$              & Read sub-tensor $x[s{:}e]$                      \\
    \hline
    $y = \upslice_{[s{:}e]}(x_1, x_2)$ & $\type(x_1), s, e$            & Update sub-tensor $x_1[s{:}e] = x_2$            \\
    \hline
    $y = \concat_{dim}(x_1, x_2)$      & $\type(x_1), \type(x_2), dim$ & Concatenate $x_1$, $x_2$ across dimension $dim$ \\
    \hline
    $y = \reshape(x)$                  & $\type(x), \type(y)$          & Reshape from $\type(x)$ to $\type(y)$           \\
    \hline
    $y = \bitcvt(x)$                   & $\type(x), \type(y)$          & Bitcast conversion between basetypes            \\
    \hline
    $y = \broadcast_{dim}(x)$          & $\type(x), dims$              & Broadcast over dimension $dim$                  \\
    \hline
    $y = \reduce_{dim}(x)$             & $\type(x), dims$              & Reduce-sum over dimension $dim$                 \\
    \hline
    $y = \transpose_{[dims]}(x)$       & $\type(x), dims$              & Permute the dimension order to $dims$           \\
    \hline
    $y = \tdot(x_1, x_2)$              & $\type(x_1), \type(x_2)$      & Matrix multiplication of $x_1$, $x_2$           \\
    \hline
    $y = \texp(x)$                     & $\type(x)$                    & Element-wise exponential                        \\
    \hline
    $y = \tdiv(x_1, x_2)$              & $\type(x_1)$                  & Element-wise divide                             \\
    \hline
    $y = \tcopy(x)$                    & $\type(x)$                    & Identical copy                                  \\
    \hline
  \end{tabular}
  \vspace{0.5em}
  \caption{Brief descriptions of tensor operators based on XLA-HLO IR~\cite{xla-hlo} discussed in the paper.\\
  $\type(x)$ refers to the tensor-type of tensor variable $x$.
  $dim$ represents the dimension number (starting from 1).
  We use NumPy-like slice notation $x[s_1{:}e_1,s_2{:}e_2,\dots]$ and ignore $\type$-based attributes in visualizations.
  }
  \label{tab:imp-ops}
  \vspace{-2em}
\end{table}

\vspace{0.2em}
\textbf{Tensor Computation Graphs.}
XLA's internal IR, called XLA-HLO IR, represents computations as a \emph{tensor computation graph} --- a directed acyclic graph with tensor operators as nodes and operands as edges.
The edges establish the data dependencies between different tensor operators.
Unless specified otherwise, a tensor computation graph consists of concrete tensor operators.

\vspace{0.2em}
\textbf{IR Properties.}
XLA-HLO IR is a \emph{data flow graph} IR, i.e., it does not have explicit imperative control flow.
Control flow is represented using higher-order functional operators such as \texttt{while}.
XLA-HLO IR only supports \emph{static tensor shapes}, i.e., the shapes of all tensor variables are known at compile time. Shape dynamism is handled by StableHLO passes~\cite{stablehlo-dynamism} prior to XLA-HLO IR.

\vspace{-0.5em}
\subsection{Tensor Rewrites in XLA Compiler}
\label{subsec:background-axioms}

The XLA compiler performs semantic-preserving transformations on tensor computation graphs using a set of algebraic rewrite rules, termed \emph{tensor rewrites}.
A tensor rewrite $P_l \rightarrow P_r$ substitutes subgraph $P_l$ with subgraph $P_r$, under certain preconditions on the operator attributes.

XLA uses 175+ tensor rewrites, with 115 of these verified to full generality by TensorRight~\cite{tensorright}.
In our work, we adopt this rewrite set, denoted as $\mathcal{R}$, as foundational axioms of the XLA-HLO IR.
One such rewrite is $\slice_{[s_1:e_1]}(\slice_{[s_2:e_2]}(x)) \rightarrow \slice_{[s_1+s_2:e_1+s_2]}(x)$.
Similar to prior works~\cite{taso, tensat, aerify, constable}, we define semantic equivalence modulo these tensor rewrites $\mathcal{R}$.

\vspace{-0.5em}
\subsection{Tensor Accelerators}
\label{subsec:background-accelerator}

Tensor accelerators (a.k.a. AI accelerators) are a class of hardware accelerators optimized for tensor computations using different microarchitectural innovations such as systolic array-based executions.
Some have programmable dataflows~\cite{gemmini, maeri, sambanova-sn10}, while others are fixed~\cite{tpuv1, eyeriss, nvdla, shidiannao, diannao}.
They have diverse functional units such as matrix transpositions~\cite{tpuv3}, matrix reshaping units, etc., with novel innovations such as intelligent on-chip routing~\cite{feather}.
\idl{}~\cite{taidl} standardizes how ISA semantics are described for a variety of diverse tensor accelerator designs.

\vspace{-0.5em}
\subsection{Term Rewriting using Equality Saturation}
\label{subsec:background-egg}

Term rewriting~\cite{term_rewriting} is classically destructive, making optimizations sensitive to the order in which rewrites are applied --- the well-known ``phase-ordering'' problem~\cite{phase-ordering}.
Equality saturation~\cite{eqsat} overcomes it using \textit{e-graphs}, which apply rewrites simultaneously and compactly represent equivalence relations over terms as equivalence classes (\textit{e-classes}) of equivalent nodes (\textit{e-nodes}).

The equality saturation workflow consists of three stages: initialization, exploration, and extraction.
First, the initial e-graph is created from the input term.
Next, the exploration stage applies rewrites until saturation is reached (i.e., applying rewrites does not add any new information) or a e-node limit is reached.
Finally, the optimal term is extracted from the e-graph.

\vspace{-0.5em}
\section{Problem Statement}
\label{sec:problem}

In this section, we formulate the problem of a compiler backend generator that generates sound and complete accelerator-specific backends from just the accelerator ISA description.
First (\S\ref{subsec:problem-backend}), we describe the input to these backends by illustrating the code generation pipeline of the XLA compiler~\cite{xla}.
Next (\S\ref{subsec:problem-isa}), we provide a high-level summary of \idl{}~\cite{taidl}, a popular ISA description language for tensor accelerators that serves as the input to our generator.
Finally (\S\ref{subsec:problem-generator}), we define soundness and completeness, and state the problem of a compiler backend generator.

\vspace{-0.5em}
\subsection{Tiled Kernels: Input to Accelerator-specific Compiler Backends}
\label{subsec:problem-backend}

\begin{figure}[!h]
  \vspace{-0.5em}
  \centering
  \includegraphics[width=0.9\textwidth]{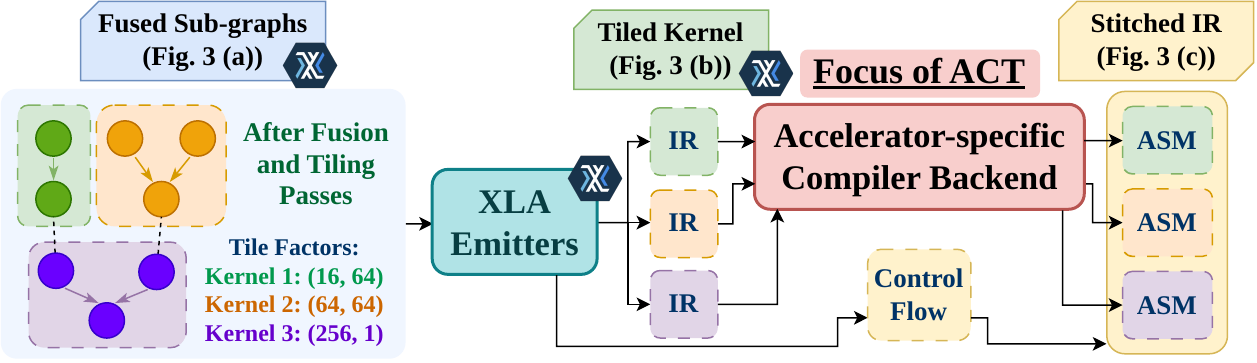}
  \vspace{-1em}
  \caption{Summary of code generation pipeline of XLA compiler~\cite{xla} after operator fusion and tiling passes. Snippets from different stages of this pipeline are shown in Fig.~\ref{fig:backend-overview-snippets}. In this paper, we focus on accelerator-specific compiler backends (shaded red) and borrow XLA's existing implementation for the rest of the pipeline.}
  \Description{Summary of code generation pipeline of XLA compiler~\cite{xla} after operator fusion and tiling passes. Snippets from different stages of this pipeline are shown in Fig.~\ref{fig:backend-overview-snippets}. In this paper, we focus on accelerator-specific compiler backends (shaded red) and borrow XLA's existing implementation for the rest of the pipeline.}
  \label{fig:backend-overview}
  \vspace{-1.2em}
\end{figure}

XLA's code generation pipeline starts after the middle-end passes of operator fusion and tiling~\cite{tpugraphs}, which partition the optimized tensor computation graph in XLA-HLO IR (\S\ref{subsec:background-tensor}) into several sub-graphs and determine the tile factors for each of these sub-graphs.
This pipeline is summarized in Fig.~\ref{fig:backend-overview}, and snippets from different stages of this pipeline are shown in Fig.~\ref{fig:backend-overview-snippets}.

\begin{figure}[!h]
  \vspace{-0.7em}
  \centering
  \begin{minipage}{0.53\textwidth}
    \centering
    \begin{lstlisting}[style={backend-snippet-a},frame=single,belowskip=-1pt]
...
\end{lstlisting}
    \begin{lstlisting}[style={backend-snippet-b},firstnumber=2,frame=single,aboveskip=-1pt,belowskip=-1pt]
%fused_subgraph.1 {
  %Q = bf16[2048,1024][!*{\bfseries\ttfamily\color{codepurple}\{:T(16,64)\}}*!]  parameter(0)
  ...
}
\end{lstlisting}
    \begin{lstlisting}[style={backend-snippet-a},firstnumber=6,frame=single,aboveskip=-1pt]
...
ENTRY %main {
  ...
  %qkv.out = bf16[2048,1024]{:T(16,64)} fusion(...), calls=%fused_subgraph.1
  ...
}
\end{lstlisting}
    \vspace{-0.2em}
    \small{(a) XLA-HLO IR after operator fusion and tiling passes.\\
      The tile factors are annotated by \texttt{\{:T(...)\}} in purple.}
  \end{minipage}
  \hfill
  \begin{minipage}{0.45\textwidth}
    \centering
    \begin{lstlisting}[style={backend-snippet-b},frame=single]
ENTRY %tiled_kernel.1 {
  %Q = bf16[!*{\bfseries\ttfamily\color{codepurple}[16,64]}*!] parameter(0)
  ...
}
\end{lstlisting}
    \vspace{-0.2em}
    \small{(b) Tiled kernel emitted for \texttt{\%fused\_subgraph.1}}
    \begin{lstlisting}[style={backend-snippet-a},frame=single,belowskip=-1pt]
...
scf.for [!*{\bfseries\ttfamily\color{codepurple} \%i = 0 to 128}*!] {
  ... {
\end{lstlisting}
    \begin{lstlisting}[style={backend-snippet-b},firstnumber=4,frame=single,aboveskip=-1pt,belowskip=-1pt]
    %x0 = llvm.inline_asm ...
    ...
\end{lstlisting}
    \begin{lstlisting}[style={backend-snippet-a},firstnumber=6,frame=single,aboveskip=-1pt]
  }
} ...
\end{lstlisting}
    \vspace{-0.2em}
    \small{(c) Stitched IR with accelerator-specific assembly code inlined in the control flow for tiling.}
  \end{minipage}
  \vspace{-1em}
  \caption{Snippet from different stages of XLA's code generation pipeline. The accelerator-specific compiler backend consumes a tiled kernel (b) and emits assembly code (lines 4--5 in (c)) inlined in the stitched IR.}
  \Description{Snippet from different stages of XLA's code generation pipeline. The accelerator-specific compiler backend consumes a tiled kernel (b) and emits assembly code (lines 4--5 in (c)) inlined in the stitched IR.}
  \label{fig:backend-overview-snippets}
  \vspace{-1em}
\end{figure}

After XLA middle-end passes, the XLA-HLO IR includes multiple sub-graphs (e.g., Fig.~\ref{fig:backend-overview-snippets} (a) lines 2--5) with tile factors annotated in purple (e.g., \texttt{\{:T(16,64)\}}).
First, XLA's \texttt{Emitter} modules~\cite{xla-emitters} emit (i) control flow for tiling (e.g., \texttt{scf.for} loop in Fig.~\ref{fig:backend-overview-snippets} (c) line 2) and (ii) single-tile versions of all sub-graphs (e.g., Fig.~\ref{fig:backend-overview-snippets} (b)). We name these single-tile versions of sub-graphs as \emph{tiled kernels}.

\begin{wrapfigure}{R}{0.34\textwidth}
  \vspace{-1.5em}
  \centering
  \includegraphics[width=0.33\textwidth]{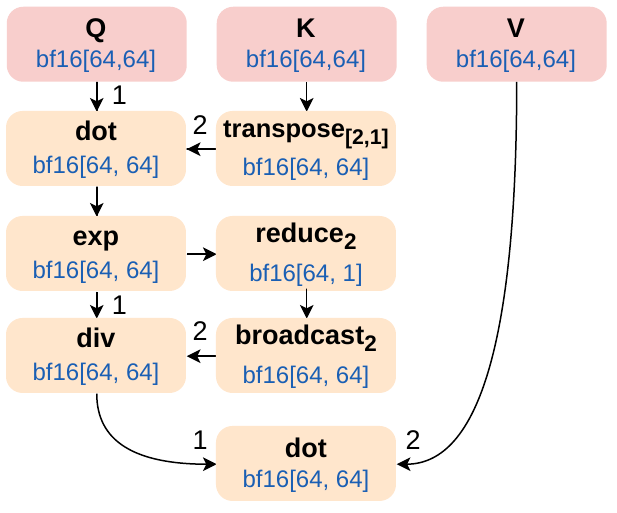}
  \vspace{-0.5em}
  \caption{Self-attention kernels in Transformer models like BERT~\cite{bert} use QKV computation $\softmax(Q \cdot K^T) \cdot V$ over 4-dimensional tensors. For illustration purposes, we assume a single-batch single-head QKV computation $G_{QKV}$ over $\mathsf{bf16}[64,64]$ tiles.}
  \Description{Self-attention kernels in Transformer models like BERT~\cite{bert} use QKV computation $\softmax(Q \cdot K^T) \cdot V$ over 4-dimensional tensors. For illustration purposes, we assume a single-batch single-head QKV computation $G_{QKV}$ over $\mathsf{bf16}[64,64]$ tiles.}
  \label{fig:qkv-graph}
  \vspace{-1.5em}
\end{wrapfigure}

A \textbf{tiled kernel} is syntactically similar to the original sub-graph, except that the tensor shapes are based on the tile factors (e.g., \texttt{bf16[16,64]} in Fig.~\ref{fig:backend-overview-snippets} (b) vs. \texttt{bf16[2048,1024]} in Fig.~\ref{fig:backend-overview-snippets} (a)).
It is a tensor computation graph in XLA-HLO IR (\S\ref{subsec:background-tensor}), which includes static tensor-types, and computes a single tile of the original sub-graph.
In this paper, we will visualize tiled kernels as directed acyclic graphs with nodes as tensor operators (from Table~\ref{tab:imp-ops}), for example, in Fig.~\ref{fig:qkv-graph}.

The \textbf{accelerator-specific compiler backend} compiles these tiled kernels individually into accelerator-specific assembly code.
Finally, the generated assembly code for different kernels is inlined and stitched together with control flow for tiling into a single \emph{stitched IR} (e.g., Fig.~\ref{fig:backend-overview-snippets} (c)).
We defer the specifics of execution of the stitched IR for various accelerators to evaluations (\S\ref{sec:evaluation}).

\medskip

\begin{center}
  \minibox[frame,c]{
    The input to an accelerator-specific compiler backend is a tiled kernel in XLA-HLO IR\\
    and the output is a sequential accelerator-specific assembly code.
  }
\end{center}

\medskip

\emph{Running Example.}
Fig.~\ref{fig:qkv-graph} visualizes a tiled kernel $G_{QKV}$ for QKV computation, which is prevalent in many transformer topologies~\cite{bert, attention}.
For illustration purposes, we assume a single-batch single-head QKV computation over $\mathsf{bf16}[64,64]$ tiles.
We use this as a running example for the rest of the paper.

\vspace{0.2em}
\textbf{Challenges}\\
Key challenges in designing accelerator-specific compiler backends are the diversity of tensor accelerator designs and the complexity of the ISA descriptions.
Different accelerators have different memory hierarchies and instruction granularities, which require different compilation strategies.
Furthermore, ISA descriptions of tensor accelerators are often parameterized by input shapes~\cite{gemmini, eva, tpuv1} and include complex data layout transformations~\cite{intel-amx, feather, trainium}, which require accelerator-specific legalization passes to map tiled kernels to the accelerator ISA.

\vspace{-0.5em}
\subsection{\idl{}: Tensor Accelerator ISA Description}
\label{subsec:problem-isa}

We choose \idl{}~\cite{taidl}, Tensor Accelerator ISA Definition Language, to describe accelerator ISAs since it supports several existing tensor accelerators \cite{intel-amx, tpuv1, nki, gemmini, feather, eva} and can precisely model the semantics of new tensor accelerator ISAs.
Here, we provide a high-level summary of \idl{}~\cite{taidl} necessary for later sections. Detailed definitions and explanations are deferred to Appendix~\ref{app:problem}.

\idl{} standardizes the way ISAs and their semantics are developed for tensor accelerators.
It builds on top of the existing formal models~\cite{atl-popl, tensorright, taso, pet} of tensors and tensor operators.
The key principle behind \idl{} is that tensor accelerators support coarse-grained instructions that work on multi-dimensional data (commonly represented as tensors).
Therefore, \idl{} descriptions use compositions of tensor operators in XLA-HLO IR to model the instruction semantics.

An \emph{ISA description} comprises (1) the software-managed storage units (collectively termed \emph{data model}) like scratchpads, and (2) the semantics of instructions that perform computations like data movement and compute instructions on these storage units.
This is analogous to the x86 ISA description comprising a data model (registers, memory) and an instruction set (\textsf{add}, \textsf{mov}, etc.).

\emph{Running Example.}
Consider a hypothetical accelerator $H_{QKV}$ (Fig.~\ref{fig:qkv-isa} (a)) with two on-chip storage units---a 16 KB scratchpad and an 8 KB intermediate buffer---and access to global memory via a DMA engine.
Its two compute units are a general matrix multiply ($\gemm$) unit and a softmax activation unit.
It supports five data movement and two compute instructions listed in Fig.~\ref{fig:qkv-isa} (b).
We use this hypothetical accelerator $H_{QKV}$ as our running example for the rest of the paper.

\noindent
\begin{minipage}{\textwidth}
  \vspace{0.5em}
  \begin{minipage}[c]{0.4\textwidth}
    \centering
    \includegraphics[width=0.95\textwidth]{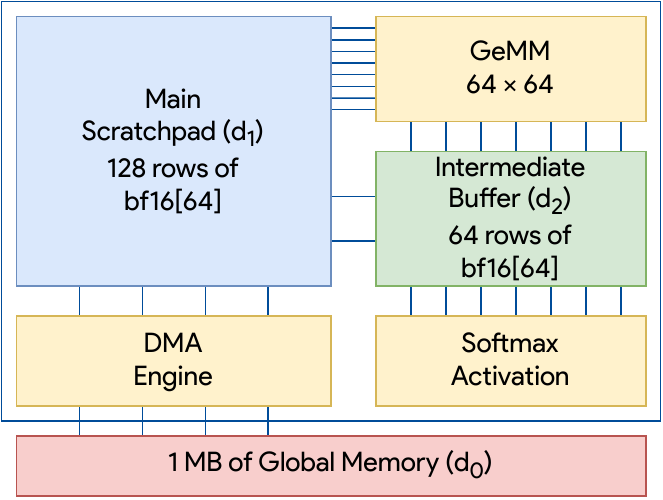}
    \begin{center}
      (a)
    \end{center}
  \end{minipage}
  \hfill
  \begin{minipage}[c]{0.59\textwidth}
    \centering
    ~\\
    \begin{tabular}{||c|l||}
      \hline
      Instruction                 & Description                                            \\
      \hline\hline
      $\loadrm$                   & $\spad{y} = \bitcvt(\reshape(\hbm{x_1}))$              \\
      \hline
      $\loadcm$                   & $\spad{y} = \transpose(\bitcvt(\reshape(\hbm{x_1})))$  \\
      \hline
      $\storerm$                  & $\hbm{y} = \reshape(\bitcvt(\spad{x_1}))$              \\
      \hline
      $\storecm$                  & $\hbm{y} = \transpose(\reshape(\bitcvt(\spad{x_1})))$  \\
      \hline
      $\mov$                      & $\spad{y} = \tcopy(\ibuf{x_1})$                        \\
      \hline
      $\gemm$                     & $\ibuf{y} = \tdot(\spad{x_1}, \spad{x_2})$             \\
      \hline
      \multirow{2}{*}{$\softmax$} & $\ibuf{y} = \tdiv(\texp(\ibuf{x_1}),$                  \\
                                  & \hspace{3em} $\broadcast(\reduce(\texp(\ibuf{x_1}))))$ \\
      \hline
    \end{tabular}
    \begin{center}
      ~\\
      (b)
    \end{center}
  \end{minipage}
  \vspace{-1em}
  \captionof{figure}{Running example. (a) High-level accelerator design for a hypothetical accelerator $H_{QKV}$.\\
    (b) Brief description of its instruction set. Tensor variables are colored based on their storage unit ($\hbm{d_0}$, $\spad{d_1}$, $\ibuf{d_2}$).}
  \label{fig:qkv-isa}
  \vspace{0.5em}
\end{minipage}

\smallskip
\textbf{(i) \idl{} constructs: Data Model}

The software-managed storage units, like scratchpads on an accelerator, are abstractly represented as \textit{tensor buffers}.
\idl{} represents a tensor buffer $\boldsymbol{d}$ as a tensor (of type $\mathbb{E}[S_0, S_1]$) with a subset of dimensions representing \emph{access dimensions} ($\boldsymbol{S_0}$) and the remaining dimensions representing the granularity of data access ($\boldsymbol{S_1}$).
\idl{} assumes that an accelerator has a byte-addressable 1-dimensional global memory (HBM) that is accessible by the host processor and is denoted $\hbm{d_0}$.

\emph{Running Example.}
Tensor buffer representations for software-managed storage units of $H_{QKV}$ are represented in lines 2--4 of Fig.~\ref{lst:qkv-isa}.
The main scratchpad ($\spad{d_1}$) consists of 128 rows ($S_0 {=} [128]$) of 64-length $\mathsf{bf16}$ vectors ($S_1 {=} [64]$).
Instructions read and/or modify the scratchpad at a row granularity.

\smallskip
\textbf{(ii) \idl{} constructs: Instruction Set}

The instruction set is a set of \textit{abstract instructions} that modify the tensor buffers and are parameterized by instruction attributes.
A \textit{concrete instruction} is an instantiation of an abstract instruction with specific values for the attributes.
For example, x86 instructions, such as \textsf{mov} and \textsf{add}, modify registers or memory and are parameterized by register names and memory addresses.

An \emph{assembly code} for a tensor accelerator consists of a stream of concrete instructions and a constant tensor.
This is analogous to x86 assembly code with code and data sections, respectively.

\idl{} models the semantics of abstract instructions of tensor accelerators as a composition of tensor operators in XLA-HLO IR~\cite{xla-hlo}.
Fig.~\ref{fig:qkv-isa} (b) provides a brief description of the instruction set of $H_{QKV}$ using tensor operators from Table~\ref{tab:imp-ops}.

In this work, we extend \idl{} descriptions by classifying the instruction attributes into two sets: (i) \emph{computational} ($\boldsymbol\alpha$) attributes, like input matrix shapes (challenge 2), that determine the core tensor computation; and rest of the attributes as (ii) \emph{addressing} ($\boldsymbol\beta$) attributes, like input addresses.
Therefore, an abstract instruction $\boldsymbol{\theta}$ which reads $n_\theta$ tensors located in buffers $d_{\theta}^{(i)} |_{i=1}^{n_\theta}$ and modifies a tensor located in buffer $d_{\theta}^{(0)}$, is represented using an \emph{abstract tensor computation} $\boldsymbol{g_{\theta}}$ concretized by the set of computational attributes $\alpha$, $(n_\theta + 1)$ \emph{data slice address maps} $\boldsymbol{h^{(i)}_{\theta}} |_{i=0}^{n_\theta}$ over the set of addressing attributes $\beta$, and a \emph{validity constraint} $\boldsymbol{e_{\theta}}$ over the set of attributes $\alpha$ and $\beta$.

\emph{Running Example.}
The abstract instruction $\loadrm$ in $H_{QKV}$ is represented in lines 7--16 of Fig.~\ref{lst:qkv-isa}.
Lines 8--9 describe the tensor buffers $\boldsymbol{d_{\loadrm}}$ and data slice address maps $\boldsymbol{h_{\loadrm}}$ of the input and output tensors.
It is a data movement instruction that loads $\nrows$ rows of data ($\nrows {\times} 128$ bytes) from the HBM ($\hbm{d_0}$) starting at address $\addin$, to the main scratchpad ($\spad{d_1}$) starting at row $\addout$.
The concrete instruction $\loadrm(\nrows {=} 4, \addin {=} 0, \addout {=} 2)$ loads 512 bytes from $\hbm{d_0[0{:}512]}$ to $\spad{d_1[2{:}6]}$.
Line 10 describes the validity constraint $\boldsymbol{e_\loadrm} = (n \le 128)$.
Finally, lines 12--16 represent the abstract tensor computation $\boldsymbol{g_\loadrm}$ using tensor operators (Table~\ref{tab:imp-ops}) and attribute $\nrows$ $\in \boldsymbol{\alpha_{\loadrm}}$.

\begin{figure}[!h]
  \vspace{0.5em}
  \begin{lstlisting}[style={qkv-isa},frame=single]
# Data Model
qkv.set_hbm("[!*\hbm{d0}*!]", size="1MB")                         # u8[2^20]
qkv.add_data_model("[!*\spad{d1}*!]", S0=[128], S1=[64], E="bf16") # bf16[128,64]
qkv.add_data_model("[!*\ibuf{d2}*!]", S0=[64], S1=[64], E="bf16")  # bf16[64,64]

# Instruction Semantics
instr = qkv.add_instr(name="[!*\textbf{\color{codegold}load\_rm}*!]", alpha=["[!*\textbf{\color{codegreen}n}*!]"], beta=["[!*\textbf{\color{codegreen}addr\_in}*!]", "[!*\textbf{\color{codegreen}addr\_out}*!]"])
instr.set_inputs([["[!*\hbm{d0}*!]", start=["[!*\textbf{\color{codegreen}addr\_in}*!]"], len=["[!*\textbf{\color{codegreen}n}*!] * 128"]]])  # u8[n*128]
instr.set_output(["[!*\spad{d1}*!]", start=["[!*\textbf{\color{codegreen}addr\_out}*!]"], len=["[!*\textbf{\color{codegreen}n}*!]"]])         # bf16[n,64]
instr.set_constraints(["[!*\textbf{\color{codegreen}n}*!] <= 128"])
# y = bitcvt(reshape(x1)) represented using XLA-HLO syntax
instr.set_abstract_computation("""ENTRY load_rm {
  x1 = u8[`[!*\textbf{\color{codegreen}n}*!] * 128`] parameter(0);
  a = u8[`[!*\textbf{\color{codegreen}n}*!]`, 64, 2] reshape(x1);
  ROOT y = bf16[`[!*\textbf{\color{codegreen}n}*!]`, 64] bitcast_convert(a);
}""")
\end{lstlisting}
  \vspace{-1.2em}
  \caption{Snippet of ISA description of $H_{QKV}$ in \idl{} with instruction attributes annotated as $\boldsymbol\alpha$ and $\boldsymbol\beta$.}
  \label{lst:qkv-isa}
  \vspace{-0.6em}
\end{figure}


\subsection{Problem Formulation: Compiler Backend Generator}
\label{subsec:problem-generator}

Finally, we formulate the problem statement of a compiler backend generator that generates sound and complete compiler backends from just the accelerator ISA descriptions in \idl{}.

A compiler backend for a tensor accelerator consumes a tiled kernel (\S\ref{subsec:problem-backend}) as the input and either (i) successfully compiles to emit an \emph{accelerator-specific assembly code} (\S\ref{subsec:problem-isa}), or (ii) does not compile, i.e., either a compilation failure or does not terminate.
We define the semantic equivalence between the tiled kernel and the compiled assembly code using bisimulation transformations, modulo the set of known tensor rewrites of XLA-HLO IR (as discussed in \S\ref{subsec:background-axioms}).

\textbf{Soundness:}
A compiler backend is said to be \emph{sound} if, for all successfully compiled tiled kernels, the emitted assembly code is semantically equivalent to the input tiled kernel.

\textbf{Completeness:}
A compiler backend is said to be \emph{complete} if it successfully compiles for all input tiled kernels that have a semantically equivalent assembly code.

The detailed definitions and formalisms of these concepts are deferred to Appendix~\ref{app:problem}.

\begin{center}
  \minibox[frame,c]{
    We design a compiler backend generator such that for any given ISA description in \idl{}, \\
    it generates an accelerator-specific compiler backend that is \textit{sound} and \textit{complete}.
  }
\end{center}

\section{Overview of \underline{A}ccelerator \underline{C}ompiler \underline{T}oolkit (\act{})}
\label{sec:overview}

We present a compiler backend generator --- \act{}, Accelerator Compiler Toolkit --- that generates a sound and complete compiler backend from a given tensor accelerator ISA description in \idl{}.\\
A key contribution of our work is a novel compilation algorithm parameterized by \idl{} constructs ($d_i, \theta, \alpha, \beta, g_\theta, h_\theta, e_\theta$, discussed in \S\ref{subsec:problem-isa}).
In this section, we provide a high-level overview of the ISA-parameterized algorithm (in \S\ref{subsec:overview-algo}-\S\ref{subsec:overview-novelty}) and outline the compiler backend generation process based on this algorithm (in \S\ref{subsec:overview-gen}).
We elaborate the details in \S\ref{sec:egraph} and \S\ref{sec:csp}.

\vspace{-0.5em}
\subsection{Overview of the ISA-parameterized Compilation Algorithm}
\label{subsec:overview-algo}

\act{}'s parameterized compilation algorithm is designed as a modular pipeline, with each module focusing on a specific aspect of the compilation.
The algorithms for each module are \textit{parameterized} by the \idl{} constructs, which are instantiated given an ISA description.

\begin{figure}[!h]
  \vspace{0.5em}
  \centering
  \includegraphics[width=0.95\textwidth]{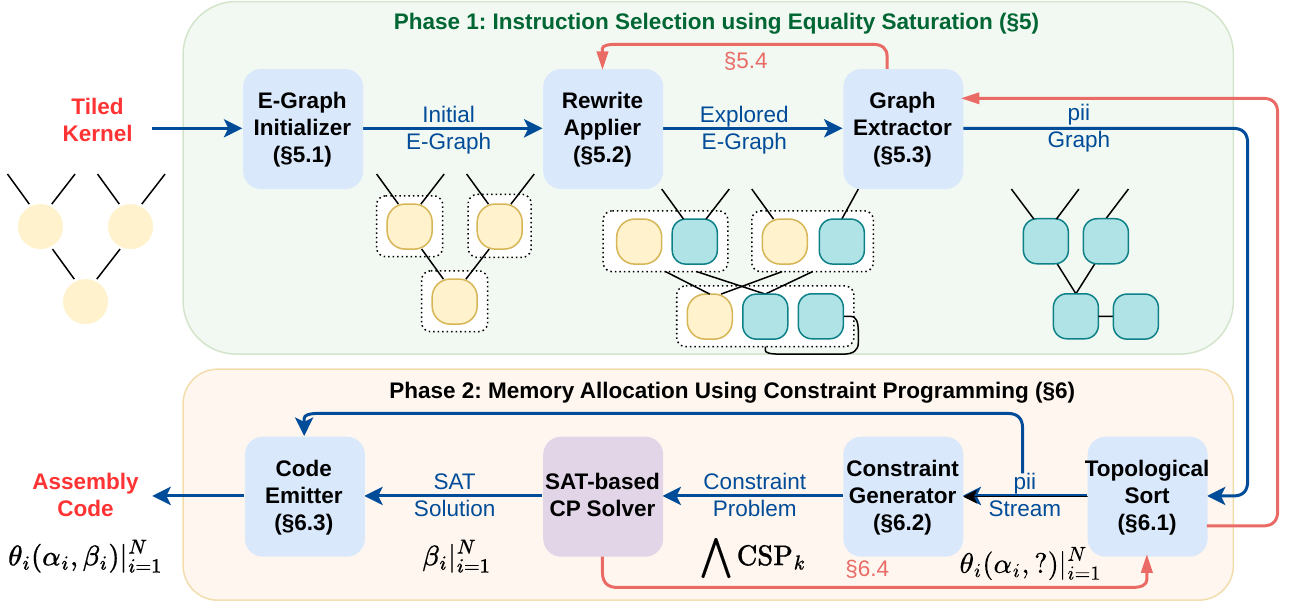}
  \vspace{-0.6em}
  \caption{Overview of \act{}'s parameterized compilation algorithm. Blue modules are parameterized by the \idl{} constructs (\S\ref{subsec:problem-isa}). The purple module is an external tool. The red edges represent the fallback mechanisms designed for completeness (proof in Appendix~\ref{app:proof}) and the enumeration pipeline (discussed later in \S\ref{sec:cost-model}).}
  \label{fig:overview}
  \vspace{-1em}
\end{figure}

Fig.~\ref{fig:overview} visualizes the high-level structure of \act{}'s ISA-parameterized compilation algorithm.
The algorithm consumes a tiled kernel and emits an assembly code.
The assembly code consists of concrete instructions $\theta_i(\alpha_i, \beta_i) |_{i=1}^N$, i.e., abstract instructions $\theta_i$ with initialized computational ($\alpha_i$) and addressing ($\beta_i$) attributes.
The algorithm consists of two phases: (1) initializing $\alpha_i$s, (i.e., \textit{instruction selection}), and (2) initializing $\beta_i$s, (i.e., \textit{memory allocation}).

\newpage

\vspace{0.2em}
\textbf{Phase 1: Tensor Instruction Selection}

\noindent
The first phase consumes the tiled kernel and emits a directed acyclic graph of \textit{partially} instantiated instructions (\pii{}), called a \pii{} graph.
A \textit{partially} instantiated instruction is an abstract instruction $\theta$ with $\alpha$ initialized and $\beta$ uninitialized, denoted as $\theta_{\alpha, ?}$.
This phase initializes $\alpha$, i.e., explores semantically equivalent instruction selection opportunities for the tiled kernel.
This is analogous to the instruction selection pass in CPU backends.
It is formulated as an equality saturation problem and utilizes an existing equality saturation engine~\cite{egg} to perform graph-level semantic-preserving substitutions.
The three modules required for equality saturation---e-graph initializer, rewrite applier, and graph extractor---are parameterized by the \idl{} constructs.

\vspace{0.2em}
\textbf{Phase 2: Tensor Memory Allocation}

\noindent
The second phase consumes the \pii{} graph and emits the final assembly code.
This phase initializes $\beta$, i.e., assigns addresses to the unmapped slices.
This is analogous to the register allocation pass in CPU backends.
First, the \pii{} graph is topologically sorted to respect instruction dependencies.
Then, these slices are mapped to the address space of the accelerator's tensor buffers.
This mapping must (1) maintain instruction dependencies, (2) assign non-overlapping address ranges to slices with overlapping live ranges, and (3) yield valid instruction attributes.
These are formulated as a constraint satisfaction problem over integers and passed to an existing SAT-based CP solver~\cite{ortools}.
Finally, the assembly code is emitted based on the solver's output.
The three modules---topological sort, constraint generator, and code emitter---are also parameterized by the \idl{} constructs.

\vspace{0.2em}
\textbf{Inter-phase fallback mechanism}

\noindent
Unlike commodity hardware, accelerator designs have constraints not captured during instruction selection, such as non-trivial addressing constraints and lack of spill-reload instructions for some tensor buffers (e.g., AWS Trainium~\cite{trainium}).
Therefore, the second phase may fail to find a valid solution for a \pii{} graph.
The compilation algorithm then falls back to the first phase, which extracts a different \pii{} graph or performs another iteration of rewrites.
This inter-phase mechanism, along with the intra-phase fallbacks, is essential for completeness.

\vspace{0.2em}
\textbf{Soundness.}
\act{}'s parameterized compilation algorithm is sound by construction (proof in Appendix~\ref{app:proof}).
Note that the soundness is defined modulo the foundational axioms of XLA-HLO IR (as discussed in \S\ref{subsec:background-axioms}).
This axiomatic approach is similar to prior works~\cite{taso, tensat, aerify, constable}.

\textbf{Completeness.}
\act{}'s parameterized compilation algorithm is also complete.
Intuitively, this is shown by proving that for any input tiled kernel with an equivalent assembly code, there exists a sequence of rewrites and a valid ordering of the \pii{} graph that leads to a successful compilation, and the fallback mechanisms ensure that all possible sequences of rewrites and orderings are explored and enumerable.
The detailed proof of completeness is provided in Appendix~\ref{app:proof}.

\textbf{Termination.}
For kernels with no equivalent assembly code, the compilation algorithm may not terminate, since completeness and guaranteed termination cannot both hold (proven by reduction from the halting problem in Appendix~\ref{app:problem}).
In our context of ML compiler backends, completeness is crucial to ensure that the middle-end does not fall back to host (CPU) code generation, avoiding the associated performance degradation (\S\ref{subsec:evaluation-perf-gemmini}).
In practice, we bound non-termination with a timeout.

\vspace{-0.5em}
\subsection{Novelty of the ISA-parameterized Compilation Algorithm}
\label{subsec:overview-novelty}

Code generation for tensor accelerators brings unique challenges (\S\ref{subsec:introduction-challenges}) that are not articulated nor solved by existing techniques.
Off-the-shelf rewrite synthesizers, e-graph extractors, and memory/register allocators do not handle parameterized tensor instructions, multi-dimensional scratchpad allocation, or complex layout-transforming ISA semantics.
Table~\ref{tab:novelty} summarizes the novel features of \act{}'s parameterized compilation algorithm that address these challenges.

\begin{table}[!ht]
  \vspace{-0.2em}
  \setlength{\tabcolsep}{4pt}
  \centering
  \begin{tabular}{||p{0.45\textwidth}|p{0.5\textwidth}||}
    \hline
    \multicolumn{1}{||c|}{\textbf{Limitations of existing techniques}} & \multicolumn{1}{c||}{\textbf{Novelty in \act{}}} \\
    \hline\hline
    Vector-ISA instruction selectors~\cite{vegen, isaria, diospyros} assume fixed-size instructions, unlike shape-parameterized tensor instructions.
     & \emph{Reduction to a two-phase algorithm} (\S\ref{subsec:overview-algo}) via a new \pii{} intermediary, splitting attributes into computational ($\alpha$, Phase 1) and addressing ($\beta$, Phase 2). \\
    \hline
    Typical rewrite synthesizers~\cite{enumo, misaal} generate rewrites \emph{without preconditions}, and cannot model parameterized instructions with computational attributes.
     & \emph{Preconditioned rewrite synthesis} (\S\ref{subsec:module-2}) that generates one IR-to-ISA rewrite per instruction, with \emph{preconditions} on computational attributes for parameterized instructions. \\
    \hline
    Existing accelerator backends and libraries do not ensure \emph{completeness} and may fail to compile kernels even when a valid assembly code exists, forcing host (CPU) fallback.
     & \emph{No-ops modeled as identity \pii{} nodes} (\S\ref{subsec:module-2}) for inner-loop tiling and padding, enabling provable discovery of \emph{all} equivalent instruction sequences and ensuring completeness. \\
    \hline
    One-shot extractors~\cite{egg, tensat, spores, smoothe} extract a \emph{single} graph, freely choosing at most one e-node per e-class, unable to handle fallback scenarios or e-node selection constraints.
     & \emph{Enumerative e-graph extraction} (\S\ref{subsec:module-3}) of \emph{all} equivalent \pii{} graphs for fallback scenarios, restricting e-nodes by previously selected ones and allowing \emph{multiple} e-nodes per e-class. \\
    \hline
    Register allocators (graph coloring~\cite{dragonbook}, puzzles~\cite{regpuzzle}) and scratchpad allocators~\cite{telamalloc, minimalloc} model only interference over 1-D sites, without accelerator addressing constraints.
     & \emph{CSP-based memory allocation} (\S\ref{subsec:module-5}) combining accelerator addressing constraints with live-range interference over multi-dimensional tensor buffers and related pruning strategies (\S\ref{subsec:fallback-csp}). \\
    \hline
  \end{tabular}
  \vspace{0.2em}
  \caption{Novel features of \act{}'s parameterized compilation algorithm.}
  \Description{Novel features of \act{}'s parameterized compilation algorithm.}
  \label{tab:novelty}
  \vspace{-2em}
\end{table}

\vspace{-0.5em}
\subsection{Overview of the Compiler Backend Generation Process}
\label{subsec:overview-gen}

\begin{figure}[!h]
  \vspace{0.5em}
  \centering
  \includegraphics[width=0.95\textwidth]{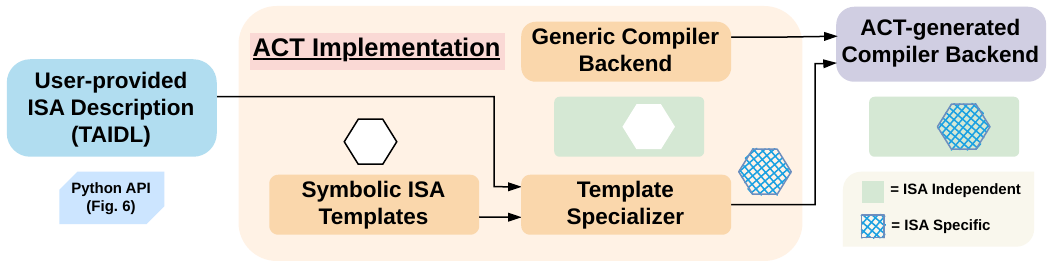}
  \vspace{-1em}
  \caption{Overview of \act{}'s compiler backend generation process. The orange region shows the implementation of the parameterized algorithm as a generic compiler backend (with missing holes) and symbolic ISA templates. At generation time, these symbolic templates are specialized using the user-provided ISA description.}
  \label{fig:generation}
  \vspace{-1em}
\end{figure}

\act{}'s ISA-parameterized compilation algorithm is implemented as a generic compiler backend with pure virtual functions and symbolic templates based on \idl{} constructs.
A user specifies an accelerator ISA in \idl{} (as shown in Fig.~\ref{lst:qkv-isa}).
\act{} processes this ISA description to instantiate a specialized compiler backend by filling in the symbolic templates with ISA-specific details.
This generation process is visualized in Fig.~\ref{fig:generation}.
This is similar to the design principle of classic compiler component generators such as Flex~\cite{flex} (for lexical analysis) and Yacc~\cite{yacc} / Bison~\cite{bison} (for parsing).

\newpage

The generation process runs once per accelerator and produces a standalone accelerator-specific compiler backend executable.
At compile-time, this generated backend executable is called once per tiled kernel.
Detailed correspondence with Flex and Bison is tabulated in Appendix~\ref{app:solution}.

More concretely, \act{} converts the \idl{} description into rewrite rules and preconditions for equality saturation (Phase 1).
It also generates ISA-specific implementations of abstract methods such as \texttt{get\_h0()} and \texttt{get\_e()}, corresponding to the \idl{} constructs $h_\theta^{(0)}$ and $e_\theta$.
The constraint problem (Phase 2) is formulated abstractly on the \idl{} constructs, without any hardcoded ISA assumptions.
This separation ensures that compiler backends are derived systematically from just the ISA description, without embedding ISA-specific logic in the compiler.

\vspace{0.2em}
\textbf{Engineering Effort per Accelerator.}
Existing \idl{} descriptions take around 10-12 lines per simple instruction (like Intel AMX~\cite{intel-amx} load/store, Intel AVX-512~\cite{intel-avx512} vector compute, Gemmini~\cite{gemmini}) and 12-15 lines per complex one (like Intel AMX~\cite{intel-amx} matrix compute, FEATHER~\cite{feather} layout reconfigurations), so generating a new accelerator-specific backend takes little effort.
These backends are also compatible with XLA, a production ML compiler, amplifying this reduction.

\vspace{-0.5em}
\section{Phase 1: Tensor Instruction Selection Using Equality Saturation}
\label{sec:egraph}

We formulate the tensor instruction selection problem as a search problem over the space of equivalent tensor computation graphs.
We use e-graphs~\cite{eqsat} to compactly represent this space of equivalent tensor computation graphs and enable efficient exploration for instruction selection.

\begin{wrapfigure}[17]{R}{0.35\textwidth}
  \vspace{-1.2em}
  \centering
  \includegraphics[width=0.35\textwidth]{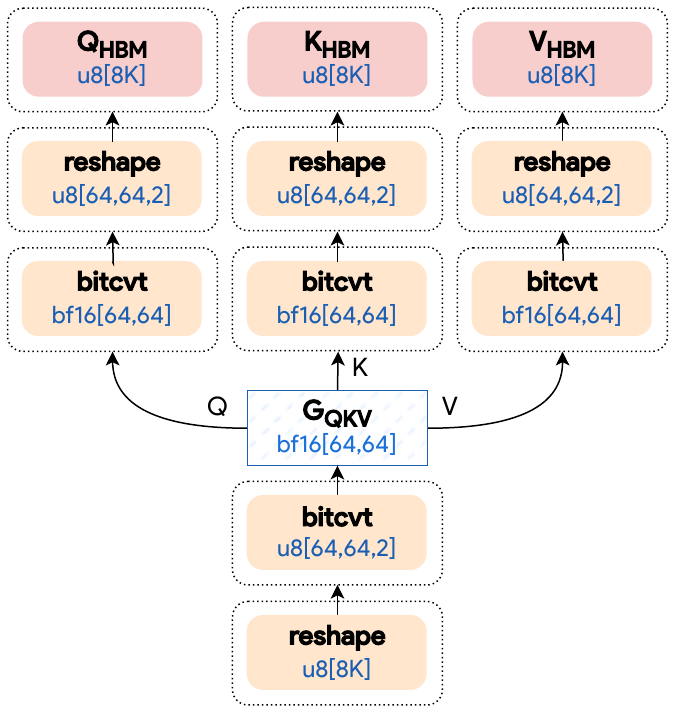}
  \vspace{-2em}
  \caption{The initial e-graph for $G_{QKV}$ [after Module 1]. $G_{QKV}$ (Fig.~\ref{fig:qkv-graph}) is not expanded here for compactness. Note that the edges are now flipped with the output tensor as the root e-node.}
  \label{fig:qkv-initial}
  \vspace{-3em}
\end{wrapfigure}

In this section, we focus on the novel features of Phase 1 highlighted in Table~\ref{tab:novelty} --- preconditioned IR-to-ISA rewrites and identity instructions (\S\ref{subsec:module-2}), as well as an enumerative e-graph extraction algorithm (\S\ref{subsec:module-3}-\S\ref{subsec:fallback-egraph}).
We use the step-wise lowering of the tiled kernel $G_{QKV}$ (Fig.~\ref{fig:qkv-graph}) for the accelerator $H_{QKV}$ (Fig.~\ref{fig:qkv-isa}) as an illustrative example.

\vspace{-0.5em}
\subsection{Module 1: E-Graph Initializer}
\label{subsec:module-1}

In our work, an e-graph is a directed graph where each e-node is a concrete tensor operator or a \emph{partially instantiated instruction} (\pii{} $\theta_{\alpha, ?}$, defined in \S\ref{subsec:overview-algo}).
An e-class is a set of e-nodes representing semantically equivalent computations rooted at the e-nodes.
By design, an e-node is optionally associated with a tensor buffer. Specifically, a \pii{} e-node $\theta_{\alpha, ?}$ is associated with the output tensor buffer of its instruction ($d_{\theta}^{(0)}$ notation from \S\ref{subsec:problem-isa}).
The e-graph is initialized with the input tiled kernel $G$ with a 1-D byte representation of the input and output tensors.
The 1-D byte representation for the $\mathsf{bf16}$ tensors requires bitcasting a $\mathsf{bf16}$ to two $\mathsf{u8}$s followed by a reshape operation.
The leaf e-nodes are assigned to the tensor buffer $\hbm{d_0}$, i.e., allocated on HBM, while the root is the output tensor, where extraction begins (\S\ref{subsec:module-3}).
Fig.~\ref{fig:qkv-initial} shows the initial e-graph for $G_{QKV}$ (Fig.~\ref{fig:qkv-graph}), without expanding the sub-graph of $G_{QKV}$.

\vspace{-0.7em}
\subsection{Module 2: Rewrite Applier}
\label{subsec:module-2}

The second module specifies the rewrites used in the exploration stage of equality saturation.
These include three types of rewrites:
(1) existing set of IR-to-IR tensor rewrites from the XLA compiler,
(2) new IR-to-ISA rewrites generated from the accelerator ISA description, and
(3) rewrites based on \emph{no-ops}, termed as \emph{identity instructions}, i.e., instructions that do not modify the memory state.

\vspace{0.2em}
\textbf{IR-to-IR rewrites from XLA compiler.}
\act{}-generated backends use the curated set of existing IR-to-IR rewrites from the XLA compiler (discussed in \S\ref{subsec:background-axioms}) to explore the space of equivalent tensor computation graphs.
Note that these 175+ rewrites are maintained by the XLA compiler developers and are not specific to any accelerator.
Furthermore, e-graphs can simultaneously explore different sequences of rewrites, avoiding the need to implement accelerator-specific IR legalization passes.

\begin{wrapfigure}[19]{R}{0.38\textwidth}
  \vspace{-1.5em}
  \centering
  \includegraphics[width=0.38\textwidth]{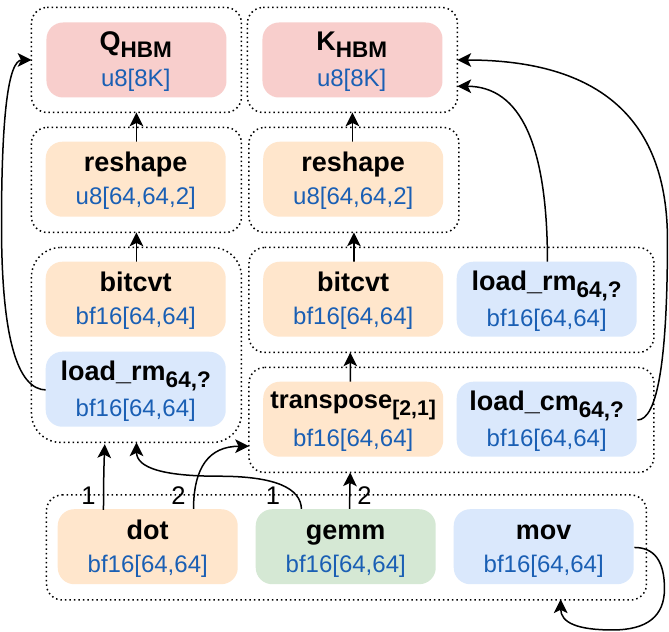}
  \vspace{-2em}
  \caption{Snippet of explored e-graph [after Module 2] after applying IR-to-ISA rewrites on the initial e-graph (Fig.~\ref{fig:qkv-initial}).
  Dotted boxes are e-classes, solid boxes are e-nodes, and edge labels \textsf{1}/\textsf{2} are operand indices.
  The new \pii{} nodes result from the \act{}-generated IR-to-ISA rewrites for $\loadrm$, $\loadcm$, $\gemm$, and $\mov$.}
  \label{fig:qkv-explored}
  \vspace{-1.5em}
\end{wrapfigure}

\vspace{0.2em}
\textbf{IR-to-ISA rewrites generated from ISA description.}
At generation time, \act{} generates new rewrites with preconditions using a pre-order traversal from the \texttt{ROOT} of the abstract tensor computation $g_\theta$ (line 15 of Fig.~\ref{lst:qkv-isa}).
These IR-to-ISA rewrites create new \pii{} nodes in the e-graph.
Combined with IR-to-IR rewrites, they replace manually-written accelerator-specific pattern-matching rules.

For instruction $\theta$, \act{} generates the rewrite $g_{\theta} {\rightarrow} \theta$ that substitutes a concrete tensor computation $g_{\theta}(\alpha)$ with \pii{} node $\theta_{\alpha, ?}$ under the precondition $\exists \beta \,, e_{\theta}(\alpha, \beta) = True$.
For example, rewrite $\bitcvt(\reshape(x)) {\rightarrow} \loadrm(x)$ is generated from Fig.~\ref{lst:qkv-isa}.
Fig.~\ref{fig:qkv-explored} shows a snippet of the explored e-graph of Fig.~\ref{fig:qkv-initial} with new \pii{} nodes colored as blue ($\spad{d_1}$) and green ($\ibuf{d_2}$).
Fig.~\ref{lst:loadrm-precondition} shows the precondition generated for $\loadrm$, which determines the computational attribute $\nrows$ at compile-time and checks the validity constraint $e_{\loadrm}$.
Most existing rewrite synthesizers~\cite{enumo, misaal} generate rewrites without such preconditions, requiring multiple rewrites to cover a parameterized instruction, whereas \act{} generates only \emph{one} IR-to-ISA rewrite per instruction.

\begin{wrapfigure}{R}{0.47\textwidth}
  \vspace{-1em}
  \begin{lstlisting}[style={loadrm-precondition},frame=single]
pub fn precond_load_rm(..) -> bool {
  // y: bf16[`@alpha.n`, 64]
  if y.shape.len() != 2
     || y.shape[1] != 64
     || y.dtype != Dtype::BF16 {
    return false;
  }
  let alpha_n = y.shape[0];
  if alpha_n > 128 {
    return false;
  }
  ... // Repeat for a, x1
}
\end{lstlisting}
  \vspace{-1.2em}
  \caption{Snippet of the rewrite precondition (Rust) generated from the semantics of $\loadrm$ in Fig.~\ref{lst:qkv-isa}. Line 8 determines the computational attribute $\nrows$.}
  \label{lst:loadrm-precondition}
  \vspace{-3em}
\end{wrapfigure}

\vspace{0.2em}
\textbf{Identity instructions (no-ops).}
We model \emph{no-ops} as identity instructions $\slice^H$ and $\concat^H$, i.e., instructions that do not modify the tensor buffers.
These are \pii{} nodes analogous to tensor operators $\slice$ and $\concat$ but with the constraint that the inputs and the output are on the same buffer.
$\slice^H$ applies only under a precondition (the $\slice$ dimensionality must match a tensor buffer, \S\ref{subsec:problem-isa}), and $\concat^H$ constrains its two operands to be allocated contiguously in the same tensor buffer so the next instruction reads them as one operand.
These identity instructions ensure exploration of all equivalent \pii{} graphs (proof in Appendix~\ref{app:proof}).

\subsection{Module 3: Graph Extractor}
\label{subsec:module-3}

The third module performs the extraction stage of the equality saturation workflow.
It extracts a directed acyclic graph of \pii{} nodes (i.e., a \pii{} graph) from the explored e-graph (after Module 2).

First, a bottom-up traversal (from leaf e-classes) is performed to detect e-classes representing compile-time constant tensors.
The e-classes with a leaf e-node, such as tensor operators \textit{eye}, \textit{iota}, are constants.
An e-class is a constant if any one of its e-nodes has all its operands as constants.
Such e-classes become base cases of the extraction below (line 6 of Fig.~\ref{lst:extraction-algo}).

\begin{wrapfigure}[20]{R}{0.5\textwidth}
  \vspace{-0.5em}
  \begin{lstlisting}[style={qkv-isa},frame=single]
def extract(ec, buf, limit):
  # node budget N_max exhausted
  if limit == 0: return {}

  # base cases: input or constant
  if ec.is_input or ec.is_const:
    return {leaf(ec)}

  graphs = {}
  for e in ec.enodes:
    # buffer-guided selection
    if e.out_buf != buf: continue
    children = []
    for i, c in enumerate(e.args):
      cbuf = e.in_buf[i]
      sub = extract(c, cbuf, limit-1)
      children.append(sub)
    # enumerate all combinations
    for combo in product(*children):
      graphs.add(pii(e, combo))
  return graphs

# Module 3 entry point
extract(egraph.root, HBM, N_max)
\end{lstlisting}
  \vspace{-1.2em}
  \caption{Top-down traversal to extract all \pii{} graphs following buffer constraints within the $N_{\mathsf{max}}$ node limit.}
  \label{lst:extraction-algo}
  \vspace{-15em}
\end{wrapfigure}

Then, a top-down traversal (from root e-class) is performed to extract the \pii{} graphs.
It is guided by the input and output tensor buffers ($d_{\theta}^{(i)} |_{i=0}^{n_\theta}$ notation from \S\ref{subsec:problem-isa}) associated with each \pii{} e-node.
Fig.~\ref{lst:extraction-algo} presents the pseudocode --- $\mathsf{extract}(EC, d, N)$ returns \emph{all} \pii{} graphs computing e-class $EC$ into tensor buffer $d$ within $N$ nodes, invoked at the root with $d = \hbm{d_0}$ (line 24).
Only the e-nodes whose output tensor buffer $d_{\theta}^{(0)}$ matches the requested $d$ are admissible (line 12), and each of them in turn requests $d_{\theta}^{(i)}$ for its $i$-th operand (lines 15--16).
For example, $\spad{\loadcm_{64,?}}$ is admissible for the buffer that $\ibuf{\gemm}$ requests for its second operand in Fig.~\ref{fig:qkv-explored}, since $d_{\loadcm}^{(0)} = \spad{d_1} = d_{\gemm}^{(2)}$.
The graphs of an admissible e-node are the Cartesian product of its operands' graph sets (line 19), and those of an e-class are the union over its admissible e-nodes.
The traversal terminates upon reaching an input variable leaf e-node or a constant e-class, or hitting a \pii{} node limit $N_{\mathsf{max}}$ (line 3).

Classical one-shot extractors~\cite{egg, tensat, spores, smoothe} freely select at most one e-node per e-class without any e-node constraints.
Our extraction differs on both counts --- selection is constrained by the buffer requested by the parent e-node, and an e-class expanded under several requested buffers contributes \emph{multiple} of its e-nodes to one \pii{} graph.
Enumerating \emph{all} of them is necessary rather than merely thorough, since Phase 2 can fail for a given \pii{} graph (\S\ref{subsec:overview-algo}) and its cost is known only after code generation (\S\ref{sec:cost-model}).
Without our novel extraction algorithm, most kernels (> 95\%) fail to compile for popular tensor accelerator ISAs such as AWS NKI~\cite{nki} and Gemmini~\cite{gemmini}.

Fig.~\ref{fig:qkv-pii} (a) shows the first extracted \pii{} graph from the explored e-graph of $G_{QKV}$ (Fig.~\ref{fig:qkv-explored}).

\vspace{-0.5em}
\subsection{Intra-phase Fallback Mechanism: Enumerating All Equivalent \pii{} Graphs}
\label{subsec:fallback-egraph}

\noindent
The exploration and extraction stages of equality saturation may not terminate for all tiled kernels.
The exploration stage iteratively applies the rewrites but may not reach saturation for all tiled kernels.
The explored e-graph may have loops, and infinite equivalent \pii{} graphs may exist.

To ensure that all equivalent \pii{} graphs are enumerable without getting stuck within an infinite loop of either of the two stages, we use a bounded approach of exploration and extraction.
We iteratively increase the \pii{} node limit $N_{\mathsf{max}}$ for the extraction stage with the iteration count $k$ of the exploration stage, i.e., $N_{\mathsf{max}} = \Gamma(k)$ where $\Gamma$ is a strictly monotonically increasing function of $k$ (e.g., $\Gamma(k) = 2^k$).
For every value of $k$, we extract a finite number of \pii{} graphs, and thus, this set is enumerable.
This is analogous to enumerating $(a,b) \in \mathbb{N}^2$ using the Cantor pairing function~\cite{cantor-pairing}.

\section{Phase 2: Tensor Memory Allocation using Integer Constraint Programming}
\label{sec:csp}

The second phase of the compilation pipeline is responsible for compiling the generated \pii{} graph down to accelerator-specific assembly code.
This phase involves scheduling the \pii{} graph and allocating the intermediate tensors on the tensor buffers.
We formulate the memory allocation as an integer constraint satisfaction problem (CSP) and solve it using a SAT-based CP solver.

In this section, we focus on the novel features of Phase 2 highlighted in Table~\ref{tab:novelty} --- the parameterized multi-dimensional CSP formulation (\S\ref{subsec:module-5}) and interference-based pruning (\S\ref{subsec:fallback-csp}).

\vspace{-0.5em}
\subsection{Module 4: Topological Sort}
\label{subsec:module-4}

\begin{figure}[!h]
  \vspace{-0.5em}
  \begin{minipage}{0.51\textwidth}
    \small
    \centering
    \includegraphics[width=\linewidth]{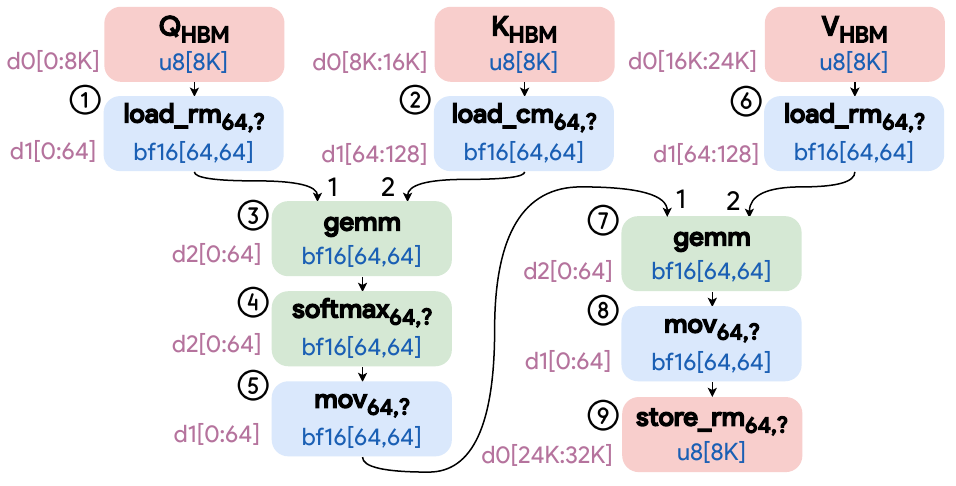}
    \begin{center}
      (a)
    \end{center}
  \end{minipage}
  \hfill
  \begin{minipage}{0.48\textwidth}
    \small
    \centering
    ~\\
    \begin{tabular}{|l|}
      \hline
      $\loadrm(\nrows = 64, \addin = 0, \addout = 0)$                 \\
      $\loadcm(\nrows = 64, \addin = 8192, \addout = 64)$             \\
      $\gemm(\mathsf{addr}_1 = 0, \mathsf{addr}_2 = 64, \addout = 0)$ \\
      $\softmax(\nrows = 64, \mathsf{addr} = 0)$                      \\
      $\mov(\nrows = 64, \addin = 0, \addout = 0)$          \\
      $\loadrm(\nrows = 64, \addin = 16384, \addout = 64)$            \\
      $\gemm(\mathsf{addr}_1 = 0, \mathsf{addr}_2 = 64, \addout = 0)$ \\
      $\mov(\nrows = 64, \addin = 0, \addout = 0)$          \\
      $\storerm(\nrows = 64, \addin = 0, \addout = 24576)$            \\
      \hline
    \end{tabular}
    \begin{center}
      ~\\
      (b)
    \end{center}
  \end{minipage}
  \vspace{-1em}
  \captionof{figure}{(a) Extracted \pii{} graph annotated with topological ordering (circled) [after Module 4] and assigned tensor data slices (purple) [after Module 5]. (b) The compiled assembly code for $G_{QKV}$ [after Module 6]. \\ Input tensors $Q, K, V$ occupy relative HBM addresses $0, 8\mathrm{K}, 16\mathrm{K}$ (as determined by the XLA middle-end).}
  \label{fig:qkv-pii}
  \vspace{-1.5em}
\end{figure}

Instruction scheduling needs to adhere to \textit{def-use} edges of the \pii{} graph and perform a topological ordering of the nodes.
Our priority is to minimize the live ranges of the intermediate tensors and reduce the possibility of memory allocation failing.
Sethi-Ullman algorithm~\cite{sethi-ullman} generates optimal code for arithmetic binary trees with minimal registers.
We extend the Sethi-Ullman numbering to multiple tensor buffers storing varying sizes of tensor variables, and order the operands of each \pii{} node by decreasing tensor buffer requirement.
We defer the extended numbering to Appendix~\ref{app:solution}.
Fig.~\ref{fig:qkv-pii} (a) shows the topological ordering of the extracted \pii{} graph annotated with circled numbers.

\vspace{-0.5em}
\subsection{Module 5: Constraint Satisfaction Problem Generator}
\label{subsec:module-5}

By this stage, we have a \pii{} graph and a topological ordering of \pii{} nodes $\theta^{(i)}_{\alpha^{(i)},?} |_{i=1}^{N}$.
Next step is to instantiate the addressing attributes $\beta^{(i)}$ of the \pii{} nodes, i.e., allocate the intermediate and constant tensors on the tensor buffers.
We formulate a constraint satisfaction problem (CSP) by computing the interference graph and incorporating addressing constraints of the tensor buffers.

The CSP is parameterized by \idl{} constructs (from \S\ref{subsec:problem-isa}) and consists of four components:
(1) instruction validity constraints ($e_\theta$),
(2) def-use edge constraints over the data slice addresses from the address map $h_\theta$,
(3) HBM addressing constraints for the input and output tensors in the input tiled kernel $G$, whose addresses the XLA middle-end pre-allocates, and
(4) live range interference constraints.
The final CSP formulation is summarized in Fig.~\ref{fig:csp}.

\begin{figure}[!h]
  \vspace{0.5em}
  \[
    \begin{aligned}
      \csp_1 &= \bigwedge_{i=1}^{N} \Big( e_{\theta^{(i)}}(\alpha^{(i)}, \beta^{(i)}) \wedge (I_s^{(i)}, I_e^{(i)}) = h^{(0)}_{\theta^{(i)}}(\beta^{(i)}) \Big) \\
      \csp_2 &= \bigwedge_{(V^{(y)}, V^{(x_i)})} (I_s^{(x_i)}, I_e^{(x_i)}) = h^{(i)}_{\theta^{(y)}}(\beta^{(y)}) \\
      \csp_3 &= \Big(\bigwedge_{i=1}^{N_V} (I_s^{(N+i)}, I_e^{(N+i)}) = (I_s^{(G_{x_i})}, I_e^{(G_{x_i})})\Big)
      \wedge (I_s^{(N)}, I_e^{(N)}) = (I_s^{(G_y)}, I_e^{(G_y)}) \\
      \mathsf{live\_range}(i) &=
      \begin{cases}
        [0, N], & V^{(i)}\ \text{is an input variable node}, \\
        [0, b_i - 1], & V^{(i)}\ \text{is a constant leaf node}, \\
        [a_i, b_i - 1], & V^{(i)}\ \text{is an intermediate node}
      \end{cases} \\
      \phi &= \{(i,j) \mid d_{\theta^{(i)}}^{(0)} = d_{\theta^{(j)}}^{(0)}\ \wedge\ \mathsf{live\_range}(i)\cap\mathsf{live\_range}(j)\neq\emptyset\} \\
      \csp_4 &= \bigwedge_{(i,j)\in\phi}\mathsf{disjoint}(i,j) \\
      \csp &= \csp_1 \wedge \csp_2 \wedge \csp_3 \wedge \csp_4
    \end{aligned}
  \]
  \vspace{-1.6em}
  \caption{Summary of the CSP formulation used in Module 5. Notations used: $N$ is the number of \pii{} nodes, $N_V$ is the number of input variables in the tiled kernel $G$, $V^{(i)}$ denotes node $i$, $e_{\theta^{(i)}}$ is the instruction validity constraint, $h_{\theta^{(i)}}$ denotes the data slice address map, $(I_s^{(i)}, I_e^{(i)})$ are data slice addresses, $a_i$ is the topological number of node $i$, $b_i$ is the maximum topological number among uses of node $i$, $d_{\theta^{(i)}}^{(0)}$ is the output tensor buffer of node $i$, and $\mathsf{disjoint}(i,j)$ enforces non-overlapping slice addresses for nodes $i$ and $j$.}
  \Description{Summary of the CSP formulation used in Module 5.}
  \label{fig:csp}
\end{figure}

The final CSP is passed to an existing SAT-based CP solver library~\cite{ortools} to find a satisfying solution over the addressing attributes $\beta^{(i)}$.
Fig.~\ref{fig:qkv-pii} (a) shows the data slice addresses assigned to \pii{} nodes as determined by solving the constraint problem.

\vspace{0.2em}
\textbf{Implementation of Module 5.}
Let's look at a concrete example of template-driven implementation (discussed in \S\ref{subsec:overview-gen}) of the \textit{parameterized} algorithm of Module 5.
It consists of a generic skeleton (Fig.~\ref{lst:csp1}) with symbolic templates (Fig.~\ref{lst:template-theta}) specialized at generation-time (Fig.~\ref{lst:template-loadrm}).
At compile time, the CSP builder calls these specialized functions at each \pii{} node.

\begin{figure}[!h]
  \vspace{-0.2em}
  \begin{lstlisting}[style={csp1},frame=single]
void CSP1(const std::vector<INSTR*>& instructions) {
  for (const auto& instr : instructions) {
    solver.AddConstraint(instr->get_e());
    solver.MakeEquality(instr->output->start_addr, instr->get_h0().first);
    solver.MakeEquality(instr->output->end_addr, instr->get_h0().second);
  }
}
\end{lstlisting}
  \vspace{-1.3em}
  \caption{Snippet of Generic implementation of Module 5 using abstract C++ class \texttt{\color{codered}INSTR} for ISA construct $\theta$ with pure virtual functions \texttt{\color{codered}get\_h0()} and \texttt{\color{codered}get\_e()} representing ISA constructs $h_\theta^{(0)}$ and $e_\theta$ respectively,}
  \label{lst:csp1}
  \vspace{-0.5em}
\end{figure}

\begin{figure}[!h]
  \begin{lstlisting}[style={template-theta},frame=single]
class INSTR_<<str [!*\textbf{\color{codegold}instr.name}*!]>> : protected INSTR { ... }
std::pair<IntExpr*, IntExpr*> INSTR_<<str [!*\textbf{\color{codegold}instr.name}*!]>>::get_h0() override {
  return {<<ortools::IntExpr* [!*\textbf{\color{codegreen}instr.output.start}*!]>>, MAKE_SUM(<<ortools::IntExpr* [!*\textbf{\color{codegreen}instr.output.start}*!]>>, <<int64 [!*\textbf{\color{codegreen}instr.output.len}*!]>>)};
}
\end{lstlisting}
  \vspace{-1.2em}
  \caption{Symbolic Template for ISA construct $h_\theta^{(0)}$.}
  \label{lst:template-theta}
  \vspace{-0.5em}
\end{figure}

\begin{figure}[!h]
  \begin{lstlisting}[style={template-loadrm},frame=single]
class INSTR_[!*\textbf{\color{codegold}load\_rm}*!] : protected INSTR { ... }
std::pair<IntExpr*, IntExpr*> INSTR_[!*\textbf{\color{codegold}load\_rm}*!]::get_h0() override {
  return {[!*\textbf{\color{codegreen}addr\_out}*!], MAKE_SUM([!*\textbf{\color{codegreen}addr\_out}*!], [!*\textbf{\color{codegreen}n}*!])};
}
\end{lstlisting}
  \vspace{-1.2em}
  \caption{Specialized Template for $h_{\loadrm}^{(0)}$ generated by processing line 9 of Fig.~\ref{lst:qkv-isa}.}
  \label{lst:template-loadrm}
  \vspace{-1em}
\end{figure}

\vspace{-0.5em}
\subsection{Module 6: Code Emitter}
\label{subsec:module-6}

The final module sequentially emits the sorted \pii{} nodes with their solved addressing attributes and discarding identity instructions (no-ops).
Fig.~\ref{fig:qkv-pii} (b) shows the emitted assembly code for $G_{QKV}$.

\subsection{Intra-phase Fallback Mechanism}
\label{subsec:fallback-csp}

The depth-first topological ordering generated by the heuristic-based algorithm (\S\ref{subsec:module-4}) is not sufficient for the completeness of Phase 2.
We prove this using a counterexample in Appendix~\ref{app:solution}.
The tensor accelerator instructions and the identity instructions ($\slice^H$ \& $\concat^H$) often have stricter constraints ($e_\theta$) on the data slice addresses than what classical register allocation algorithms support.
Theoretically, we need to search through all topological orderings for completeness.

\vspace{0.2em}
\textbf{Pruning technique.}
We observe that $\csp_4$ is monotonic w.r.t. the interference graph, i.e., $\phi_1 \subseteq \phi_2 \implies \neg \csp_4(\phi_1) \rightarrow \neg \csp_4(\phi_2) \implies \neg \csp(\phi_1) \rightarrow \neg \csp(\phi_2)$.
Therefore, we discard any topological ordering with a sub-interference graph of a previously failed topological ordering.
This reduces the exponential search space of topological orderings to a small number of unique interference graphs, which is tractable to search through if the heuristic-based ordering fails.
Exhaustively searching these topological orderings guarantees the completeness of Phase 2.

\vspace{-0.5em}
\section{Code Generation with Cost Models}
\label{sec:cost-model}

In the previous sections, we have discussed the core algorithm of \act{} with fallback mechanisms that guarantee completeness, i.e., find an equivalent assembly code if one exists.
We observe that the first equivalent assembly code may not be performant enough.
Therefore, we designed a code generator with a simple enumeration pipeline using cost models to generate performant code.

\begin{figure}[!h]
  \vspace{-0.5em}
  \centering
  \begin{minipage}{0.43\textwidth}
    \centering
    \begin{lstlisting}[style={outer-algo},frame=single]
def codegen(G, n=2):
  # (asm, cost, enumeration time)
  final = (None, inf, 0)
  for asm in [!*{\bfseries\ttfamily\color{codegreen}act\_enum}*!](G):
    # Compare candidate asm
    time = now()
    cost = cost_model(asm)
    if cost < final.cost:
      # Select candidate asm
      final = (asm, cost, time)
    elif time > n * final.time:
      # Termination heuristic
      return final
\end{lstlisting}
    \vspace{-0.2em}
    \small{(a) Candidate selection using \\
      performance cost model}
  \end{minipage}
  \hfill
  \begin{minipage}{0.56\textwidth}
    \centering
    \begin{lstlisting}[style={act-algo},frame=single]
def [!*{\bfseries\ttfamily\color{codegreen}act\_enum}*!](G):
  egraph = init(G);             # Module1
  while not egraph.saturated:   # [!*{\color{codepurple}\S5.4}*!]
    egraph = egraph.applier(k)  # Module2
    piis = egraph.extract([!*$\Gamma$*!](k)) # Module3
    for pii in piis:            # [!*{\color{codepurple}\S4.1}*!]
      schedules = topo(pii);    # Module4
      for schd in schedules:    # [!*{\color{codepurple}\S6.4}*!]
        csp = csp_gen(schd);    # Module5
        sol = solve(csp)        # CSP Solver
        if sol.SAT:
          asm = emit(pii, sol)  # Module6
          yield asm             # Enumerate
\end{lstlisting}
    \vspace{-0.2em}
    \small{(b) Candidate enumeration function using\\
      \act{}-generated compiler backend}
  \end{minipage}
  \vspace{-0.8em}
  \caption{Pseudocode for code generator using cost models to generate performant code}
  \label{fig:autotuning}
  \vspace{-1em}
\end{figure}

The code generator (Fig.~\ref{fig:autotuning}) uses \act{}-generated compiler backend as an enumerator to generate multiple equivalent assembly candidates.
A simple candidate selection loop iterates through these candidates and selects the final assembly code with the least cost using a performance cost model.

\textbf{Candidate enumeration.}
Fig.~\ref{fig:autotuning} (b) shows the pseudocode of the enumerator function based on the \act{} algorithm (described in Fig.~\ref{fig:overview}) with \textsf{yield} statement\footnote{\url{https://docs.python.org/3/reference/expressions.html\#yield-expressions}} at line 13.
The fallback mechanisms (lines 3, 6, 8) give control back to the previous module.
This allows the compiler backend to continue after a successful compilation and repeatedly \emph{yield} equivalent assembly code.

\textbf{Candidate selection.}
Fig.~\ref{fig:autotuning} (a) shows the code generator that iterates through the candidate assembly codes (line 4) enumerated by the \act{}-generated compiler backend (Fig.~\ref{fig:autotuning} (b)).
Performance statistics of these candidates are collected (line 7), and the candidate with the best performance (least execution cost) is selected (line 8) as the final assembly code.

\textbf{Termination heuristic.}
We design a termination heuristic that stops the enumeration if the performance of the enumerated candidates doesn't improve for a considerable time (Fig.~\ref{fig:autotuning} (a) line 11).
Such a heuristic is required since there are, theoretically, infinite equivalent assembly code candidates with redundant load-store pairs.
However, redundant data movement makes such candidates longer than ideal and less performant.
The first phase prioritizes shorter assembly codes (\S\ref{subsec:fallback-egraph}), and thus, such candidates are likely to be enumerated after their ideal counterparts.
Therefore, we expect the performance to saturate after the ideal candidates are enumerated.
In \S\ref{subsec:evaluation-perf-amx}, we show that the code generators match the performance of state-of-the-art kernel libraries like the oneDNN library~\cite{onednn-library}, with the observed compilation time less than 100 ms.

\textbf{Choice of Cost Model.}
Performance statistics can be collected using different fidelity cost models, including cycle-accurate simulators~\cite{verilator}, analytical models~\cite{timeloop}, and learned cost models~\cite{ithemal}.
Note that the enumeration pipeline and the \act{} algorithm are \emph{agnostic to the choice of cost model}.

\section{Evaluation}
\label{sec:evaluation}

We performed three sets of evaluations to demonstrate the effectiveness of \act{}.
\begin{itemize}[noitemsep, nolistsep]
  \item \textbf{Supported Accelerators (\S\ref{subsec:evaluation-support}):} We generated seven compiler backends for diverse tensor accelerator designs from industry and academia, demonstrating generality of our approach.
  \item \textbf{Comparative Evaluations (\S\ref{subsec:evaluation-perf}--\S\ref{subsec:evaluation-perf-gemmini}):} We evaluated (a) the compilation coverage of \act{}-generated compiler backends and (b) performance of their compiled code, against existing accelerator-specific backends and expert-written kernel libraries for tensor accelerators.
  \item \textbf{Compilation Time Analysis (\S\ref{subsec:evaluation-time}):} We performed breakdown analyses of the compilation time for varying tiled kernel sizes from synthetic and real-world workloads.
\end{itemize}

\vspace{-0.5em}
\subsection{Implementation}

We used egg~\cite{egg} for equality saturation (\S\ref{sec:egraph}) and Google OR-Tools~\cite{ortools} for solving CSPs (\S\ref{sec:csp}).
\act{} is implemented as a generic compiler backend (1.9 KLOC of C++ \& 6.7 KLOC of Rust), symbolic ISA templates (64 LOC of C++ \& 68 LOC of Rust), and a template specializer (1.7 KLOC of Python).
The ISA descriptions are parsed by the ANTLR-based \idl{} front-end~\cite{taidl}, which we reuse.

\vspace{-0.5em}
\subsection{Supported Accelerators}
\label{subsec:evaluation-support}

\begin{figure}[!h]
  \centering
  \vspace{1em}
  \begin{minipage}{0.57\textwidth}
    \centering
    \begin{tabular}{||c|c|c|c||}
      \hline
      Tensor                       & Existing & Existing & \act{}  \\
      Accelerator                  & Backend  & Library  & Backend \\
      \hline\hline
      AWS Trainium~\cite{trainium} & \cmark   & \cmark   & \cmark  \\
      Intel AMX~\cite{intel-amx}   & \cmark   & \cmark   & \cmark  \\
      Google TPUv1~\cite{tpuv1}    & \xmark   & \xmark   & \cmark  \\
      Gemmini~\cite{gemmini}       & \xmark   & \cmark   & \cmark  \\
      FEATHER~\cite{feather}       & \xmark   & \xmark   & \cmark  \\
      EVA~\cite{eva}               & \xmark   & \xmark   & \cmark  \\
      $H_{QKV}$                    & \xmark   & \xmark   & \cmark  \\
      \hline
    \end{tabular}
    \vspace{0.6em}
    \captionof{table}{The seven tensor accelerators supported by \act{}.
      \xmark ~ refers to the absence of publicly available accelerator-specific compiler backends or expert-written kernel libraries.}
    \label{tab:accelerators}
    \vspace{-1.2em}
  \end{minipage}
  \hfill
  \begin{minipage}{0.38\textwidth}
    \centering
    \vspace{-3em}
    \includegraphics[width=\textwidth]{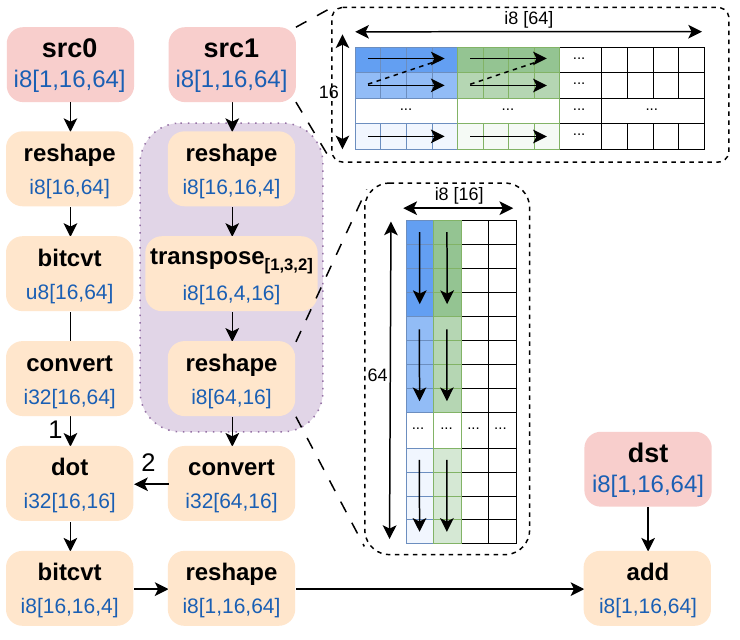}
    \vspace{-2em}
    \caption{Highlighted region shows the layout transformation applied to tile register \textsf{src1} in Intel AMX instruction \tmul{}.}
    \label{fig:amx-tmul}
    \vspace{-1.2em}
  \end{minipage}
\end{figure}

\idl{}~\cite{taidl} provides ISA descriptions for diverse accelerator designs from industry~\cite{trainium, intel-amx, tpuv1} and academia~\cite{tpuv1, gemmini, feather}.
We extend these \idl{} descriptions by annotating the instruction attributes as computational ($\alpha$) or addressing ($\beta$) --- a one-line change to the 10--15 lines of the original \idl{} descriptions for each instruction (e.g., line 7 of Fig.~\ref{lst:qkv-isa}).
Including the running example $H_{QKV}$ implemented using Allo~\cite{allo}, we have a total of \textbf{seven} accelerators tabulated in Table~\ref{tab:accelerators}.

We use \act{} to generate compiler backends for these seven accelerators, showing \emph{generality} of our approach.
Since \act{}'s compilation algorithm is parameterized by \idl{} constructs, it handles diverse design features of these accelerators with no changes to the core compilation algorithm.

\noindent
We list the notable design features below.

\begin{enumerate}[noitemsep, nolistsep]
  \item \textbf{AWS Trainium}~\cite{trainium} includes two on-chip scratchpads with 2-dimensional memory~\cite{nki-dm}, organized in 128 partitions. AWS provides a virtual ISA, AWS NKI~\cite{nki}, to program Trainium.
  \item \textbf{Intel AMX}~\cite{intel-amx} includes 2-dimensional registers (tiles) and supports instructions with complex layout transformations like \tmul{} (visualized in Fig.~\ref{fig:amx-tmul}), which are semantically modeled in \idl{} as a series of reshape and transpose operators (purple region).
  \item \textbf{Google TPUv1}~\cite{tpuv1} supports instructions with variable-sized inputs ($B {\times} 256$).
  \item \textbf{Gemmini}~\cite{gemmini} supports instructions to reconfigure dataflow (output vs weight-stationary).
  \item \textbf{FEATHER}~\cite{feather} supports instructions to reconfigure input data layouts.
  \item \textbf{EVA}~\cite{eva} is a CGRA-based accelerator with evolvable ISA and 3-dimensional addressing.
  \item $\boldsymbol{H_{QKV}}$ is implemented using an accelerator design language (ADL), Allo~\cite{allo}.
\end{enumerate}

\vspace{0.5em}
\noindent
\textbf{Execution of Compiled Assembly Code}

Revisiting \S\ref{subsec:problem-backend}, we discussed that \act{}-generated compiler backends emit code in the form of a sequence of \idl{} instructions, which are then stitched together with control flow for tiling into a single \emph{stitched} IR.
This stitched IR includes MLIR dialects~\cite{mlir} such as \texttt{scf} and \texttt{arith}.

Depending on the host-accelerator interface, the stitched IR is compiled and deployed differently.

(i) For built-in accelerators such as Intel AMX~\cite{intel-amx} and Gemmini~\cite{gemmini}, the stitched IR is compiled into an executable using the host CPU's code generation (e.g., x86 backend). In this case, the \texttt{scf.for} construct is compiled to the host's control flow instructions (e.g., branch and jump instructions).

(ii) For dedicated accelerators such as AWS Trainium~\cite{trainium} and FEATHER~\cite{feather}, the stitched IR is translated to the accelerator's kernel programming interface (e.g., AWS Neuron Kernel Interface (NKI)~\cite{nki} for AWS Trainium~\cite{trainium}). In this case, the \texttt{scf.for} construct is directly mapped to the kernel programming interface's control flow constructs (e.g., \texttt{for} loop construct in AWS NKI~\cite{nki-cf}).

This mapping is automatically inferred from the \idl{} description of the accelerator's loop construct. The stitching implementation is otherwise generic across accelerators.

Note that regardless of the host-accelerator interface, \act{}-generated compiler backends focus on tiled kernels in XLA-HLO IR and emit assembly code as per the ISA description in \idl{}.

\vspace{-0.5em}
\subsection{Comparative Evaluations}
\label{subsec:evaluation-perf}

Out of the seven tensor accelerators supported by \act{}, only three have a publicly available accelerator-specific compiler backend or expert-written kernel library for comparative evaluations.

\begin{itemize}[noitemsep, nolistsep]
  \item \textbf{AWS Trainium (\S\ref{subsec:evaluation-perf-nki}):} \act{}-generated backend (\act{}-NKI) vs. \texttt{neuronx-cc}~\cite{neuronx-cc} (backend)
  \item \textbf{Intel AMX (\S\ref{subsec:evaluation-perf-amx}):} \act{}-generated backend (\act{}-AMX) vs. oneDNN library~\cite{onednn-library}
  \item \textbf{Gemmini (\S\ref{subsec:evaluation-perf-gemmini}):} \act{}-generated backend (\act{}-Gemmini) vs. Exo~\cite{exo} and SW library~\cite{gemmini-github}
\end{itemize}

\begin{figure}[!h]
  \vspace{-0.5em}
  \centering
  \includegraphics[width=0.98\textwidth]{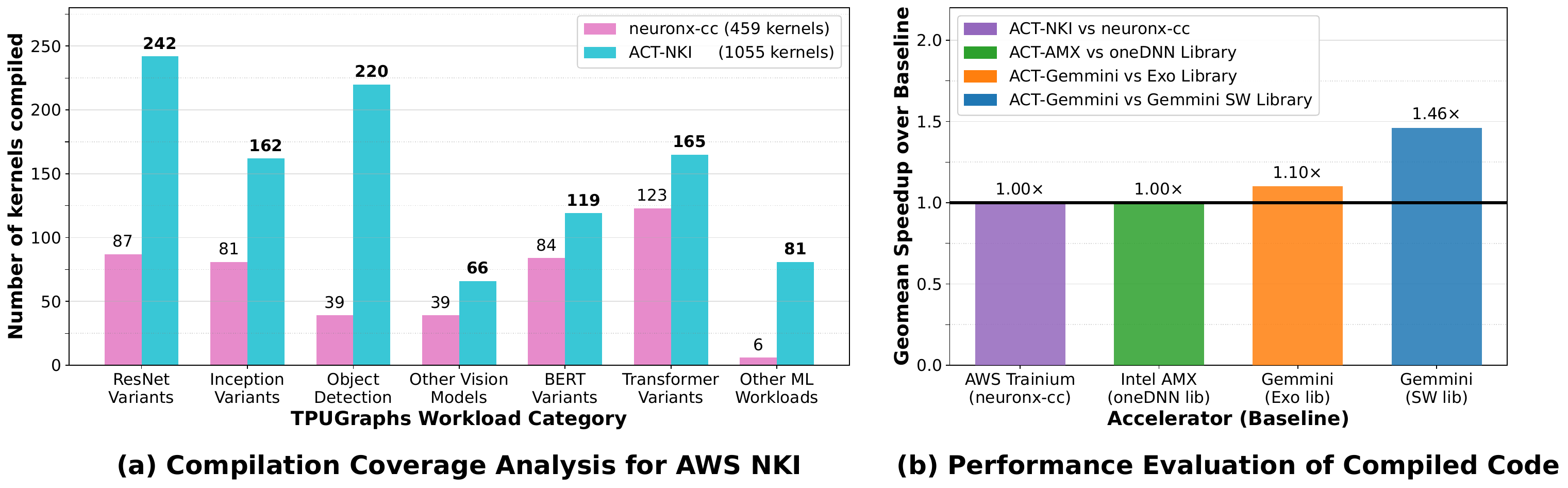}
  \vspace{-1em}
  \caption{
    Summary of comparative evaluations. (a) Number of TPUGraphs kernels correctly compiled by \act{}-NKI and \texttt{neuronx-cc}, grouped by workload category. (b) Geomean speedup of the compiled code by \act{}-generated compiler backends against their corresponding baselines.
  }
  \label{fig:perf-summary}
\end{figure}

\noindent
Here, we present the notable results with detailed analyses for each accelerator in \S\ref{subsec:evaluation-perf-nki}--\S\ref{subsec:evaluation-perf-gemmini}.

\textbf{(a) Compilation Coverage}\\
Fig.~\ref{fig:perf-summary} (a) shows the number of tiled kernels in TPUGraphs~\cite{tpugraphs} Tile dataset that can be compiled to AWS NKI by \act{}-NKI and AWS's compiler \texttt{neuronx-cc}.
More details are in \S\ref{subsec:evaluation-perf-nki}.

\vspace{-1em}
$$
  \begin{array}{c}
    \text{Compilation} \\
    \text{Coverage}
  \end{array} =
\frac{
  \begin{array}{c}
    \text{Number of kernels that (i) are successfully compiled to AWS NKI and} \\
    \text{(ii) the compiled NKI code is functionally correct (randomized testing)}
  \end{array}
}{\text{Total number of tiled kernels in the benchmark suite (= 1055)}}
$$

\vspace{0.2em}
\noindent
TPUGraphs kernels are only compatible with AWS Trainium since Google TPUv3 (the original target for TPUGraphs) and AWS Trainium have similar compute capabilities and the same tiling factor (128).
Such an extensive dataset of tiled kernels doesn't exist for Intel AMX and Gemmini.
Thus, the compilation coverage study is limited to AWS Trainium.

\vspace{0.2em}
\begin{center}
  \minibox[frame,c]{
    \act{}-NKI achieves \textbf{2.3x} better compilation coverage for AWS NKI than \texttt{neuronx-cc}.
  }
\end{center}

\textbf{(b) Performance of Compiled Code}\\
Fig.~\ref{fig:perf-summary} (b) summarizes the geomean speedup of the assembly code compiled by \act{}-generated compiler backends against their respective baselines~\cite{neuronx-cc, onednn-library, exo, gemmini-github}.
More details are in \S\ref{subsec:evaluation-perf-nki}--\S\ref{subsec:evaluation-perf-gemmini}.

\vspace{0.2em}
\begin{center}
  \minibox[frame,c]{
    \act{}-generated compiler backends generate performant assembly code that
    matches or\\
    outperforms commercial compiler backends and expert-written kernel libraries.
  }
\end{center}

\subsection{Comparative Evaluations: AWS Trainium~\cite{trainium}}
\label{subsec:evaluation-perf-nki}

We evaluate \act{}-NKI against AWS's production compiler \texttt{neuronx-cc}.

\textbf{Benchmarks}\\
The TPUGraphs~\cite{tpugraphs} Tile dataset is a benchmark suite of tiled kernels generated by Google's TPUv3 compiler from real-world ML workloads.

\textbf{Evaluation Methodology}\\
We use AWS's compiler \texttt{neuronx-cc}~\cite{neuronx-cc} with \texttt{--print-nki} flag to emit AWS NKI code for the benchmark kernels.
We evaluate correctness and performance on an AWS EC2 instance of \texttt{trn1}.

\textbf{Results}\\
Fig.~\ref{fig:perf-summary} (a) shows that \act{}-NKI achieves \textbf{2.3x} better compilation coverage than \texttt{neuronx-cc}.
Since \act{}-NKI has the completeness property, it correctly compiles all 1055 benchmark kernels.

For kernels compiled by both, we observe identical performance, i.e., speedup of \textbf{1x}.
Note that achieving identical performance is a positive result in itself since \texttt{neuronx-cc} is a production compiler, whereas \act{}-NKI is automatically generated by \act{} from just the \idl{} description.

\textbf{Observations \& Analysis}\\
The three main reasons for the difference in compilation coverage are:
\begin{enumerate}[noitemsep, nolistsep]
  \item[(i)] \act{}-NKI transforms operators not directly supported in AWS NKI like \texttt{shift-left} using IR-to-IR rewrites ($\mathcal{R}$ in \S\ref{subsec:background-tensor}) during equality-saturation-based instruction selection (\S\ref{subsec:module-2}).
  \item[(ii)] \texttt{neuronx-cc} doesn't support certain single-node tiled kernels, leading to empty NKI code.
  \item[(iii)] \act{}-NKI detects and allocates compile-time constant tensors (\S\ref{subsec:module-3}), unlike \texttt{neuronx-cc}, which fails to initialize constants, leading to unsound assembly code (fails correctness tests).
\end{enumerate}

\noindent
We reported these results to the AWS Trainium team, who have acknowledged these issues with \texttt{neuronx-cc}. Furthermore, AWS has adopted our tooling, showing the practical impact of our work.

\vspace{-1.5em}
\subsection{Comparative Evaluations: Intel AMX~\cite{intel-amx}}
\label{subsec:evaluation-perf-amx}

We evaluate \act{}-AMX against the Intel oneDNN library, the state-of-the-art kernel library used by production tensor compilers like XLA~\cite{xla} and TorchInductor~\cite{pytorch} to compile to Intel AMX.

\textbf{Benchmarks}\\
The oneDNN library~\cite{onednn-library} consists of common DNN kernels heavily optimized for Intel x86 ISA, including AMX.
We selected five tiled kernels commonly observed in oneDNN examples~\cite{onednn-examples}, also used in AMX-related evaluations in \cite{taidl}.
These kernels are broadly categorized into three types: (i) only int8 matrix operations (\texttt{cnn\_inf\_amx} and \texttt{rnn\_inf\_amx}), (ii) only fp32 matrix operations (\texttt{mem\_format\_avx} and \texttt{sgemm\_avx}), and (iii) a mixture of int8 and fp32 matrix operations (\texttt{cnn\_inf\_mix}).
Detailed statistics and visualization for these kernels are present in Appendix~\ref{app:amx}.

\textbf{ISA Description}\\
The \idl{} description we used to generate \act{}-AMX included 4 AMX-INT8 instructions and 24 AVX512-FP32 instructions.
Thus, it supports compilation to both AMX and AVX512.

\textbf{Evaluation Methodology}\\
The enumeration pipeline (\S\ref{sec:cost-model}) of \act{}-AMX uses a simple analytical cost model based on individual instruction costs.
For fair comparison, \act{}-AMX uses the same fixed tile size as the oneDNN library.
We evaluate correctness and performance on an Intel Xeon Gold 5415+ CPU.

\textbf{Results}\\
Final assembly codes by \act{}-AMX \emph{match} the performance of oneDNN library, i.e., speedup of \textbf{1x}.
The int8 computations are mapped to AMX instructions by both \act{}-AMX and oneDNN.

\textbf{Observations \& Analysis}\\
The hand-optimized oneDNN assembly code is one of the candidates enumerated by \act{}-AMX, by virtue of \act{}'s completeness guarantee.
This candidate has the best performance, as per the cost model, and thus selected as the final assembly code by \act{}-AMX's enumeration pipeline.

Similar to AWS NKI results, achieving identical performance is a positive result since oneDNN is an expert-written kernel library with hand-written pattern-matching rules that perform complicated IR legalization due to the complex layout transformations in the instruction semantics (see Fig.~\ref{fig:amx-tmul}), whereas \act{}-AMX is automatically generated by \act{} from just the \idl{} description.

\act{}-AMX performs the same IR legalization using a sequence of IR-to-IR rewrites: 22 rewrites over 6 iterations in Phase 1 for \texttt{cnn\_inf\_amx}, and 14 rewrites over 4 iterations for \texttt{rnn\_inf\_amx}.
This shows the importance of our design choice to integrate IR-to-IR rewrites with the instruction selection phase of \act{} (\S\ref{subsec:module-2}), with no hand-written accelerator-specific legalization rules.

\vspace{-0.5em}
\subsection{Comparative Evaluations: Gemmini~\cite{gemmini}}
\label{subsec:evaluation-perf-gemmini}

We evaluate \act{}-Gemmini against the Exo library~\cite{exo} and Gemmini SW library~\cite{gemmini-github}

\textbf{Baselines}

\emph{Exo Library.}
Exo~\cite{exo} is a kernel programming DSL based on an orthogonal paradigm of exocompilation, i.e., the user writes both the input program and its schedule.
This includes specifying tile sizes (line 106 of \footnote{\label{exo}\url{https://github.com/exo-lang/exo/blob/v1.0.0/src/exo/platforms/gemmini.py}}), loop transformations (line 158 of \footnotemark[\value{footnote}]), instruction selection (line 149 of \footnotemark[\value{footnote}]), and buffer selection (line 207 of \footnotemark[\value{footnote}]).
Exo developers have manually written hand-tuned schedules for six shapes of GEMM targeting Gemmini.
This is the state-of-the-art kernel library for these shapes.

\emph{Gemmini SW Library.}
The Gemmini architecture ships with a custom kernel library~\cite{gemmini-github} hand-optimized by Gemmini's designers.
The Gemmini SW library supports three common DNN kernels---MAC: $A \times B + C$, GEMM: $A \times B$, and ADD: $A + B$, where $A$, $B$, and $D$ are i8 matrices.

\textbf{Benchmarks}

\begin{figure}[!h]
  \vspace{-0.5em}
  \centering
  \includegraphics[width=0.95\textwidth]{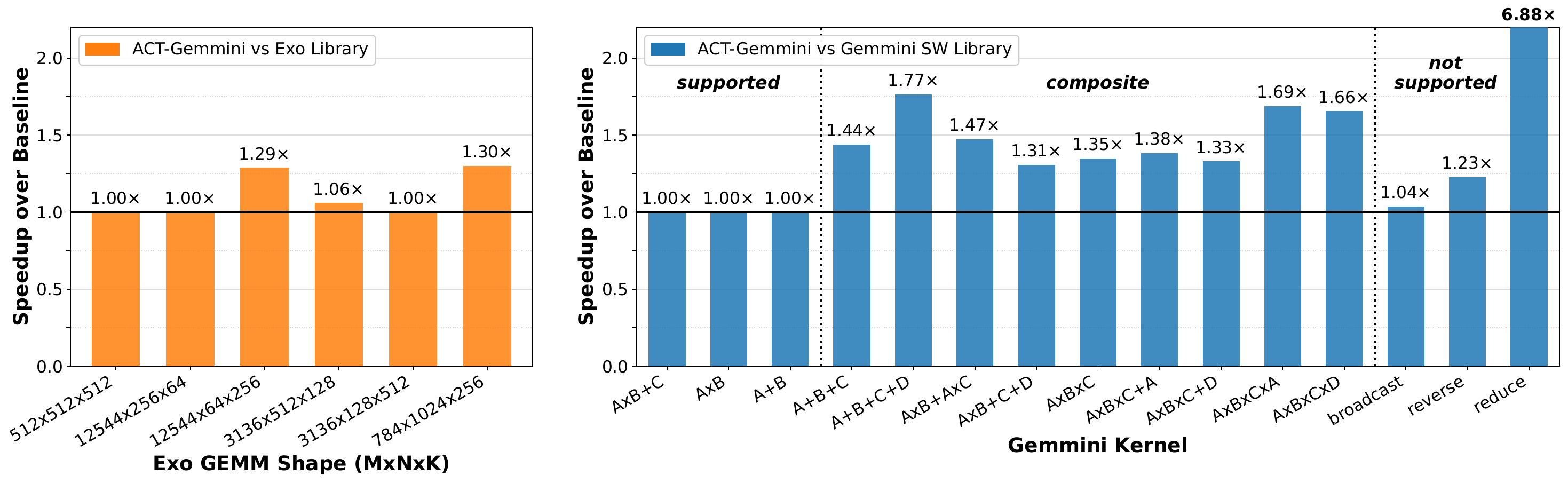}
  \vspace{-1.2em}
  \caption{
    Speedup of compiled code by \act{}-Gemmini over Exo library (left) and Gemmini SW library (right).
    Note that the ``not supported'' evaluation (right) is a \emph{cross-architecture} comparison: these operators have no Gemmini implementation in the SW library, so the baseline numbers were measured against the host CPU.
  }
  \label{fig:perf-gemmini}
  \vspace{-0.8em}
\end{figure}

\emph{Exo Library.}
We selected the six shapes of GEMM (Fig.~\ref{fig:perf-gemmini}, left) with hand-tuned Exo schedules.

\emph{Gemmini SW Library.}
We selected 15 tiled kernels (Fig.~\ref{fig:perf-gemmini}, right) categorized into three classes---\\ (i) \emph{``supported''}: three kernels directly supported by the Gemmini SW library (MAC, GEMM, ADD),
(ii) \emph{``composite''}: nine compositions of GEMM ($\times$) and ADD ($+$) (fusion of upto 3 library calls), and (iii) \emph{``not supported''}: three tensor operators not supported by the library (broadcast, reverse, reduce).

\textbf{ISA Description}\\
The \idl{} description we used to generate \act{}-Gemmini included 11 RISC-type instructions for Gemmini. Gemmini also includes experimental CISC-type instructions that use an on-chip FSM. Here, we focus on the RISC-type instructions, with the rest discussed in Appendix~\ref{app:gemmini}.

\textbf{Evaluation Methodology}\\
The enumeration pipeline (\S\ref{sec:cost-model}) of \act{}-Gemmini uses a simple analytical cost model based on individual instruction costs.
The reported performance is measured using RTL simulations.
Since \act{} focuses on backend passes, we delegate middle-end optimizations to XLA's autotuner~\cite{xla-autotune}.
Against the Exo library, we let \act{}-Gemmini search across tiling and loop fission factors. Against the Gemmini SW library, we use a fixed tiling factor, same as the SW library.

\textbf{Results}\\
Fig.~\ref{fig:perf-gemmini} shows the speedup of compiled code by \act{}-Gemmini over the expert-written kernel libraries.
\act{}-Gemmini with XLA autotuning \emph{matches or outperforms} the hand-tuned Exo library with a speedup of up to \textbf{1.30x}.
\act{}-Gemmini achieves speedups up to \textbf{1x}, \textbf{1.77x}, and \textbf{6.88x} over the Gemmini SW library for ``supported'', ``composite'', and ``not supported'' kernels, respectively.
For ``not supported'' kernels, the SW library has no Gemmini implementation, so this speedup is measured against host (Gemmini's Rocket CPU) execution --- the practical cost of host fallback.

\textbf{Observations \& Analysis}\\
Here, we summarize the key observations with individual kernels discussed in Appendix~\ref{app:exo} \& \ref{app:gemmini}.

\emph{Exo Library.}
\act{}-Gemmini enumerates candidates with different tuning factors, including Exo's hand-tuned factors.
Similar to \S\ref{subsec:evaluation-perf-amx}, the hand-tuned baseline code is one of the enumerated candidates, by virtue of \act{}'s completeness guarantee.
For three shapes, this candidate has the best performance, as per the cost model, and thus selected as the final assembly code by \act{}-Gemmini.
For other three shapes, candidates with better loop fission factors were selected by \act{}-Gemmini.

\emph{Gemmini SW Library (``supported'').}
Similarly, we observed that the final assembly codes compiled by \act{}-Gemmini were the same as the Gemmini SW Library for the three ``supported'' kernels.

\emph{Gemmini SW Library (``composite'').}
\act{}-Gemmini eliminates the redundant load-store instructions between library calls for ``composite'' kernels by storing the intermediates in the accelerator scratchpads.
This increases data reuse and reduces data movement, thus better performance.
We observed that speedup increases with the kernel size (\# nodes).
For GEMM ($\times$), speedup goes from 1x ($A \times B$) to 1.35x ($A \times B \times C$) to 1.66x ($A \times B \times C \times D$).
Likewise for ADD ($+$), 1x to 1.44x to 1.77x.
We limit our evaluations to a maximum of 3 library calls to avoid exaggerated performance results.

\emph{Gemmini SW Library (``not supported'').}
\act{}-Gemmini is able to compile previously unsupported tensor operators like \texttt{reduce} to Gemmini using IR-to-IR rewrites (\S\ref{subsec:background-tensor}) and compile-time constant detection (\S\ref{subsec:module-3}). This leads to large speedups (up to 6.88x) over host execution fallback.

\subsection{Compilation Time Analysis}
\label{subsec:evaluation-time}

Next, we perform breakdown analyses of the compilation time of \act{}-generated compiler backends for varying tiled kernel sizes from synthetic and real-world workloads.

Tiled kernels typically contain 10--50 nodes, as seen in TPUGraphs dataset~\cite{tpugraphs}.
However, the Gemmini kernels used in \S\ref{subsec:evaluation-perf-gemmini} were limited to at most 10 nodes since the baselines were kernel libraries.
Therefore, to analyze the compilation time of \act{}-Gemmini on representative tiled kernels, we generated 100 synthetic kernels (size 7--89 nodes) using NNSmith~\cite{nnsmith}, a random DNN generator (more details in Appendix~\ref{app:nnsmith}).
We also analyze the compilation time of \act{}-NKI for the three largest TPUGraphs kernels, labeled as T1 (150 nodes), T2 (270 nodes), and T3 (390 nodes).

\textbf{Evaluation methodology}

\noindent
The compilation time reported is \textsf{final.time} in Fig.~\ref{fig:autotuning} (a) and was measured on a 20-core Intel i7-13700H CPU, averaged over 20 trials.
This is the wall clock time of the \emph{entire} compilation, including all \pii{} candidates, fallback re-attempts (\S\ref{subsec:overview-algo}), and the cost model loop (\S\ref{sec:cost-model}).
\act{}-Gemmini and \act{}-NKI were set to an overall timeout after 5 seconds, which was never reached (max $\sim$500 ms).

\begin{figure}[!h]
  \vspace{-0.5em}
  \centering
  \includegraphics[width=0.95\textwidth]{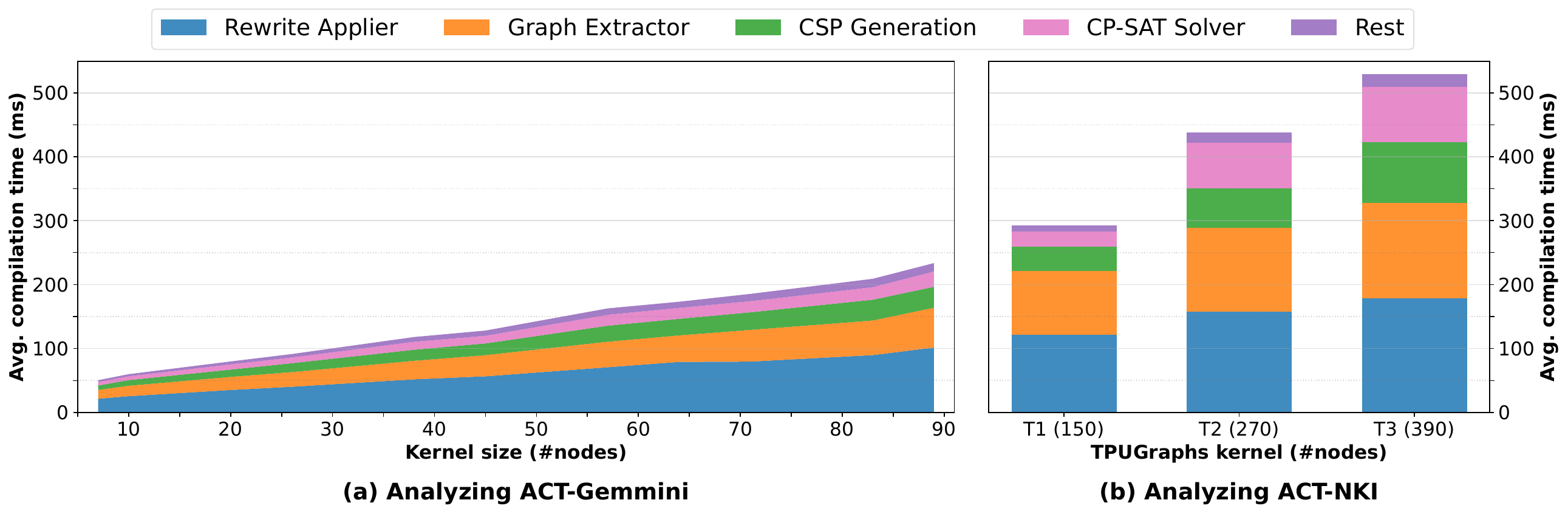}
  \vspace{-1.1em}
  \caption{
    Breakdown of compilation time. (a) \act{}-Gemmini on synthetic kernels generated using NNSmith~\cite{nnsmith} and (b) \act{}-NKI on the three largest kernels in TPUGraphs~\cite{tpugraphs} Tile dataset.
  }
  \label{fig:compile-time}
  \vspace{-0.6em}
\end{figure}

\begin{wrapfigure}{R}{0.28\textwidth}
  \centering
  \includegraphics[width=0.28\textwidth]{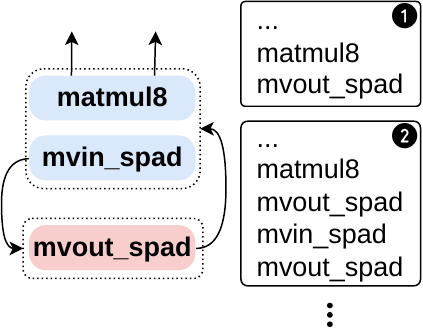}
  \vspace{-2em}
  \caption{Snippet of e-graph (left) with cycles between \textsf{mvin} and \textsf{mvout} e-classes. This represents infinite equivalent candidates (right), and the candidate with no redundant data movement is enumerated first.}
  \label{fig:compact}
  \vspace{-1.8em}
\end{wrapfigure}

\textbf{Results \& Observations}

\noindent
Fig.~\ref{fig:compile-time} shows the breakdown of compilation time of \act{}-Gemmini and \act{}-NKI.
We observed small compilation times ($< 233$ ms) for a wide range of kernel sizes and even for large kernel sizes (529 ms for 390 nodes).
The main reasons for the small compilation time are:
\begin{itemize}[noitemsep, nolistsep]
  \item Compact representation of infinite equivalent assembly code by modeling instruction selection as an equality saturation problem. Fig.~\ref{fig:compact} shows a simple snippet of an e-graph with cycles representing infinite equivalent assembly code.
  \item Bounded approach of graph extraction (in \S\ref{subsec:fallback-egraph}) prioritizes shorter assembly code, thereby enumerating candidates with no redundant data movement first. This leads to faster saturation of performance cost and shorter compilation times.
  \item Instruction scheduling heuristic (in \S\ref{subsec:module-4}) aims to minimize the interference graph. This results in fewer constraints in the generated CSP and, thereby, results in short solver time, even for large pii graphs (86 ms for 390 nodes).
\end{itemize}

\vspace{0.5em}
\begin{center}
  \minibox[frame,c]{
    \act{}-generated compiler backends maintain low compilation overhead\\
    for a wide range of tiled kernel sizes.
  }
\end{center}

\section{Related Works}
\label{sec:related}

~

\textbf{ISA description languages.}
VADL~\cite{vadl} and ILA~\cite{ila} describe ISAs at the \emph{scalar} level, using constructs such as \texttt{forall}, \texttt{fold}, and explicit array indexing, and focus primarily on CPU ISAs.
\idl{}~\cite{taidl} instead describes instruction semantics at the \emph{tensor} level (\S\ref{subsec:problem-isa}), and its tensor operators and types can themselves be parameterized by instruction attributes.
Together, these features enable \idl{} to model the complex, parameterized instruction sets of tensor accelerators. We adopt \idl{} since it already supports several existing accelerator platforms~\cite{trainium, intel-amx, tpuv1, gemmini, feather, eva, allo}.

\vspace{0.2em}
\textbf{Vectorizer generators and synthesis-based compilation.}
Vectorizer generators such as Vegen~\cite{vegen}, Diospyros~\cite{diospyros}, and Isaria~\cite{isaria} target CPUs like x86 and DSPs like Hexagon HVX with fixed-length vector instructions.
Synthesis-based instruction selectors such as Hydride~\cite{hydride} and Rake~\cite{rake} also target fixed-width instructions, which are amenable to SMT-based reasoning.
Compared to these techniques that only focus on instruction selection, \act{} generates the entire compiler backend, including memory allocation for tensor accelerators.
Furthermore, prior works cannot directly handle the complex parameterized instructions present in tensor accelerator ISAs.

\vspace{0.2em}
\textbf{Automated compiler backend generation.}
Prior work on automated CPU backend generation includes declarative machine-description approaches~\cite{isel-dias}, CEGIS-based instruction-selector synthesis~\cite{rulesynth}, processor-description-based generation~\cite{compgen-adl}, and LLM-based approaches~\cite{vega}.
OpenVADL~\cite{openvadl, openvadl-llvm} generates TableGen-style LLVM backends from VADL~\cite{vadl}, assuming little semantic gap between VADL and LLVM IR and thus requiring no complex legalization.
Another related work in this area is a backend generator for the xADL processor-description language~\cite{brandner-backend, brandner-cases}, which targets only scalar register-based CPU ISAs and leaves vector ISAs to future work.
Tensor accelerators instead have parameterized instructions, software-managed scratchpads, and layout-transforming semantics (e.g., Intel AMX~\cite{intel-amx}, Fig.~\ref{fig:amx-tmul}) that require accelerator-specific legalization.
\act{} therefore interleaves legalization with instruction selection in an e-graph rather than using a pattern-based selector, which also avoids phase ordering (\S\ref{subsec:background-egg}).

\vspace{0.2em}
\textbf{Compiler support for novel accelerators.}
3LA \cite{3la} is a recent work that makes commendable progress in developing basic compiler support by enabling simulation testing for end-to-end ML workloads on novel hardware.
However, it requires the user to provide tensor IR-to-accelerator IR mapping for instruction selection and an accelerator IR-to-assembly code generation driver, and does not model scratchpad reuses.
In contrast, \act{} focuses on automatically generating a compiler backend just from the accelerator ISA description.

\vspace{0.2em}
\textbf{MLIR-based compiler stacks.}
MLIR~\cite{mlir} provides a multi-level compiler infrastructure with extensible dialects, and many modern ML compilers are built around this ecosystem, including the XLA compiler~\cite{xla} that uses the StableHLO dialect~\cite{stablehlo}.
MLIR ecosystem includes various lowering passes at a machine-independent IR level such as memref bufferization (note that this is different from the on-chip memory allocation discussed in \S\ref{sec:csp}), as well as hand-written target-specific lowering passes~\cite{mlir-amx, mlir-tpu} for code generation.
\act{} is complementary to these efforts --- it focuses on automatically generating the accelerator-specific compiler backend itself (instruction selection, scheduling, and on-chip tensor memory allocation), and these generated backends can be integrated as a target-specific lowering pass in a larger MLIR-based pipeline.

\vspace{0.2em}
\textbf{Kernel programming languages}
like Exo~\cite{exo}, Pallas~\cite{pallas}, Triton~\cite{triton}, and NKI~\cite{nki} externalize target-specific code generation to user code, simplifying accelerator programming.
These languages offer high-level constructs to program hardware, rather than directly coding in assembly, increasing user productivity.
Exo, in particular, offers an intuitive scheduling DSL for exploring optimizations.
However, these languages do not automatically generate compiler backends; instead, they rely on users to manually write instructions and schedules, and require a device-specific code generator.

\vspace{0.2em}
\textbf{Equality Saturation in Compilers.}
Prior works like Tensat~\cite{tensat}, SPORES~\cite{spores}, Cranelift~\cite{cranelift} use equality saturation to optimize machine learning computation graphs, linear algebra, and Rust functions, respectively.
These works focus only on middle-end optimizations using IR-to-IR rewrites, whereas \act{} uses both IR-to-IR rewrites and auto-generated IR-to-ISA rewrites to concurrently explore instruction selection opportunities.
MISAAL~\cite{misaal} uses equality saturation to lower Halide IR using synthesized rewrite rules, and LIAR~\cite{idiom-eqsat} likewise applies synthesized rewrites to recognize latent idioms in a functional array language.
However, their rewrites carry no preconditions, and thus can not model parameterized instructions prevalent in tensor accelerator ISAs~\cite{nki}~(\S\ref{subsec:module-2}).

\vspace{0.2em}
\textbf{E-graph extraction algorithms.}
Prior equality-saturation workflows typically perform one-shot extraction using either greedy heuristics (the default strategy in egg~\cite{egg}) or ILP-based optimal extraction~\cite{tensat, spores}, with SmoothE~\cite{smoothe} recently relaxing the same one-shot problem into a differentiable one.
These methods are typically tied to a local objective integrated into the extraction algorithm.
In our setting, this is insufficient because a candidate \pii{} graph selected in Phase 1 can still fail in Phase 2 (as discussed in \S\ref{subsec:overview-algo}).
Therefore, \act{} uses an enumerative extraction algorithm (\S\ref{subsec:module-3}) with inter-phase fallback ensuring all equivalent \pii{} graphs are enumerated and remains independent of any built-in cost objectives, allowing integration of external black-box cost models.

\vspace{0.2em}
\textbf{Instruction scheduling}
is the closest problem to the topological ordering discussed in \S\ref{subsec:module-4}.
A common technique for scheduling instructions for pipelined processors is list scheduling~\cite{gibbons-list-sch}.
But unlike our problem, these algorithms minimize the critical path length rather than the interference graphs.
The Sethi–Ullman algorithm~\cite{sethi-ullman} generates optimal code for arithmetic expressions using minimal registers.
Prior works have generalized this to 1-load binary DAGs~\cite{sethi-ullman-dag} and non-binary trees~\cite{sethi-ullman-non-binary}.
However, any depth-first traversal of the expression tree is not always optimal for more than one register file (counterexample in Appendix~\ref{app:solution}).
We extend the Sethi-Ullman algorithm to multiple multi-dimensional tensor buffers with a fallback pruning strategy (\S\ref{subsec:fallback-csp}).

\vspace{0.2em}
\textbf{On-chip scratchpad and register allocation}
are the closest problems to the tensor memory allocation discussed in \S\ref{subsec:module-5}.
A widely adopted solution for register allocation is graph coloring~\cite{dragonbook} over the interference graph, while the puzzle board model~\cite{regpuzzle} handles registers of different sizes.
Unison~\cite{unison, unison-cc} formulates register allocation and instruction scheduling as a single combinatorial problem, and bit-aware register binding~\cite{bit-aware-regbind} admits a polynomial-time exact solution when re-allocation is permitted.
None of these, however, extend to our setting, since we pack multi-dimensional tensor buffers into multi-dimensional scratchpads (\S\ref{subsec:module-5}) rather than one-dimensional registers, and not all buffers are spillable~\cite{nki}.
On-chip scratchpad memory allocation is often related to the rectangle packing problem, which can be modeled as a constraint satisfaction problem (CSP), with improved scalability by pruning techniques proposed in meta-CSP~\cite{metacsp}, TelaMalloc~\cite{telamalloc}, and MiniMalloc~\cite{minimalloc}.
However, these solutions only model interference graphs and, thus, can not represent complex addressing constraints prevalent in tensor accelerator ISAs~\cite{nki} and the identity instructions (discussed in \S\ref{subsec:module-2}), both of which \act{} encodes in its generated CSP.

\vspace{-0.5em}
\section{Conclusion}
\label{sec:conclusion}

In this paper, we propose the first accelerator compiler backend generator, \act{}, that automatically generates a sound and complete compiler backend from a tensor accelerator ISA description.
We provide a novel ISA-parameterized compilation algorithm and demonstrate the generality of our approach by generating backends for seven tensor accelerators.
These backends generate performant code with large compilation coverage and small compilation times.

\act{} fills an important gap in software development for accelerator platforms left under-explored due to the lack of compiler backends targeting them.
It provides an agile and evolvable methodology to quickly build compiler backends that bridge the hardware and software communities.

\section{Data Availability Statement}
Our artifact is a repository of code (10+ KLOC) written in C++, Rust, and Python that implements \act{} compatible with XLA-HLO IR.
The artifact also includes benchmarks and docker environments used for performance evaluation, fuzz testing, stress testing, and scripts to reproduce the plots.
Zenodo archive of the artifact is available at \url{https://doi.org/10.5281/zenodo.21926082}.

ACT is part of a larger open-source ecosystem, built around our ISA description language TAIDL, that automatically generates essential software tools, such as test oracles and compiler backends, from ISA descriptions of tensor accelerators.
Our tooling has been adopted by multiple academic and industry teams designing novel tensor accelerators.
The ecosystem and its associated tutorials are accessible at \url{https://github.com/act-compiler/act}; we welcome contributions and feedback.

\section{Acknowledgments}
We thank the anonymous reviewers for their constructive feedback, as well as Stefanos Baziotis and Damitha Lenadora for their valuable comments on the paper.
Special thanks to Zhihao Wang, Advait Tahilyani, Alan Xie, Pedro Couto, Tanmay Garudadri, Nitin Krishna, Mihir Tandon, and Aarav Khanna for their contributions to the ecosystem and its tooling.
We are grateful to Niansong Zhang, Jianming Tong, and Minsik Kim for their guidance on using their respective accelerator designs.
This work was supported in part by ACE, one of the seven centers in JUMP 2.0, a Semiconductor Research Corporation (SRC) program sponsored by DARPA, and by NSF under grant CCF-2338739.


\bibliographystyle{ACM-Reference-Format}
\bibliography{refs}


\renewcommand{\topfraction}{0.9}
\renewcommand{\bottomfraction}{0.7}
\renewcommand{\textfraction}{0.07}
\renewcommand{\floatpagefraction}{0.8}
\setcounter{topnumber}{3}
\setcounter{bottomnumber}{2}
\setcounter{totalnumber}{5}
\raggedbottom

\appendix

\clearpage

\begin{center}
  {\Large\bfseries Appendix}
\end{center}

\medskip

\noindent
This appendix collects the material deferred from the paper: the detailed problem formalization, an expanded account of the six compilation modules, and the soundness and completeness proofs (Appendices~\ref{app:problem}--\ref{app:proof}); visualizations of the tensor operators (Appendix~\ref{app:viz}); and the per-accelerator data behind the evaluation of \S\ref{sec:evaluation} (Appendices~\ref{app:nnsmith}--\ref{app:amx}).
Each section restates the definitions and running examples it relies on, and pointers such as \S\ref{subsec:module-1} or Fig.~\ref{fig:perf-gemmini} refer to the paper above.

\bigskip

\begingroup
\let\actoldcontentsline\contentsline
\renewcommand{\contentsline}[4]{\ifactapxtoc\actoldcontentsline{#1}{#2}{#3}{#4}\fi}
\linespread{0.88}\footnotesize
\setlength{\parskip}{0pt}
\let\actoldaddvspace\addvspace
\renewcommand{\addvspace}[1]{\actoldaddvspace{0.1em}}
\tableofcontents
\endgroup

\addtocontents{toc}{\protect\actapxtocstart}
\setcounter{tocdepth}{2}

\clearpage
\section{Detailed Problem Formulation}
\label{app:problem}

This section formalizes the problem statement of a compiler backend generator that produces sound and complete compiler backends from an accelerator ISA description alone.

We proceed in three steps.
First, we formalize the tensor accelerator ISA description (\S\ref{app:isa}--\S\ref{app:theta}) written in \idl{}, using existing formal models of tensors and tensor operators~\cite{atl-popl, tensorright, taso, pet}.
The formalism is driven by \idl{}'s key observation that tensor accelerators support coarse-grained instructions operating on multi-dimensional data, commonly represented as tensors.
Second, we define the equivalence condition (\S\ref{app:equivalence}) between a tiled kernel and an accelerator-specific assembly code.
Third, we use that equivalence to define soundness and completeness of a compiler backend, completing the problem formulation (\S\ref{app:generator}) of a compiler backend generator that addresses the challenges and goals of \S\ref{subsec:introduction-challenges}.

Throughout, the object a compiler backend consumes is a \emph{tiled kernel}: the single-tile version of a fused sub-graph produced by the XLA middle-end, syntactically an ordinary tensor computation graph except that its tensor shapes are those of one tile rather than of the whole operand.

\noindent
\begin{minipage}{\textwidth}
  \begin{minipage}[c]{0.37\textwidth}
    \centering
    \includegraphics[width=\textwidth]{figures/qkv-isa.pdf}
    \begin{center}
      (a)
    \end{center}
  \end{minipage}
  \hfill
  \begin{minipage}[c]{0.62\textwidth}
    \centering
    \begin{tabular}{||c|l||}
      \hline
      Instruction                 & Description                                            \\
      \hline\hline
      $\loadrm$                   & $\spad{y} = \bitcvt(\reshape(\hbm{x_1}))$              \\
      \hline
      $\loadcm$                   & $\spad{y} = \transpose(\bitcvt(\reshape(\hbm{x_1})))$  \\
      \hline
      $\storerm$                  & $\hbm{y} = \reshape(\bitcvt(\spad{x_1}))$              \\
      \hline
      $\storecm$                  & $\hbm{y} = \transpose(\reshape(\bitcvt(\spad{x_1})))$  \\
      \hline
      $\mov$                      & $\spad{y} = \tcopy(\ibuf{x_1})$                        \\
      \hline
      $\gemm$                     & $\ibuf{y} = \tdot(\spad{x_1}, \spad{x_2})$             \\
      \hline
      \multirow{2}{*}{$\softmax$} & $\ibuf{y} = \tdiv(\texp(\ibuf{x_1}),$                  \\
                                  & \hspace{3em} $\broadcast(\reduce(\texp(\ibuf{x_1}))))$ \\
      \hline
    \end{tabular}
    \begin{center}
      (b)
    \end{center}
  \end{minipage}
  \captionof{figure}{Running example. (a) High-level accelerator design for a hypothetical accelerator $H_{QKV}$.
    (b) Brief description of its instruction set $\Theta^{H_{QKV}}$. Tensor variables are colored based on their storage unit ($\hbm{d_0}$, $\spad{d_1}$, $\ibuf{d_2}$).}
  \label{app:fig-qkv-isa}
  \Description{A block diagram of a hypothetical QKV accelerator with a main scratchpad, an intermediate buffer, a GeMM engine, a softmax activation unit and a DMA engine over global memory, beside a table listing its seven instructions and the tensor computation each performs.}
\end{minipage}

\subsection{Tensor Accelerator ISA Description $\isa^{H}$}
\label{app:isa}

An ISA description comprises (1) the software-managed storage units, collectively termed the \emph{data model}, such as scratchpads; and (2) the semantics of instructions that perform computations, such as data movement and compute instructions, on those storage units.

\vspace{0.2em}
\textbf{Running example.}
Fig.~\ref{app:fig-qkv-isa} reproduces the running example of \S\ref{subsec:problem-isa} (Fig.~\ref{fig:qkv-isa}) alongside the formalization that follows.
$H_{QKV}$ is a hypothetical accelerator with two on-chip storage units --- a 16\,KB scratchpad and an 8\,KB intermediate buffer --- and access to global memory through a DMA engine.
Its two compute units are a general matrix multiply ($\gemm$) unit and a softmax activation unit.
It supports five data movement instructions and two compute instructions, described in Fig.~\ref{app:fig-qkv-isa}~(b).

\subsection{Data Model $\dm^{H}$}
\label{app:dm}

The software-managed storage units, like the scratchpads on an accelerator, are abstractly represented as \textit{tensor buffers}.
A tensor buffer is a tensor with a subset of its dimensions representing \emph{access dimensions} and the remaining dimensions representing the granularity of data access.
Accelerator instructions modify sub-tensors of these tensor buffers, called \textit{data slices}.
We assume that an accelerator has a byte-addressable 1-D global memory (HBM) accessible by the host processor, denoted $\hbm{d_0}$.
We now define these terms using tensor notation.
Definitions are numbered in one series throughout this section; the boxed ones fix the \emph{objects} an ISA description denotes, and the unboxed ones that follow in \S\ref{app:equivalence} onwards define functions and relations over those objects.

\begin{definitionbox}[def:tensor-buffer]{Tensor Buffer $d$}
  A \textbf{tensor buffer} $d$ is a tensor of type $\mathbb{E}[S_0, S_1]$ in which the first $\mydim(S_0)$ dimensions are \emph{access dimensions} ($S_0$) and the remaining dimensions are the \emph{granularity of data access} ($\mathbb{E}[S_1]$).
  Equivalently, $d$ is a multi-dimensional storage unit of shape $S_0$ whose elements are tensors of uniform type $\mathbb{E}[S_1]$; we write its type as $(S_0, \mathbb{E}[S_1])$ when the access structure is what matters.
\end{definitionbox}

\begin{definitionbox}[def:data-slice]{Data Slice $d[I_s{:} I_e]$}
  A \textbf{data slice} $d[I_s{:} I_e]$ of tensor buffer $d$ is a tensor of type $\mathbb{E}[(I_e - I_s), S_1]$ where $I_s, I_e \in \mathbb{N}_0^{|S_0|}$ and $I_s < I_e \le S_0$.
  $I_s$ and $I_e$ are the start (inclusive) and end (exclusive) addresses of the data slice under zero-based multi-dimensional addressing.
\end{definitionbox}

\begin{definitionbox}[def:memory-state]{Memory State $M$}
  A \textbf{memory state} $M = (d_0, d_1, \dots)$ is a tuple of tensors, one for each tensor buffer.
\end{definitionbox}

\noindent
Tensor buffer representations for the software-managed storage units of $H_{QKV}$ are annotated in Fig.~\ref{app:fig-qkv-isa}.
The 16\,KB scratchpad ($\spad{d_1}$) consists of 128 rows of 64-length $\mathsf{bf16}$ vectors, and instructions read or modify it at row granularity.
It is therefore represented as $\spad{d_1} = ([128], \mathsf{bf16}[64])$.

\subsection{Instruction Set $\Theta^{H}$}
\label{app:theta}

The instruction set $\Theta^{H}$ is a set of \textit{abstract instructions} that modify the tensor buffers in $\dm^H$ and are parameterized by instruction attributes.
This is analogous to x86 instructions such as \textsf{mov} and \textsf{add}, which modify registers or memory and are parameterized by immediate values, register names, and memory addresses.
A \textit{concrete instruction} is an instantiation of an abstract instruction with specific values for those attributes, and is a \emph{deterministic} modification of the tensor buffers.
Fig.~\ref{app:fig-qkv-isa}~(b) briefly describes the instruction set $\Theta^{H_{QKV}}$ of $H_{QKV}$.

Consider the abstract instruction $\loadrm$ in $\Theta^{H_{QKV}}$, parameterized by three attributes: $\nrows \in [1, 128]$, $\addin \in [0, 2^{20})$, and $\addout \in [0, 128)$.
It is a data movement instruction that loads $\nrows$ rows of data ($\nrows \times 128$ bytes) from HBM ($\hbm{d_0}$) starting at address $\addin$, into the main scratchpad ($\spad{d_1}$) starting at row $\addout$.
The concrete instruction $\loadrm(\nrows = 4, \addin = 0, \addout = 2)$ reads 512 bytes from data slice $\hbm{d_0[0{:}512]}$ of tensor-type $\mathsf{u8}[512]$ into data slice $\spad{d_1[2{:}6]}$ of tensor-type $\mathsf{bf16}[4,64]$.
Each $\mathsf{bf16}$ value (2 bytes) is a bitwise cast of 2 contiguous $\mathsf{u8}$ values.
This transformation is a concrete tensor computation from $\mathsf{u8}[512]$ to $\mathsf{bf16}[4,64]$ built from the tensor operators of Table~\ref{app:tab-ops}.

\begin{figure}[htbp]
  \centering
  \includegraphics[width=0.55\textwidth]{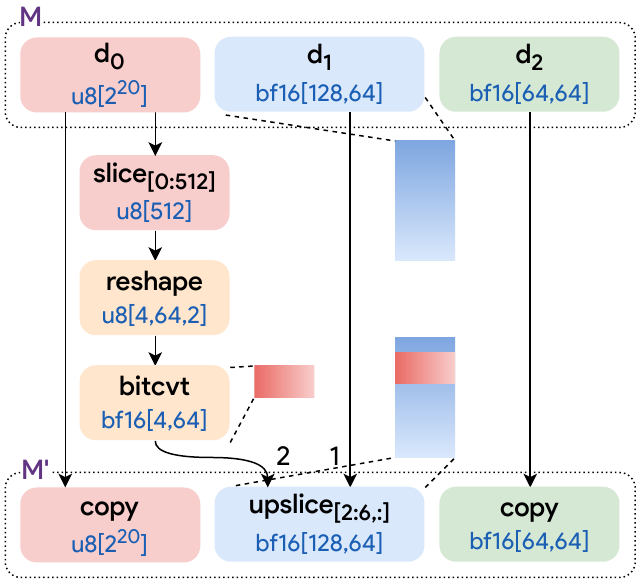}
  \caption{Execution of the concrete instruction $\loadrm(\nrows = 4, \addin = 0, \addout = 2)$ as a concrete tensor computation graph.}
  \label{app:fig-qkv-exec}
  \Description{A tensor computation graph showing a data slice read from HBM, passed through slice, reshape and bitcvt operators, and upsliced into the scratchpad, with the other buffers copied unchanged.}
\end{figure}

More generally, the semantics of $\loadrm$ can be written as an abstract tensor computation $\spad{y} = \bitcvt(\reshape(\hbm{x_1}))$ whose input and output tensors are data slices of $\hbm{d_0}$ and $\spad{d_1}$ respectively.
The addresses of these data slices are a function of the attributes $\addin$ and $\addout$, while the number of rows loaded is a function of the attribute $\nrows$.
This observation is what \act{} builds on: we \emph{extend} \idl{} descriptions by classifying the instruction attributes into two sets, $\alpha = \{\nrows\}$ and $\beta = \{\addin, \addout\}$, termed \textit{computational} and \textit{addressing} attributes respectively.
The computational attributes determine the core tensor computation; every remaining attribute is an addressing attribute.
This split is not present in \idl{} itself, and it is what makes the two phases of \act{}'s algorithm separable --- Phase~1 reasons about $\alpha$ and Phase~2 about $\beta$.

\vspace{0.2em}
\textbf{Generalized representation for instruction set semantics.}
An abstract instruction is parameterized by two sets of integer attributes $\alpha$ and $\beta$.
Its semantics are represented as (1) an abstract tensor computation concretized by the set of \emph{computational attributes} $\boldsymbol\alpha$, where (2) the data slices for the input and output tensors of that computation lie on the tensor buffers, and their addresses are parameterized by the set of \emph{addressing attributes} $\boldsymbol\beta$.

Kernel programming languages like Exo~\cite{exo} model custom hardware instructions under similar assumptions, albeit defining semantics with scalar operators and Python-like syntax.
Intuitively, a scalar pseudo-code description with \textsf{FOR} and \textsf{IF} statements, as used by the Intel Intrinsics Guide~\cite{intel-isa}, can be represented with element-wise tensor operators --- scalars being 0-D tensors --- together with the \textsf{while}\footnote{\url{https://openxla.org/xla/operation_semantics\#while}} and \textsf{select}\footnote{\url{https://openxla.org/xla/operation_semantics\#select}} operators.

\noindent
We now formalize this generalized representation using the notation defined above.

\begin{definitionbox}[def:theta-abstract]{Abstract Instruction $\theta$}
  An \textbf{abstract instruction} $\theta \in \Theta^{H}$, which reads $n_\theta$ tensors located in buffers $d_{\theta}^{(i)} |_{i=1}^{n_\theta}$ and modifies a tensor located in buffer $d_{\theta}^{(0)}$ (by convention, superscript $(0)$ denotes the output and $(1),\dots,(n_\theta)$ the inputs), is represented using an abstract tensor computation $g_{\theta}$ concretized by the set of computational attributes $\alpha$, together with $(n_\theta + 1)$ \emph{data slice address maps} $h^{(i)}_{\theta} |_{i=0}^{n_\theta}$ over the set of addressing attributes $\beta$, and a \emph{validity constraint} $e_{\theta}$ over the attributes $\alpha$ and $\beta$.
\end{definitionbox}

\begin{definitionbox}[def:theta-concrete]{Concrete Instruction $\theta_{\alpha, \beta}$}
  A \textbf{concrete instruction} $\theta_{\alpha, \beta}$ represents a valid modification if $e_{\theta}(\alpha, \beta)$ is true.
  It reads data slices $x_i = d_{\theta}^{(i)} [I_s^{(i)}{:}I_e^{(i)}] |_{i=1}^{n_\theta}$ and modifies data slice $y = d_{\theta}^{(0)} [I_s^{(0)}{:}I_e^{(0)}]$ with the corresponding concrete tensor computation $y = f(x_i |_{i=1}^{n_\theta})$, where $f = g_{\theta}(\alpha)$ and $\forall i \in [0, n_{\theta}] \, (I_s^{(i)}, I_e^{(i)}) = h_{\theta}^{(i)}(\beta)$.
\end{definitionbox}

\begin{definitionbox}[def:theta-execution]{Execution of a Concrete Instruction $\mathcal{E}(\theta_{\alpha, \beta})$}
  \textbf{Execution} $\mathcal{E}(\theta_{\alpha, \beta})$ of a concrete instruction $\theta_{\alpha, \beta}$ is a concrete tensor computation over a tuple of tensors that updates the memory state $M = (d_0, d_1, \dots)$ to $M' = (d_0', d_1', \dots)$, where
  \[
    d_i' =
    \begin{cases}
      \upslice_{[h_{\theta}^{(0)}(\beta)]}\Bigl(d_{\theta}^{(0)},\ g_{\theta}(\alpha)\Bigl(\slice_{[h_{\theta}^{(i)}(\beta)]}(d_{\theta}^{(i)}) \Bigr\vert_{i=1}^{n_\theta}\Bigr) \Bigr) & \text{for } d_i = d_{\theta}^{(0)}    \\[0.5em]
      \copyop(d_i)                                                                                                                                                                      & \text{for } d_i \neq d_{\theta}^{(0)}
    \end{cases}
  \]
\end{definitionbox}

\noindent
Here $\upslice$, $\slice$, and $\copyop$ are the tensor operators of Table~\ref{app:tab-ops}, visualized in Appendix~\ref{app:viz}.
Intuitively, execution slices each input operand out of its buffer at the addressed range, applies the concretized computation $g_\theta(\alpha)$ to produce the output sub-tensor, and upslices that back into the output buffer $d_{\theta}^{(0)}$ at its addressed range; every other buffer is copied unchanged.

Revisiting the concrete instruction $\loadrm(\nrows = 4, \addin = 0, \addout = 2)$, which loads 512 bytes from $\hbm{d_0[0{:}512]}$ to $\spad{d_1[2{:}6]}$: Fig.~\ref{app:fig-qkv-exec} visualizes its execution $\mathcal{E}$, where the orange nodes represent $g_{\loadrm}(\nrows = 4)$ with input and output tensor-types $\mathsf{u8}[512]$ and $\mathsf{bf16}[4,64]$ respectively.
The data slice addresses for the input and output tensors, that is the operator attributes of $\slice$ and $\upslice$, are $h_{\loadrm}^{(1)}(\addin {=} 0, \addout {=} 2) = \hbm{(0, 512)}$ and $h_{\loadrm}^{(0)}(\addin {=} 0, \addout {=} 2) = \spad{(2, 6)}$ respectively.

\subsection{Semantic Equivalence of a Tiled Kernel and Compiled Assembly Code}
\label{app:equivalence}

An assembly code $\asm^{H}$ for a tensor accelerator $H$ consists of a stream of concrete instructions, defined in the ISA description $\isa^{H}$, together with a constant tensor.
This is analogous to x86 assembly code with its code and data sections.

Semantic equivalence of a compiled assembly code and a tiled kernel $G$ is defined over the output produced by sequentially executing the concrete instructions, compared against the golden output $G(X)$ computed by $G$ for given input tensors $X$.
Since the multi-dimensional input and output tensors of $G$ are stored in HBM, which is a 1-D byte-addressable memory, we first define a byte encoding for tensors.

\begin{definition}[$\boldsymbol{\bflat}$]
  \label{def:bflat}
  The \textit{byte-flattening} operator $\bflat$ converts a tensor $t$ of tensor-type $\mathbb{E}[S]$ into a 1-D tensor $\bflat(t)$ of type $\textsf{u8}[\mem(t)]$, where $\mem(t) = \mem(\mathbb{E}) \times \prod S$ and $\mem(\mathbb{E})$ is the memory size in bytes of the scalar basetype $\mathbb{E}$.
\end{definition}

\noindent
$\bflat$ names what \S\ref{subsec:module-1} calls the 1-D byte representation of a tensor: for a $\mathsf{bf16}$ tensor it is a $\bitcvt$ of each $\mathsf{bf16}$ into two $\mathsf{u8}$ values followed by a $\reshape$.

Because equivalence is stated modulo rewrites, we also need the underlying relation on tensor computation graphs.

\begin{definition}[$\boldsymbol{\equiv_{\mathcal{R}}}$]
  \label{def:equivalence-r}
  Tensor computation graphs $G_1$ and $G_2$ are \textbf{semantically equivalent modulo $\mathcal{R}$}, written $G_1 \equiv_{\mathcal{R}} G_2$, if $G_1$ can be transformed into $G_2$ by applying rewrite rules in $\mathcal{R}$.
  We take $\mathcal{R}$ to be the XLA compiler's curated set of 175+ tensor rewrites, treated as the foundational axioms of the \hbox{XLA-HLO} IR.
  Following prior work~\cite{taso, tensat, aerify, constable}, we assume the axioms in $\mathcal{R}$ are closed under inversion and composition, so that $\equiv_{\mathcal{R}}$ is reflexive, symmetric and transitive.
\end{definition}

\noindent
The closure assumption in Def.~\ref{def:equivalence-r} is what licenses chaining equivalences, which the proofs of Appendix~\ref{app:proof} rely on.

Intuitively, semantic equivalence is a sequence of two events: (1) initialization of HBM with the byte-flattened input tensors and the compile-time constant tensor of the assembly code, and (2) sequential execution of the concrete instructions of the assembly code according to the ISA description.
The compiled assembly code is semantically equivalent to the input tiled kernel $G$ if, for all input tensors $X$, HBM ends up holding the byte-flattened input tensors $X$ followed by the output tensor $G(X)$.
We formalize this as a transformation $\Lambda_{LHS}$ of $\asm^{H}_G$ and the expected HBM state as a transformation $\Lambda_{RHS}$ of $G$.

Fix $\mem_{in} = \mem(X)$ and $\mem_{out} = \mem(X) + \mem(G(X))$.
The initial HBM layout places the byte-flattened inputs at offset $0$, reserves the region $[\mem_{in}, \mem_{out})$ for the output, and places the constant tensor at offset $\mem_{out}$, so the constant never overlaps the region that equivalence is checked over.

\begin{definition}[$\boldsymbol{\Lambda_{LHS}}$]
  \label{def:lambda-lhs}
  Given an assembly code $\asm^{H}_G = ((\theta_i(\alpha_i, \beta_i) |_{i=1}^{N}), \mathsf{const})$, $\Lambda_{LHS}(\asm^{H}_G)$ is a concrete tensor computation over a memory state $M = (d_0, d_1, \dots)$ and input tensors $X$.
  The memory state after initialization of HBM is
  \[
    M_0 = \bigl(\upslice_{[\mem_{out}:\mem_{out}+\mem(\mathsf{const})]}(\upslice_{[0:\mem_{in}]}(d_0, \bflat(X)),\ \bflat(\mathsf{const})),\ d_1, \dots\bigr).
  \]
  The final memory state is
  \[
    M_N = \mathcal{E}(\theta_N(\alpha_N, \beta_N)) \circ \cdots \circ \mathcal{E}(\theta_1(\alpha_1, \beta_1)) (M_0),
  \]
  of which only the first $\mem_{out}$ bytes of HBM, written $M_N[0]$, are relevant for equivalence. Finally,
  \[
    \Lambda_{LHS}(\asm^{H}_G)(M,X) = \slice_{[0:\mem_{out}]}(M_{N}[0]).
  \]
\end{definition}

\begin{definition}[$\boldsymbol{\Lambda_{RHS}}$]
  \label{def:lambda-rhs}
  Given an input tiled kernel $G$,
  \[
    \Lambda_{RHS}(G)(M,X) = \concat_1(\bflat(X), \bflat(G(X))).
  \]
\end{definition}

\begin{figure}[htbp]
  \centering
  \includegraphics[width=0.52\textwidth]{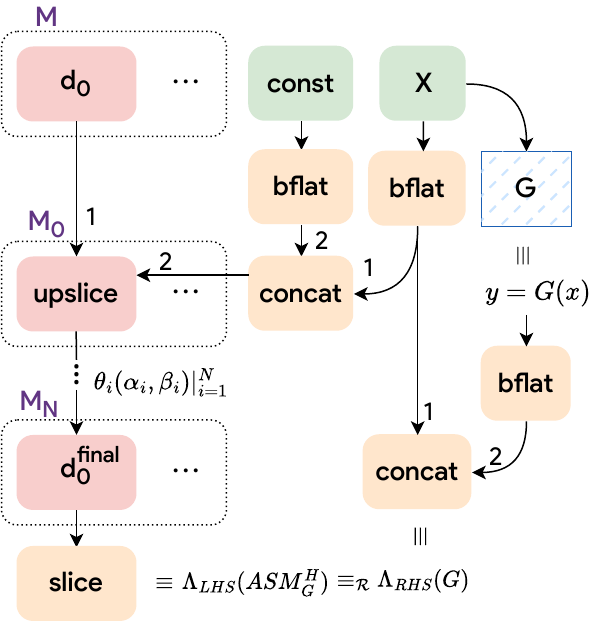}
  \caption{Visualization of $\equiv^{H}_\mathcal{R}$. $M$ and $X$ are the inputs to the tensor computations. The leaf nodes represent the LHS and RHS of the semantic equivalence.}
  \label{app:fig-qkv-equivalence}
  \Description{A diagram with two branches meeting at an equivalence symbol: on the left, HBM initialized and evolved by the concrete instructions then sliced; on the right, the byte-flattened inputs concatenated with the byte-flattened golden output.}
\end{figure}

\noindent
Both $\Lambda_{LHS}$ and $\Lambda_{RHS}$ map a memory state and input tensors to a single $\mathsf{u8}[\mem_{out}]$ tensor, so they can be compared directly.
We now define semantic equivalence using these bisimulation transformations.

\begin{definition}[$\boldsymbol{\equiv^H_\mathcal{R}}$]
  \label{def:equivalence}
  Given the foundational axioms $\mathcal{R}$ of the IR and the ISA description $\isa^{H}$, the semantic equivalence $\equiv^H_\mathcal{R}$ is defined as
  \[
    \asm^{H}_G \equiv^{H}_\mathcal{R} G \iff \forall M ,\, \forall X ,\; \Lambda_{LHS}(\asm^{H}_G)(M, X) \equiv_{\mathcal{R}} \Lambda_{RHS}(G)(M, X),
  \]
  where $\equiv_{\mathcal{R}}$ is the relation of Def.~\ref{def:equivalence-r}.
  $\Lambda_{LHS}$ and $\Lambda_{RHS}$ are visualized in Fig.~\ref{app:fig-qkv-equivalence}.
\end{definition}

\noindent
Since $\equiv_{\mathcal{R}}$ is an equivalence relation (Def.~\ref{def:equivalence-r}) and Def.~\ref{def:equivalence} quantifies over all $M$ and $X$, $\equiv^{H}_\mathcal{R}$ inherits reflexivity, symmetry and transitivity.

\subsection{Compiler Backend Generator}
\label{app:generator}

We can now define the soundness and completeness of a compiler backend in terms of the formalism developed above.
A compiler backend $\compiler^{H}$ consumes a tiled kernel $G$ and may (i) emit an assembly code, (ii) return $\fail$, denoting compilation failure, or (iii) not terminate, denoted $\nterm$.

\begin{definition}[\textbf{Soundness}]
  \label{def:sound}
  A compiler backend $\compiler^{H}$ is \textbf{sound} iff
  $$\forall G ,\, \compiler^{H}(G) \notin \{\fail, \nterm\} \implies \compiler^{H}(G) \equiv^{H}_{\mathcal{R}} G$$
\end{definition}

\begin{definition}[\textbf{Completeness}]
  \label{def:complete}
  A compiler backend $\compiler^{H}$ is \textbf{complete} iff
  $$\forall G ,\, \exists \asm^{H}_G \equiv^{H}_{\mathcal{R}} G \implies \compiler^{H}(G) \notin \{\fail, \nterm\}$$
\end{definition}

\vspace{0.2em}
\textbf{Discussion: semi-decidability versus totality.}
Soundness and completeness as defined above do not require the compiler backend to be a total function: it may fail to terminate on tiled kernels that have no semantically equivalent assembly code.
This is a deliberate design choice, because requiring all three properties at once is impossible.

\begin{theorem}
  \label{thm:no-total-generator}
  There exists no compiler backend generator $\compgen$ such that for every tensor accelerator ISA description $\isa^H$ in \idl{}, the generated backend $\compgen(\isa^H)$ is sound, complete, and total.
\end{theorem}

\begin{proof}
  Assume for contradiction that such a generator $\compgen$ exists.
  Then for every ISA description $\isa^H$, the backend $\compiler^H = \compgen(\isa^H)$ is sound (Def.~\ref{def:sound}), complete (Def.~\ref{def:complete}), and total, meaning it always terminates and so never returns $\nterm$.

  Let $M$ be an arbitrary Turing machine and $w$ an arbitrary input.
  Construct an ISA description $\isa^{H(M,w)}$ with one abstract instruction $\theta$ parameterized by one computational attribute $\alpha = k \in \mathbb{N}$.
  Define each concrete instantiation $\theta(k)$ by bounded simulation:
  \begin{enumerate}[noitemsep, nolistsep]
    \item Simulate $M$ on $w$ for exactly $k$ steps.
    \item If $M$ halts within $k$ steps, $\theta(k)$ computes the identity on a unit tensor.
    \item Otherwise, $\theta(k)$ computes the constant-zero function.
  \end{enumerate}
  Each $\theta(k)$ has finite, computable semantics, and this encoding is expressible in \idl{}~\cite{taidl}.

  Let $G$ be the tiled kernel computing the identity on a unit tensor.

  \emph{If $M$ halts on $w$}, there exists $k^*$ such that the one-instruction assembly $\asm = (\theta(k^*), \emptyset)$ satisfies $\asm \equiv^H_{\mathcal{R}} G$.
  By completeness, $\compiler^H(G) \notin \{\fail, \nterm\}$, so it compiles successfully.

  \emph{If $M$ does not halt on $w$}, then for all $k$ the instruction $\theta(k)$ computes the constant-zero function, which is not equivalent under $\mathcal{R}$ to the identity computed by $G$; hence no assembly built from these instructions is equivalent to $G$.
  By soundness, $\compiler^H(G)$ cannot return a successful assembly, and by totality it cannot return $\nterm$; therefore it returns $\fail$.

  Hence we can decide whether $M$ halts on $w$ by running $\compiler^H(G)$ and checking whether the result is a successful compilation or $\fail$.
  This decides the halting problem, a contradiction.
  Therefore no sound, complete, and total generator exists.
\end{proof}

\noindent
In practice, semi-decidability is not a concern, since a timeout can always be imposed on the compiler backend and treated as a compilation failure.
We therefore focus on soundness and completeness without requiring totality.
\S\ref{subsec:evaluation-time} reports that our generated compiler backends have reasonable compilation times and that we encountered no non-termination in practice.

\clearpage
\section{Detailed Discussion of the Solution}
\label{app:solution}

This section expands on the six modules of \act{}'s compilation algorithm in the places where the presentation above had to be brief.
It is organized to mirror \S\ref{sec:egraph} and \S\ref{sec:csp}: one subsection per module, in the order the modules run.
We begin with an orientation that is easy to lose sight of in the module-by-module presentation, namely what kind of artifact \act{} actually is.

\subsection{Analogy to Classical Compiler Component Generators}
\label{app:analogy}

\act{} is a \emph{generator}, not a compiler, and the distinction is the source of most of its design.
The closest familiar analogues are Flex~\cite{flex} and Bison~\cite{bison}, which are also generators: each takes a declarative description of an artifact, instantiates a fixed parameterized algorithm against that description, and emits a working component.
Table~\ref{app:tab-analogy} lays the three side by side.

Reading the table row by row makes the correspondence concrete.
The \emph{core parameterized algorithm} is the piece of theory that is fixed once and for all and never regenerated: NFA-to-DFA construction for Flex, LALR(1) table generation for Bison, and \act{}'s ISA-parameterized compilation algorithm.
The \emph{generic skeleton} is the code that is identical across every generated instance --- a table-driven scanning loop, a shift-reduce parser loop, and in our case a compiler backend whose accelerator-specific decisions are left as pure virtual functions.
The \emph{specialized templates} are what the generator actually writes: transition tables, ACTION/GOTO tables, and for \act{} the rewrite rules and the concrete implementations that override those virtual functions.
This is why \act{} needs no accelerator-specific algorithmic work: as with Flex and Bison, all the accelerator-specific content lives in the specialized templates, and the algorithm above them never changes.

\begin{table}[htbp]
  \centering
  \begin{tabular}{||c|c|c|c||}
    \hline
                        & Flex~\cite{flex}               & Bison~\cite{bison}           & \act{} (Our work)     \\
    \hline\hline
    User-provided input & Regular expressions            & Context-free grammar         & ISA description       \\
    \hline
    Generated output    & Lexer/Scanner                  & LALR(1) Parser               & Compiler backend      \\
    \hline
    Core Parameterized  & NFA-to-DFA                     & LALR(1) table                & ISA-parameterized     \\
    Algorithm (Theory)  & construction~\cite{dragonbook} & generation~\cite{dragonbook} & compilation algorithm \\
    \hline
    Generic             & Generic table-driven           & Generic shift-reduce         & Generic backend with  \\
    Skeleton            & scanning loop                  & parser loop                  & pure virtual functions \\
    \hline
    Specialized         & DFA transition table           & ACTION/GOTO tables           & Rewrites \& overridden \\
    Templates           & with accept states             & with lookahead sets          & virtual functions     \\
    \hline
    Specialization      & GNU M4                         & GNU M4                       & Python-based          \\
    Approach            & macro processor                & macro processor              & regex processor       \\
    \hline
  \end{tabular}
  \caption{Template-driven compiler component generators based on a parameterized algorithm.
    The core parameterized algorithm and the generic skeleton are fixed once, when the generator itself is built; the specialized templates are what the generator writes anew for each user-provided input, using the specialization approach of the last row.}
  \label{app:tab-analogy}
  \Description{Table comparing Flex, Bison and ACT across user-provided input, generated output, core parameterized algorithm, generic skeleton, specialized templates, and specialization approach.}
\end{table}

\subsection{Module 1: E-Graph Initializer}
\label{app:module-1}

\S\ref{subsec:module-1} describes Module~1 as lowering the tiled kernel into an initial e-graph, without detailing how XLA-HLO's tiled memory formats are represented.
We expand on that here, since tiled layouts are what make \act{}-generated backends drop-in compatible with the XLA compiler.

\begin{figure}[htbp]
  \centering
  \includegraphics[width=0.95\textwidth]{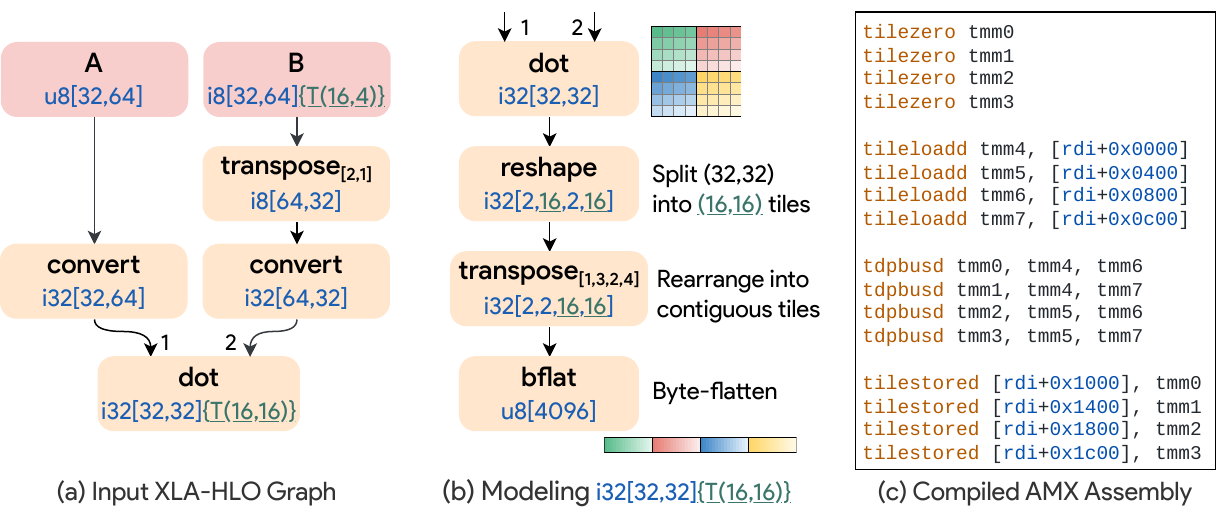}
  \caption{Modeling a tiled memory layout in Module~1.
    (a) The input XLA-HLO IR, with tiled layouts on the input tensor $B$ and on the output tensor.
    (b) The layout \textsf{i32[32,32]\{T(16,16)\}} as modeled by \act{}, using $\reshape$ and $\transpose$ to gather each tile into contiguous order and $\bflat$ to flatten it to bytes.
    (c) The final assembly code generated by \act{}-AMX.}
  \label{app:fig-amx-k1}
  \Description{Three panels: an input XLA-HLO graph, the reshape-transpose-bflat chain that models a tiled layout, and the resulting AMX assembly listing.}
\end{figure}

The XLA compiler often stores tensors in a tiled memory format,\footnote{\url{https://openxla.org/xla/tiled_layout}} written $\mathbb{E}[S_t]\textsf{\{T($S_T$)\}}$, in which a tensor of type $\mathbb{E}[S_t]$ is broken into tiles of shape $S_T$ that are each stored contiguously.
Fig.~\ref{app:fig-amx-k1}~(a) shows the XLA-HLO IR for a tiled kernel commonly observed in oneDNN examples: a matrix multiplication whose input tensor $B$ and output tensor both carry tiled layouts.
\act{} models such a layout as a composition of $\reshape$ and $\transpose$ operators together with the byte-flattening operator $\bflat$ (Def.~\ref{def:bflat}), so that no new operator or IR extension is required.
Fig.~\ref{app:fig-amx-k1}~(b) shows this modeling for the layout \textsf{i32[32,32]\{T(16,16)\}}: the $[32,32]$ tensor is split into $[2,16,2,16]$, the tile axes are rearranged into contiguous order by a $\transpose$, and $\bflat$ then yields the flat byte sequence actually held in memory.
Because the layout is expressed in the same operator vocabulary as the rest of the graph, equality saturation can rewrite across it: \act{}-AMX applied 22 IR-to-IR rewrites over 6 iterations to compile this kernel into the assembly of Fig.~\ref{app:fig-amx-k1}~(c).

\subsection{Module 2: Rewrite Applier}
\label{app:module-2}

Module~2 supplies the rewrites used during the exploration stage.
Two implementation details are worth recording beyond that description.

\vspace{0.2em}
\textbf{Accelerator-specific rewrite filtering.}
The XLA compiler's 175+ IR-to-IR rewrites are not specific to any accelerator, and applying all of them wastes exploration budget on operators the target cannot express.
We therefore filter the rewrite set against the target ISA before saturation begins.
A rewrite is retained only if it can eventually contribute to an IR-to-ISA rewrite: starting from the operators that appear in some abstract computation $g_\theta$, we add any rewrite that produces such an operator, then add the operators that rewrite consumes, and iterate.
For example, if operator $C$ appears in some $g_\theta$, then a rewrite producing $C$ is added to the set; if that set now contains a rewrite $B \to C$, operator $B$ is added in turn, and the process repeats until a fixed point.
Iterating rather than taking a single pass is what makes the filter handle transitive rewrite chains.
The result is that the operator set potentially supported by an accelerator is pruned at backend-generation time rather than at compile time.

\begin{figure}[htbp]
  \centering
  \includegraphics[width=0.62\linewidth]{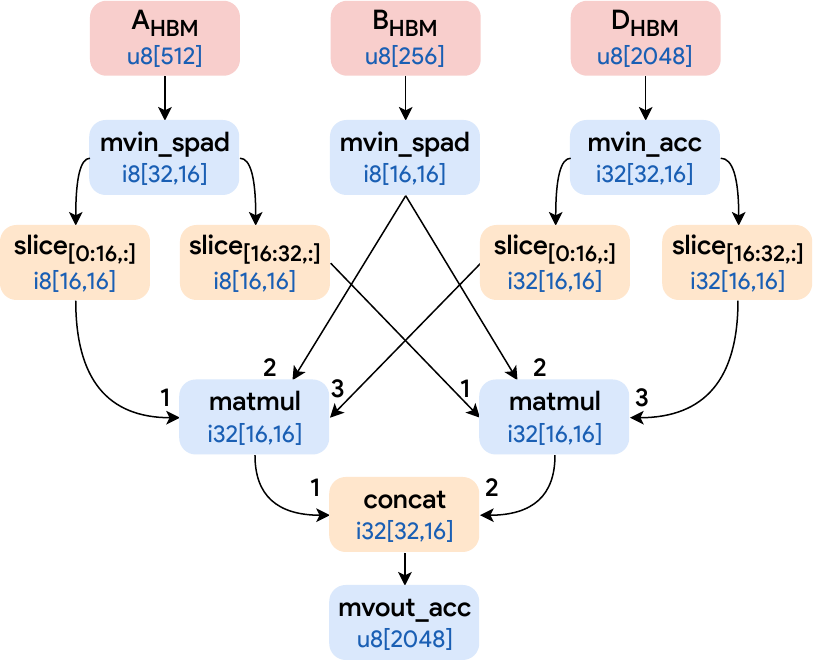}
  \caption{A \pii{} graph generated by \act{}-Gemmini that uses the identity instructions $\slice^H$ and $\concat^H$ to split an operand across two \textsf{matmul} instructions and reassemble the results.}
  \label{app:fig-slice-concat}
  \Description{A pii graph in which a tensor loaded into the scratchpad is split by two slice nodes, multiplied by two matmul nodes, and recombined by a concat node before being written out.}
\end{figure}

\vspace{0.2em}
\textbf{Identity instructions.}
Beyond their role in completeness, the identity instructions $\slice^H$ and $\concat^H$ are what let \act{} tile an inner loop and pad tensor variables without any dedicated machinery: both are expressible as no-ops on the memory state, so the e-graph can introduce them freely wherever a shape needs adjusting.
Fig.~\ref{app:fig-slice-concat} shows a small \pii{} graph generated by \act{}-Gemmini in which exactly this happens: two $[32,16]$ operands are each split by a pair of $\slice^H$ nodes into $[16,16]$ halves, fed through two \textsf{matmul} instructions, and the results reassembled by a $\concat^H$.
Appendix~\ref{app:proof} proves that the availability of these two instructions is what makes Phase~2's reconstruction of a \pii{} graph from an assembly code possible, and with it the completeness of the algorithm as a whole.

\subsection{Module 3: Graph Extractor}
\label{app:module-3}

Module~3 extracts \pii{} graphs from the saturated e-graph by a top-down traversal.
We expand here on its termination condition, which the pseudocode of \S\ref{subsec:module-3} states compactly.

The traversal terminates at an e-class when it is expecting a tensor in $d_0$ (HBM) and that e-class either is marked as constant, or contains an e-node representing a flattened input tensor variable.
In the second case the flattened input tensor is added directly to the \pii{} graph, since it is already available in HBM and needs no instruction to produce it.
In the first case the tensor expression that computes the constant value is added to the \pii{} graph instead, so that the constant is materialized by the accelerator rather than fetched from the host.
Constant e-classes are detected by a bottom-up traversal, with the base case appearing in the extraction pseudocode of \S\ref{subsec:module-3}.
The two cases are mutually exclusive and can never both apply to the same e-class, which is what makes the termination condition well defined.

\subsection{Module 4: Topological Sort}
\label{app:module-4}

Module~4 orders the \pii{} nodes before memory allocation.
\S\ref{subsec:module-4} states that we extend Sethi-Ullman numbering to multiple tensor buffers; we give the extension in full here, together with a counterexample showing why the heuristic cannot be relied on alone.

\vspace{0.2em}
\textbf{Extending Sethi-Ullman numbering.}
Classic Sethi-Ullman numbering~\cite{dragonbook} labels each node of an expression tree with the minimum number of registers needed to evaluate it without spilling: a leaf needs one register, and a binary node with child labels $\ell_1, \ell_2$ needs $\max(\ell_1,\ell_2)$ registers if $\ell_1 \neq \ell_2$ and $\ell_1{+}1$ otherwise, evaluating the costlier child first.
Two things differ in our setting: there are several tensor buffers rather than one register file, and the values occupying them have widely varying sizes.
We therefore track occupancy in bytes, per buffer.
For a tensor buffer $d$ and a node $y$ with operands $x_i$, we define
\[
  \esun_d(y) = \max\bigl(\mem_d(y),\ \min_i \tau_d(x_i)\bigr),
  \qquad
  \tau_d(x_i) = \esun_d(x_i) + \sum_{j \neq i} \mem_d(x_j),
\]
where $\mem_d(x_j)$ is the memory size in bytes of the tensor variable represented by node $x_j$, and is $0$ whenever that tensor does not reside on buffer $d$.
For a leaf node $y$, an input or constant tensor, $\esun_d(y) = \mem_d(y)$.
Intuitively $\tau_d(x_i)$ is the peak occupancy of buffer $d$ when operand $x_i$ is evaluated last, after every sibling $x_j$ has already been materialized; the $\min_i$ picks the operand order that minimizes this peak.
We compute $\esun_d$ for every tensor buffer in $\dm^{H}$ except $d_0$ (HBM), and select operands in decreasing order of $\max_d \bigl(\tau_d(x_i) / \mem(d)\bigr)$, where $\mem(d)$ is the total capacity of buffer $d$.
Normalizing by $\mem(d)$ is what makes pressure on buffers of very different sizes comparable.

\begin{figure}[htbp]
  \centering
  \includegraphics[width=0.42\textwidth]{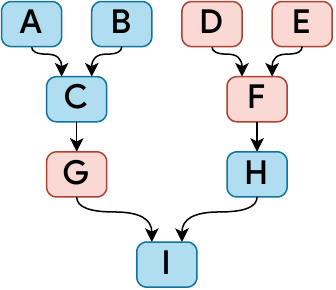}
  \caption{A counterexample with two register files (blue, red) showing that depth-first topological orderings are not optimal for more than one register file.
    Every depth-first ordering uses $(3,2)$ or $(2,3)$ registers, whereas the interleaved ordering $A$--$I$ uses only $(2,2)$.}
  \label{app:fig-esun-limit}
  \Description{A directed acyclic graph of nine nodes in two colours forming two interleaved subtrees that meet at a single sink node.}
\end{figure}

\vspace{0.2em}
\textbf{Why the heuristic is not sufficient.}
We chose Sethi-Ullman numbering as the base algorithm because its objective matches ours: to improve the likelihood that the CSP of Module~5 is satisfiable we want the interference graph to be as small as possible, which is analogous to minimizing register pressure.
But while Sethi-Ullman numbering is provably optimal for a single register file, it is not optimal for more than one, and Fig.~\ref{app:fig-esun-limit} is a counterexample.
The graph has two branches that meet at a common sink, and each branch produces values in \emph{both} register files, drawn blue and red --- that crossover is what defeats a depth-first order.
Any depth-first topological ordering evaluates one subtree fully before starting the other, and is therefore forced to hold three live values in one file, yielding $(3,2)$ or $(2,3)$.
The ordering $A$--$I$ instead interleaves the two subtrees and keeps at most two live values in each file, using $(2,2)$.
No depth-first traversal produces this ordering.

This is a genuine limitation rather than an artifact of our extension, and it is precisely why Phase~2 does not rely on the heuristic ordering for completeness but falls back to an exhaustive search over topological orderings (\S\ref{subsec:fallback-csp}).
In practice the heuristic is nonetheless effective: it guides the search toward low-pressure schedules, and we rarely observe the fallback being needed.

\subsection{Module 5: Constraint Satisfaction Problem Generator}
\label{app:module-5}

Module~5 encodes memory allocation as a constraint satisfaction problem and discharges it with the Google OR-Tools CP-SAT solver.
One practical consequence deserves recording.
CP-SAT is multi-threaded, and its search is therefore non-deterministic: the same tiled kernel can compile to different --- though equally valid --- assembly across runs.
Compiler engineers usually expect a backend to be deterministic, so \act{} exposes a boolean flag that forces it.
When the flag is set, we invoke CP-SAT with \texttt{NumSearchWorkers = 1}, which serializes the search and makes the solution reproducible, at some cost in solving time (see the OR-Tools discussion at \url{https://groups.google.com/g/or-tools-discuss/c/ZGj5vhhZX48}).

\subsection{Module 6: Code Emitter}
\label{app:module-6}

Module~6 is the simplest of the six and needs little expansion: it walks the \pii{} nodes in the order fixed by Module~4 and emits each one with the addressing attributes solved by Module~5.
The single point of interest is that identity instructions are \emph{discarded} at this step rather than emitted.
This is sound because by the time Module~6 runs, whatever selection or reassembly an identity instruction expressed has already been folded into the data slice addresses that Module~5 solved for its neighbours: the value each consumer reads is fixed by its own addressing attributes, so dropping the intervening $\slice^H$ or $\concat^H$ changes nothing observable.
It is also what keeps their use free: Modules~2 and~3 may introduce as many of them as the search finds convenient without any of them reaching the generated assembly.

\clearpage
\section{Theoretical Guarantees of \act{}'s ISA-parameterized Compilation Algorithm}
\label{app:proof}

The claims and proofs in this section rest on the formalization of Appendix~\ref{app:problem} and are independent of any implementation detail.
Their purpose is to establish a rigorous foundation for the soundness and completeness of \act{}'s core algorithm for automatically generating compiler backends from tensor accelerator ISA descriptions.

\subsection{Notation and Standing Assumptions}
\label{app:proof-notation}

The proofs are stated in the vocabulary of the compilation algorithm, which we restate here to fix the notation used throughout.

Phase~1 is built on \emph{equality saturation}.
The tiled kernel is loaded into an \emph{e-graph}: a data structure whose nodes (\emph{e-nodes}) are operators and whose \emph{e-classes} are sets of e-nodes known to compute the same value.
Applying a rewrite does not replace anything --- it adds the rewritten form to the same e-class, so one e-graph represents every graph reachable by the rewrites applied so far.
\act{} alternates an \emph{exploration} stage, which applies rewrites, with an \emph{extraction} stage, which reads candidate \pii{} graphs back out.
Three kinds of rewrite are used (\S\ref{app:module-2}): the \hlo{} IR-to-IR axioms, the IR-to-ISA rewrites generated from the ISA description, and the identity-instruction rewrites.

Four further symbols recur throughout.

\begin{description}[style=unboxed, leftmargin=1em, itemsep=0.4em, parsep=0pt]
  \item[$\phase_1$, $\phase_2$.] The two phases of the compilation algorithm.
    $\phase_1(G)$ denotes the sequence of \pii{} graphs enumerated by tensor instruction selection on the tiled kernel $G$, with $\phase_1(G)[i]$ its $i^{\text{th}}$ element.
    $\phase_2(\piig^H)$ denotes the assembly code produced by tensor memory allocation on the \pii{} graph $\piig^H$, or $\fail$ if the constraint problem is unsatisfiable.
  \item[$\Gamma$.] The \pii{} node budget used by the extraction stage.
    \act{} interleaves exploration and extraction, raising the node limit $N_{\mathsf{max}} = \Gamma(k)$ with the exploration iteration count $k$, where $\Gamma$ is strictly monotonically increasing and unbounded (for example $\Gamma(k) = 2^k$).
  \item[$\csp_4$.] The interference constraint of Module~5, $\csp_4 = \bigwedge_{(i,j) \in \phi} \mathsf{disjoint}(i,j)$, where the interference graph $\phi$ pairs two nodes exactly when they write the \emph{same} tensor buffer and their live ranges overlap. It forces such a pair to be given non-overlapping addresses within that buffer; nodes on different buffers are never constrained against each other.
\end{description}

\noindent
We also use freely that $\equiv_{\mathcal{R}}$ is an equivalence relation, which Def.~\ref{def:equivalence-r} establishes from the closure of $\mathcal{R}$ under inversion and composition.
This is what licenses chaining equivalences in \S\ref{app:proof-theorem}.
Lemma~\ref{lemma:egraph-equivalent} additionally requires $\equiv_{\mathcal{R}}$ to be a \emph{congruence} --- that replacing a sub-graph by an equivalent one leaves the enclosing graph equivalent --- which is what makes reasoning e-class by e-class sound.

We provably guarantee that \act{} generates \emph{sound} and \emph{complete} compiler backends:

\begin{theorem}
  \label{theorem:act}
  For every ISA description $\isa^{H}$, the backend $\act{}(\isa^{H})$ is sound (Def.~\ref{def:sound}) and complete (Def.~\ref{def:complete}).
\end{theorem}

\noindent
We break the proof of Theorem~\ref{theorem:act} into five lemmas, combined in \S\ref{app:proof-theorem}.

\subsection{The Intermediary \pii{} Graph}
\label{app:proof-pii}

The proofs are organized around an intermediary representation that sits between the tiled kernel and the assembly code: a graph whose nodes are instructions with their computational attributes fixed but their addressing attributes still unknown.
Writing the unknown as $?$, we call these \emph{partially instantiated instruction} (\pii{}) nodes.
Splitting the problem at exactly this point is what lets Phase~1 reason about $\alpha$ alone and Phase~2 about $\beta$ alone.

\begin{definition}
  \label{def:pii-graph}
  A \textbf{\pii{} graph} $\piig^H$ is a directed acyclic graph of $N$ \pii{} nodes $\theta_i(\alpha_i,?)|_{i=1}^{N}$, which can be transformed into a concrete tensor computation by the substitutions $\theta_i(\alpha_i,?) \rightarrow g_{\theta_i}(\alpha_i)$.
  A \pii{} node is either an instruction $\theta_{\alpha,?}$ from the instruction set $\Theta^H$, or one of the identity instructions $\slice^H_{\alpha,?}$ and $\concat^H_{\alpha,?}$.
\end{definition}

\begin{definition}[$\boldsymbol{\equiv^H_\mathcal{R}}$ \textbf{lifted to \pii{} graphs}]
  \label{def:pii-equivalence}
  Semantic equivalence between a \pii{} graph $\piig^H$ and a tiled kernel $G$ is defined as
  \begin{align*}
    \piig^H \equiv^H_\mathcal{R} G &\iff \Lambda_{pii}(\piig^H) \equiv_\mathcal{R} \Lambda_{RHS}(G),                    \\
    \text{where}\quad \Lambda_{pii}(\piig^H)(M,X) &= \concat_1(\bflat(X), \piig^H(X)).
  \end{align*}
  Similarly, $\asm^H_G \equiv^H_\mathcal{R} \piig^H \iff \Lambda_{LHS}(\asm^H_G) \equiv_\mathcal{R} \Lambda_{pii}(\piig^H)$.
  Here $\piig^H(X)$ denotes the byte-level output the \pii{} graph produces in HBM, so that it has the same tensor-type as $\bflat(G(X))$ in Def.~\ref{def:lambda-rhs}, and $\Lambda_{LHS}$, $\Lambda_{RHS}$ are as defined in \S\ref{app:equivalence}.
  The memory-state argument $M$ is carried so that $\Lambda_{pii}$, $\Lambda_{LHS}$ and $\Lambda_{RHS}$ share one input signature.
  It is inert in $\Lambda_{pii}$ and $\Lambda_{RHS}$, which read only $X$; $\Lambda_{LHS}$ does read it, which is why Def.~\ref{def:equivalence} quantifies over all $M$.
\end{definition}

\noindent
Def.~\ref{def:pii-equivalence} reuses the symbol $\equiv^H_\mathcal{R}$ of Def.~\ref{def:equivalence} on two further pairs of objects.
All three readings unfold to $\equiv_\mathcal{R}$ on $\mathsf{u8}[\mem_{out}]$ tensors, so they compose, which is what \S\ref{app:proof-theorem} relies on.

\subsection{Theoretical Guarantees of Phase 1}
\label{app:proof-egraph}

Phase~1 performs tensor instruction selection. Its two guarantees are that everything it enumerates is correct, and that it enumerates everything.

\begin{lemma}
  \label{lemma:egraph-equivalent}
  Every enumerated \pii{} graph is equivalent to the input tiled kernel $G$, i.e.,
  \[
    \boldsymbol{\forall i ,\, \phase_1(G)[i] \equiv^H_\mathcal{R} G}.
  \]
\end{lemma}

\begin{proof}
  Equality saturation only ever replaces a term by one the rewrite set declares equal, so it suffices that every rewrite used in Module~2 is semantics-preserving.
  There are three kinds.
  The IR-to-IR rewrites are a subset of the foundational axioms $\mathcal{R}$, hence valid by definition of $\equiv_\mathcal{R}$.
  The IR-to-ISA rewrites are derived mechanically from the ISA description: each rewrites a sub-graph matching the abstract tensor computation $g_\theta$ into the corresponding \pii{} node $\theta_{\alpha,?}$, and is therefore valid by Def.~\ref{def:theta-abstract}.
  The identity-instruction rewrites are valid because $\slice^H$ and $\concat^H$ are by construction equal to their tensor-operator counterparts $\slice$ and $\concat$.
  Since each rewrite preserves $\equiv_\mathcal{R}$ and $\equiv_\mathcal{R}$ is transitive, every e-class of the saturated e-graph contains only terms equivalent to the initial one, and hence every extracted \pii{} graph is equivalent to $G$.
\end{proof}

\begin{lemma}
  \label{lemma:egraph-enumerate}
  All equivalent \pii{} graphs are enumerated, i.e.,
  \[
    \boldsymbol{\forall \piig^H \equiv^H_\mathcal{R} G ,\, \exists i \text{ \textbf{s.t.} } \phase_1(G)[i] {=} \piig^H}.
  \]
\end{lemma}

\begin{proof}
  Let $\piig^H \equiv^H_\mathcal{R} G$ have $N_0$ nodes.
  By Def.~\ref{def:pii-equivalence} this means $\Lambda_{pii}(\piig^H) \equiv_\mathcal{R} \Lambda_{RHS}(G)$, so the two are joined by a finite sequence of rewrites; since $\Lambda_{pii}$ and $\Lambda_{RHS}$ differ from $\piig^H$ and $G$ only by the common $\concat_1(\bflat(X), \cdot)$ wrapper, the same sequence joins $\piig^H$ to $G$.
  Every rewrite in it is drawn from the three kinds Module~2 supplies, and the exploration stage applies every applicable rewrite at each iteration, so there is a finite $k_0$ after which $\piig^H$ is represented in the e-graph.
  Because $\Gamma$ is strictly monotonically increasing and unbounded, the set $\{k : N_0 \le \Gamma(k)\}$ is non-empty; let $k_1 = \min\{k : N_0 \le \Gamma(k)\}$ be its least element.
  At iteration $k' = \max(k_0, k_1)$ the graph is both present in the e-graph and within the extraction node budget, so it is among the graphs extracted at $k'$.
\end{proof}

\subsection{Theoretical Guarantees of Phase 2}
\label{app:proof-csp}

Phase~2 performs tensor memory allocation. It contributes one soundness lemma, one completeness lemma, and one structural lemma relating assembly codes back to \pii{} graphs.

\begin{lemma}
  \label{lemma:csp-equivalent}
  Whenever Phase~2 succeeds, the assembly code it emits is equivalent to the input \pii{} graph, i.e.,
  \[
    \boldsymbol{\phase_2(\piig^H) \neq \fail \implies \phase_2(\piig^H) \equiv^H_\mathcal{R} \piig^H}.
  \]
\end{lemma}

\begin{figure}[htbp]
  \centering
  \includegraphics[width=0.942\textwidth]{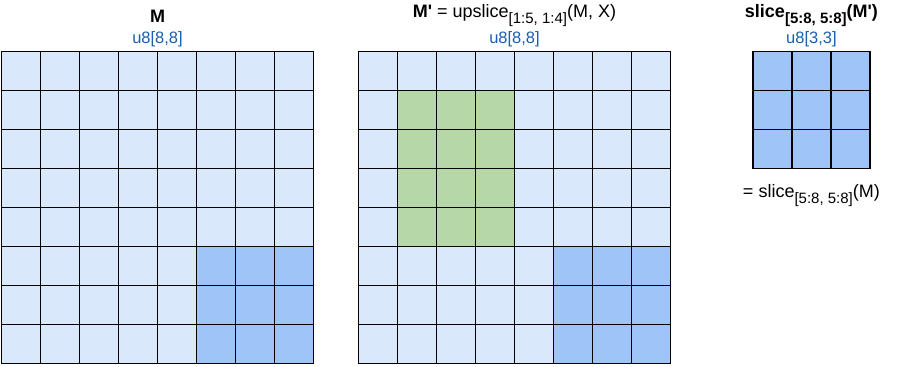}
  \caption{Visualization of the rewrite rule $\slice_{[s_1{:}e_1]}(\upslice_{[s_2{:}e_2]}(M, X)) \equiv_{\mathcal{R}} \slice_{[s_1{:}e_1]}(M)$, valid under the precondition $\mathsf{disjoint}([s_1{:}e_1], [s_2{:}e_2])$.
    Because the written region (green) and the read region are disjoint, the read returns the same bytes whether or not the write happened, so the $\upslice$ can be discarded.}
  \label{app:fig-upslice-slice}
  \Description{Three grids: an 8 by 8 tensor M, the same tensor after a disjoint region has been overwritten, and a 3 by 3 read window whose contents are identical in both.}
\end{figure}

\begin{proof}
  Write $\asm^H_G = \phase_2(\piig^H)$ and transform $\Lambda_{LHS}(\asm^H_G)$ into $\Lambda_{pii}(\piig^H)$.
  By Def.~\ref{def:theta-execution} each executed instruction contributes an $\upslice$ of its result into its output buffer, and each later instruction that consumes a value contributes a $\slice$ of some buffer.
  Working backwards from the last instruction, every $\slice$-of-$\upslice$ pair that arises falls into exactly one of two cases.

  \emph{The read is the consumer of that write.}
  Module~5's def-use constraint assigns the consumer the same data slice addresses the producer was given, so the pair is $\slice_{[s{:}e]}(\upslice_{[s{:}e]}(M, X))$, which reduces to $X$: the consumer receives precisely the tensor the producer computed, and the intermediate buffer write disappears while the value it carried is retained.

  \emph{The read belongs to some other value.}
  Then the two data slices have overlapping live ranges, so the interference constraint $\csp_4$ has assigned them non-overlapping addresses within their common buffer, discharging the precondition of
  \[
    \slice_{[s_1{:}e_1]}(\upslice_{[s_2{:}e_2]}(M, X)) \rightarrow \slice_{[s_1{:}e_1]}(M)
    \quad\text{under}\quad
    \mathsf{disjoint}([s_1{:}e_1], [s_2{:}e_2]),
  \]
  which steps the read back past a write it cannot observe.

  Alternating the two cases eliminates every intermediate buffer write while preserving each consumed value, leaving exactly the composition of $g_\theta$ sub-graphs that constitutes $\Lambda_{pii}(\piig^H)$.
  Both rewrites preserve $\equiv_\mathcal{R}$, so the whole transformation does.
  The second has been verified in full generality using TensorRight~\cite{tensorright}, and Fig.~\ref{app:fig-upslice-slice} visualizes it.
\end{proof}

\begin{lemma}
  \label{lemma:csp-completeness}
  Phase~2 does not fail if an equivalent assembly code exists, subject to the completeness of the CP-SAT solver, i.e.,
  \[
    \boldsymbol{\exists \asm^H_G \equiv^H_\mathcal{R} \piig^H \implies \phase_2(\piig^H) \neq \fail}.
  \]
\end{lemma}

\begin{proof}
  Phase~2 exhaustively searches topological orderings, using interference-graph pruning (\S\ref{subsec:fallback-csp}) which discards only orderings whose interference graph is a supergraph of one already shown unsatisfiable, and so preserves completeness.\footnote{This is the direction implied by the monotonicity of $\csp_4$ under $\phi$ stated in \S\ref{subsec:fallback-csp}, namely $\phi_1 \subseteq \phi_2 \implies \neg\csp(\phi_1) \rightarrow \neg\csp(\phi_2)$: a supergraph of an unsatisfiable interference graph is itself unsatisfiable, so pruning supergraphs never discards a satisfiable ordering. Pruning in the opposite direction would not be sound, since a subgraph carries fewer constraints and may be satisfiable.}
  The search therefore covers the ordering in which the instructions of $\asm^H_G$ appear.
  For that ordering, the data slice addresses already present in $\asm^H_G$ constitute a satisfying assignment to the constraint problem, so the CSP is satisfiable and Phase~2 does not fail.
  The qualification on CP-SAT is necessary because \act{} delegates satisfiability to an external solver; the argument shows a model exists, not that the solver finds it.
\end{proof}

\begin{lemma}
  \label{lemma:csp-existence}
  Every assembly code has an equivalent \pii{} graph, i.e.,
  \[
    \boldsymbol{\forall \asm^H_G ,\, \exists \piig^H ,\, \asm^H_G \equiv^H_\mathcal{R} \piig^H}.
  \]
\end{lemma}

\begin{figure}[htbp]
  \centering
  \includegraphics[width=0.782\textwidth]{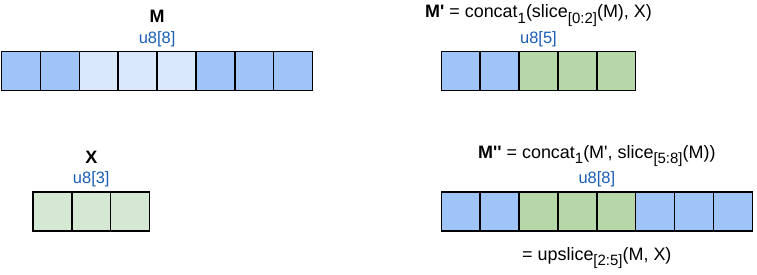}
  \caption{Visualization of the rewrite rule $\upslice_{[s{:}e]}(M,X) \equiv_{\mathcal{R}} \concat_1(\concat_1(\slice_{[0{:}s]}(M), X), \slice_{[e{:}]}(M))$.
    Writing $X$ into the middle of $M$ is the same as concatenating the untouched prefix of $M$, then $X$, then the untouched suffix --- which is what lets an $\upslice$ be re-expressed using only the identity instructions $\slice^H$ and $\concat^H$.}
  \label{app:fig-concat-concat-slice}
  \Description{A one-dimensional tensor M and a shorter tensor X, shown being recombined into a tensor equal to M with X written into its middle range.}
\end{figure}

\begin{proof}
  We construct the \pii{} graph from $\Lambda_{LHS}(\asm^H_G)$.
  By Def.~\ref{def:theta-execution} and Def.~\ref{def:lambda-lhs}, $\Lambda_{LHS}(\asm^H_G)$ consists of $g_\theta$ sub-graphs together with $\bflat$, $\slice$ and $\upslice$ nodes.
  The $\bflat$ nodes coincide with those of $\Lambda_{pii}$ and can be ignored.
  Each $g_\theta$ sub-graph is replaced by the corresponding \pii{} node $\theta_{\alpha,?}$.
  Each $\slice$ node is replaced by a $\slice^H$ node, and each $\upslice$ node is decomposed into $\slice^H$ and $\concat^H$ nodes using
  \[
    \upslice_{[s{:}e]}(M,X) \rightarrow \concat_1(\concat_1(\slice_{[0{:}s]}(M), X), \slice_{[e{:}]}(M)).
  \]
  The result satisfies Def.~\ref{def:pii-graph}, so it is a \pii{} graph, and each step preserves $\equiv_\mathcal{R}$.
  Note where the identity instructions earn their place: without $\slice^H$ and $\concat^H$ as admissible \pii{} nodes there would be no way to express the decomposition above, and this lemma --- and with it completeness --- would fail.
  The generalized rewrite has been verified using TensorRight~\cite{tensorright}, and Fig.~\ref{app:fig-concat-concat-slice} visualizes it.
\end{proof}

\subsection{Proving Soundness and Completeness}
\label{app:proof-theorem}

We now assemble Theorem~\ref{theorem:act} from the lemmas of \S\ref{app:proof-egraph} and \S\ref{app:proof-csp}.

\begin{proof}[Proof of Theorem~\ref{theorem:act}, soundness]
  Suppose $\compiler^H(G) \notin \{\fail, \nterm\}$, so the backend emitted an assembly code $\asm^H_G = \phase_2(\piig^H)$ for some enumerated \mbox{$\piig^H = \phase_1(G)[i]$}.
  Lemmas~\ref{lemma:csp-equivalent} and~\ref{lemma:egraph-equivalent} give the two halves of the chain:
  \[
    \Lambda_{LHS}(\asm^H_G) \equiv_\mathcal{R} \Lambda_{pii}(\piig^H)
    \qquad\text{and}\qquad
    \Lambda_{pii}(\piig^H) \equiv_\mathcal{R} \Lambda_{RHS}(G).
  \]
  By transitivity of $\equiv_\mathcal{R}$ (Def.~\ref{def:equivalence-r}) it follows that $\Lambda_{LHS}(\asm^H_G) \equiv_\mathcal{R} \Lambda_{RHS}(G)$, which is exactly $\compiler^H(G) \equiv^H_\mathcal{R} G$.
  Hence $\compiler^H$ is sound in the sense of Def.~\ref{def:sound}.
\end{proof}

\begin{proof}[Proof of Theorem~\ref{theorem:act}, completeness]
  Suppose there exists $\asm^H_G \equiv^H_\mathcal{R} G$.
  Def.~\ref{def:complete} requires $\compiler^H(G) \notin \{\fail, \nterm\}$, so we must exclude both outcomes.

  \emph{Not $\fail$.}
  By Lemma~\ref{lemma:csp-existence} there is a \pii{} graph $\piig^H$ with $\asm^H_G \equiv^H_\mathcal{R} \piig^H$.
  Unfolding both hypotheses, $\Lambda_{LHS}(\asm^H_G) \equiv_\mathcal{R} \Lambda_{pii}(\piig^H)$ and $\Lambda_{LHS}(\asm^H_G) \equiv_\mathcal{R} \Lambda_{RHS}(G)$; by symmetry and transitivity of $\equiv_\mathcal{R}$ we get $\Lambda_{pii}(\piig^H) \equiv_\mathcal{R} \Lambda_{RHS}(G)$, that is $\piig^H \equiv^H_\mathcal{R} G$.
  By Lemma~\ref{lemma:egraph-enumerate}, Phase~1 enumerates this $\piig^H$ at some finite index.
  By Lemma~\ref{lemma:csp-completeness}, Phase~2 does not fail on it.
  So the backend emits an assembly code rather than returning $\fail$.

  \emph{Not $\nterm$.}
  Let $k'$ be the finite iteration at which Lemma~\ref{lemma:egraph-enumerate} places $\piig^H$ among the extracted graphs.
  Each exploration iteration applies finitely many rewrites to a finite e-graph and each extraction is bounded by the node budget $\Gamma(k)$, so iterations $1$ through $k'$ complete in finite time.
  For each of the finitely many \pii{} graphs enumerated up to that point, Phase~2 searches a finite set of topological orderings and solves a finite constraint problem.
  The backend therefore reaches $\piig^H$ and succeeds on it after finitely many steps, and so terminates.

  Hence $\compiler^H(G) \notin \{\fail, \nterm\}$ and $\compiler^H$ is complete in the sense of Def.~\ref{def:complete}.
\end{proof}

\noindent
The termination argument above applies only under the hypothesis of Def.~\ref{def:complete}, namely that an equivalent assembly code exists.
When none exists, the algorithm may enumerate indefinitely and return $\nterm$ --- and by Theorem~\ref{thm:no-total-generator} no generator can do better while remaining sound and complete.

\subsection{Visualization of the Relevant Rewrite Rules}
\label{app:proof-viz}

The two rewrite rules that carry the Phase~2 proofs are visualized below on concrete tensors, using the colour convention of Appendix~\ref{app:viz}.

\clearpage
\section{Visualization of Tensor Operators}
\label{app:viz}

The tensor operators of Table~\ref{app:tab-ops} are the vocabulary in which \act{} states everything else: tiled kernels are graphs over these operators, the abstract computation $g_\theta$ of every instruction is a graph over these operators, and the foundational axioms $\mathcal{R}$ are rewrites between such graphs.
Their operational semantics are specified normatively by \hlo{}~\cite{xla-hlo, stablehlo}, and denotational semantics for a subset are given by TensorRight~\cite{tensorright}.
This section restates the operator list and then visualizes each operator on a small concrete example, since the shape-manipulating operators in particular are far easier to read as pictures than as index arithmetic.

The visualizations share one colour convention throughout:

\begin{description}[style=unboxed, leftmargin=1.5em, itemsep=0.15em, parsep=0pt, topsep=0.4em]
  \item[Blue] marks the first operand,
  \item[Green] marks the second operand, and
  \item[Peach] marks a tensor whose element values are \emph{computed} rather than moved.
\end{description}

\noindent
Colouring the operands separately means that when an operator merely rearranges elements, the reader can trace each output cell back to the operand it came from.
A darker shade marks a region that is only \emph{part} of a larger tensor --- the window a $\slice$ reads, or the range an $\upslice$ writes --- so that partial access is distinguishable at a glance from whole-tensor access.
Cells are annotated with element values only where the values themselves carry information; for the purely structural operators the grid geometry is the whole story.

\begin{table}[htbp]
  \centering
  \begin{tabular}{|| c | c | c ||}
    \hline
    Tensor Operator                                    & Operator Attributes                                 & Description                                     \\
    \hline\hline
    $y = \slice_{[s{:}e]}(x)$                          & $\type(x), s, e$                                    & Read sub-tensor $x[s{:}e]$                      \\
    \hline
    $y = \upslice_{[s{:}e]}(x_1, x_2)$                 & $\type(x_1), s, e$                                  & Update sub-tensor $x_1[s{:}e] = x_2$            \\
    \hline
    $y = \concat_{dim}(x_1, x_2)$                      & $\type(x_1), \type(x_2), dim$                       & Concatenate $x_1$, $x_2$ across dimension $dim$ \\
    \hline
    $y = \reshape(x)$                                  & $\type(x), \type(y)$                                & Reshape from $\type(x)$ to $\type(y)$           \\
    \hline
    $y = \bitcvt(x)$                                   & $\type(x), \type(y)$                                & Bitcast conversion between basetypes            \\
    \hline
    $y = \broadcast_{dim}(x)$                          & $\type(x), dim$                                     & Broadcast over dimension $dim$                  \\
    \hline
    $y = \reduce_{dim}(x)$                             & $\type(x), dim$                                     & Reduce-sum over dimension $dim$                 \\
    \hline
    $y = \transpose_{[dims]}(x)$                       & $\type(x), dims$                                    & Permute the dimension order to $dims$           \\
    \hline
    \multirow{2}{*}{$y = \tdot_{(c_1,c_2)}(x_1, x_2)$} & \multirow{2}{*}{$\type(x_1), \type(x_2), c_1, c_2$} & Dot product of $x_1$, $x_2$ over                \\
                                                       &                                                     & contracting dimensions $c_1$, $c_2$             \\
    \hline
    $y = \texp(x)$                                     & $\type(x)$                                          & Element-wise exponential                        \\
    \hline
    $y = \tdiv(x_1, x_2)$                              & $\type(x_1), \type(x_2)$                            & Element-wise divide                             \\
    \hline
    $y = \tcopy(x)$                                    & $\type(x)$                                          & Identical copy                                  \\
    \hline
  \end{tabular}
  \caption{Brief descriptions of the tensor operators discussed in this paper (detailed operational semantics in \cite{xla-hlo, stablehlo} and denotational semantics for a subset in \cite{tensorright}).
    $\type(x)$ refers to the tensor-type of tensor variable $x$.
    $dim$ represents the dimension number (starting from 1).
    We use NumPy-like slice notation $x[s_1{:}e_1,s_2{:}e_2,\dots]$ and ignore $\type$-based attributes in the visualizations.}
  \label{app:tab-ops}
  \Description{Table listing the twelve tensor operators used in this paper, each with its operator attributes and a one-line description.}
\end{table}

\subsection{Addressing Operators}

$\slice$ and $\upslice$ are the two operators that name a sub-region of a tensor by address, and they are the pair on which \act{}'s memory allocation rests: a data slice written by one instruction and read by another is expressed as an $\upslice$ followed by a $\slice$, and Appendix~\ref{app:proof} turns exactly that composition into the rewrite rule that discharges Phase~2's correctness.
Note that $\upslice$ returns a tensor of the shape of its \emph{first} operand, not of the sub-tensor being written.

\begin{figure}[htbp]
  \centering
  \includegraphics[width=0.567\textwidth]{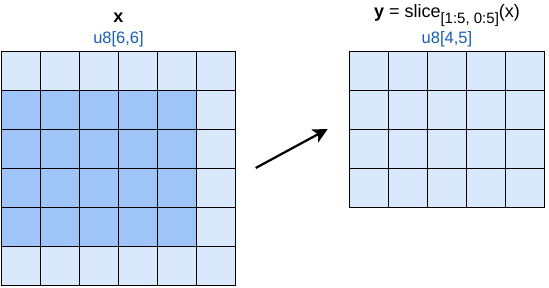}\\
  \small (a)

  \vspace{1.2em}

  \includegraphics[width=0.621\textwidth]{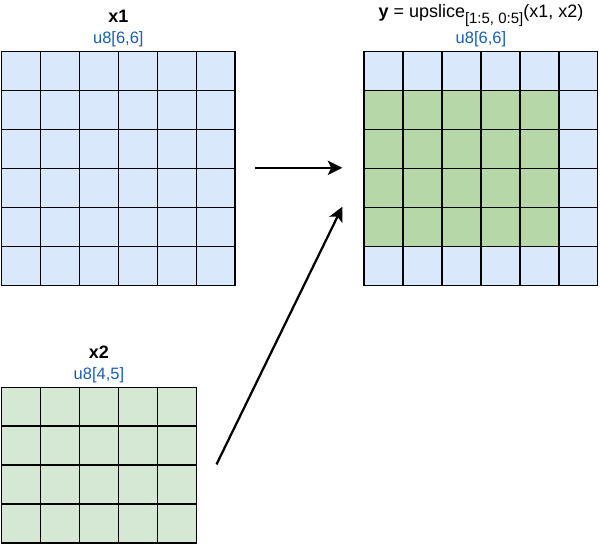}\\
  \small (b)

  \caption{The addressing operators.
    (a) $\slice$ extracts the $4 \times 5$ sub-tensor of $x$ spanning rows $1$--$4$ and columns $0$--$4$; the darker region marks the elements read.
    (b) $\upslice$ returns a copy of $x_1$ with the $4 \times 5$ tensor $x_2$ written into the range $[1{:}5, 0{:}5]$.
    The output therefore has the shape of $x_1$, and the green region records which of its elements came from $x_2$.}
  \label{app:fig-addressing}
  \Description{Two panels showing slice extracting a four-by-five window from a six-by-six grid, and upslice writing a four-by-five grid back into a six-by-six grid.}
\end{figure}

\subsection{Structural Operators}

Every output element of these operators is a copy of some input element: they rearrange, replicate or reinterpret the shape of a tensor without doing arithmetic.
This is exactly the class whose outputs stay blue or green in the visualizations, since each one can be traced back to the operand it came from.
$\concat$ and $\reshape$ are of particular interest to \act{}: $\reshape$ together with $\transpose$ is how \act{} models tiled memory layouts (Appendix~\ref{app:solution}), and $\concat$ is one of the two identity instructions whose availability makes Phase~1 complete.

\begin{figure}[htbp]
  \centering
  \includegraphics[width=0.485\textwidth]{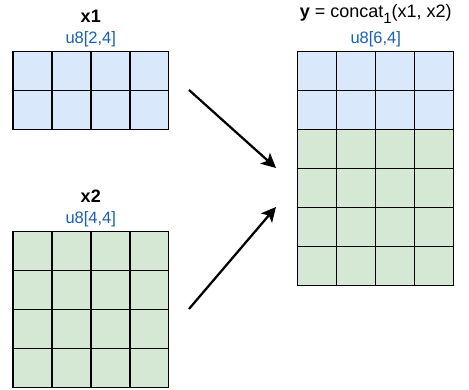}\\
  \small (a)

  \vspace{1.2em}

  \includegraphics[width=0.442\textwidth]{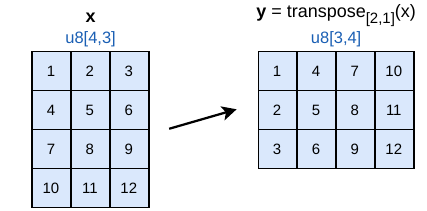}\\
  \small (b)

  \caption{Concatenation and transposition.
    (a) $\concat$ composes a $[6,4]$ tensor from $x_1$ of shape $[2,4]$ and $x_2$ of shape $[4,4]$ by stacking them across dimension~1; the output colours record which operand contributed each row.
    (b) $\transpose$ permutes the dimension order, taking $x$ from $[4,3]$ to $[3,4]$.
    The element values are shown so that the permutation can be read off directly.}
  \label{app:fig-structural}
  \Description{Two panels showing concat stacking a two-by-four and a four-by-four grid into a six-by-four grid, and transpose turning a four-by-three grid into a three-by-four grid.}
\end{figure}

\begin{figure}[htbp]
  \centering
  \includegraphics[width=0.767\textwidth]{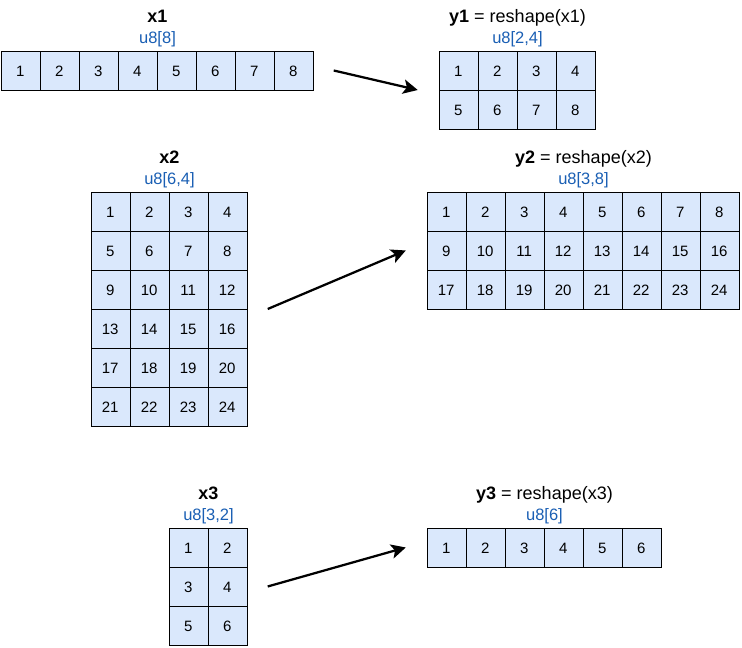}
  \caption{$\reshape$ preserves the row-major element order and changes only the tensor-type, as shown by the element values carrying across each of the three examples: $[8] \to [2,4]$, $[6,4] \to [3,8]$, and $[3,2] \to [6]$.
    Because the element order is fixed, a $\reshape$ is fully determined by its input and output types, which is what allows \act{} to treat it as a layout annotation rather than as data movement.}
  \label{app:fig-reshape}
  \Description{Three examples of reshape, each showing a tensor of one shape and the same numbered elements laid out in a different shape.}
\end{figure}

\begin{figure}[htbp]
  \centering
  \begin{minipage}[t]{0.467\textwidth}
    \centering
    \includegraphics[width=\linewidth]{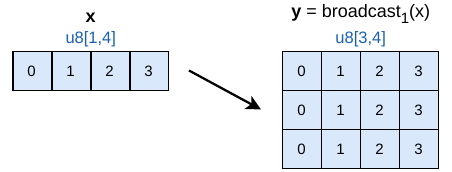}\\
    \small (a)
  \end{minipage}
  \hfill
  \begin{minipage}[t]{0.467\textwidth}
    \centering
    \includegraphics[width=\linewidth]{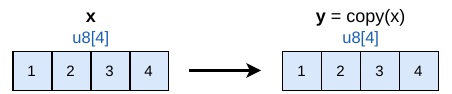}\\
    \small (b)
  \end{minipage}
  \caption{Replication and copying.
    (a) $\broadcast$ replicates $x$ across dimension~1, turning $[1,4]$ into $[3,4]$; every output row is a copy of the single input row.
    (b) $\tcopy$ copies $x$ into $y$ unchanged.
    \act{} keeps $\tcopy$ as an explicit operator even though it changes nothing, because a copy between two tensor buffers is real data movement that the cost model has to account for.}
  \label{app:fig-broadcast-copy}
  \Description{Two panels showing broadcast replicating a one-by-four row into a three-by-four grid, and copy reproducing a four-element vector unchanged.}
\end{figure}

\subsection{Computational Operators}

The remaining operators produce element values that are not present in any operand, and so are drawn in peach.
$\bitcvt$ belongs here even though it invents no information: reinterpreting the bytes of a \texttt{u32} as four \texttt{u8} values yields elements that did not exist before, and its output cannot be traced back cell-for-cell to its input.

\begin{figure}[htbp]
  \centering
  \includegraphics[width=0.614\textwidth]{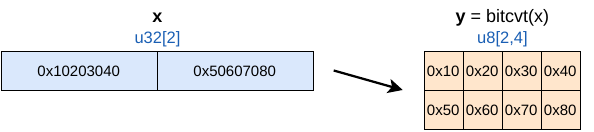}
  \caption{$\bitcvt$ performs an element-wise bitcast between basetypes without changing the underlying bytes.
    Converting from \texttt{u32} to \texttt{u8} therefore turns each input element into four output elements, and the cell widths in the figure are drawn in proportion to the basetype width to make that four-to-one correspondence visible.
    $\bitcvt$ is how \act{} reconciles the byte-addressable view of global memory with the typed view used by tensor operators.}
  \label{app:fig-bitcvt}
  \Description{A bitcast conversion showing two 32-bit hexadecimal values expanding into eight 8-bit hexadecimal values.}
\end{figure}

\begin{figure}[htbp]
  \centering
  \includegraphics[width=0.388\textwidth]{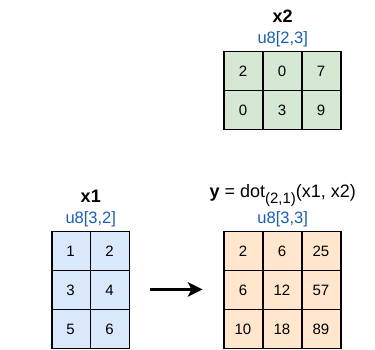}
  \caption{$y = \tdot_{(2,1)}(x_1, x_2)$ matrix-multiplies $x_1$ of shape $[3,2]$ with $x_2$ of shape $[2,3]$, contracting dimension~2 of $x_1$ against dimension~1 of $x_2$ to give $y$ of shape $[3,3]$.
    The operands are placed so that each output element sits at the intersection of the row of $x_1$ and the column of $x_2$ that produce it.}
  \label{app:fig-dot}
  \Description{A matrix multiplication of a three-by-two and a two-by-three tensor, laid out so each output cell sits at the intersection of its contributing row and column.}
\end{figure}

\begin{figure}[htbp]
  \centering
  \begin{minipage}[t]{0.467\textwidth}
    \centering
    \includegraphics[width=\linewidth]{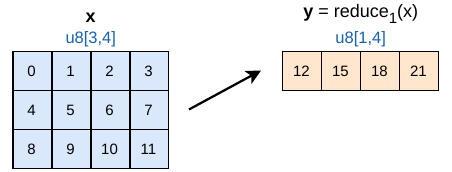}\\
    \small (a)
  \end{minipage}
  \hfill
  \begin{minipage}[t]{0.467\textwidth}
    \centering
    \includegraphics[width=\linewidth]{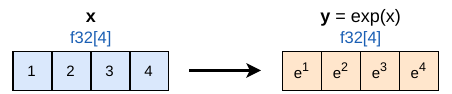}\\
    \small (b)

    \vspace{1.2em}

    \includegraphics[width=\linewidth]{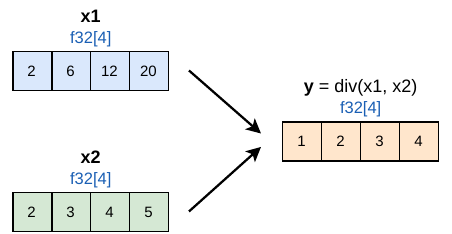}\\
    \small (c)
  \end{minipage}
  \caption{Reduction and element-wise arithmetic.
    (a) $\reduce$ sum-reduces $x$ across dimension~1, turning $[3,4]$ into $[1,4]$; each output element is the sum of the corresponding input column.
    (b) $\texp$ is the element-wise exponential.
    (c) $\tdiv$ divides $x_1$ by $x_2$ element-wise.
    $\reduce$ is the counterpart of the $\broadcast$ of Fig.~\ref{app:fig-broadcast-copy}: both act along one dimension, but only $\reduce$ combines values.}
  \label{app:fig-elementwise}
  \Description{Three panels showing a sum reduction of a three-by-four grid to a one-by-four row, an element-wise exponential, and an element-wise division of two vectors.}
\end{figure}

\clearpage
\section{NNSmith Kernel Generation Setup}
\label{app:nnsmith}

The compilation-time analysis of \S\ref{subsec:evaluation-time} needs tiled kernels that are representative in \emph{size} rather than in workload.
Real Gemmini kernels do not serve: the ones used for the performance comparison were capped at about ten nodes because the baselines were hand-written kernel libraries, which exist only for small, common shapes.
Tiled kernels in practice run to 10--50 nodes, as seen in the TPUGraphs dataset~\cite{tpugraphs}.
We therefore generated 100 synthetic kernels with NNSmith~\cite{nnsmith}, a random DNN generator, and used them both for that compilation-time analysis and for fuzz testing \act{}-Gemmini.
This section records the generation setup so the kernel population can be reproduced exactly.

\subsection{Operator Set}

The generated kernels exercise the 10 \hbox{XLA-HLO} operators listed in Table~\ref{app:tab-nnsmith}, but only six of them --- $\tdot$, $\add$, $\sub$, $\negative$, $\tmin$ and $\tmax$ --- are requested directly by the include-list of Fig.~\ref{app:lst-nnsmith}.
The other four arrive indirectly.
$\treverse$ is not in NNSmith's TensorFlow operator set, so we request \texttt{core.Abs}, which lowers to it; \texttt{core.Abs} is therefore the one include-list entry with no counterpart in Table~\ref{app:tab-nnsmith}.
$\broadcast$ and $\reduce$ are introduced by XLA's own lowering and tiling passes rather than being generated, which is the same way they arise in real kernels.
$\tclamp$ likewise appears only after lowering, when XLA fuses an adjacent $\tmax$ and $\tmin$ pair into a single operator.
We consider this indirection a feature rather than a limitation of the setup: it means the kernel population contains operators in the same combinations that XLA itself produces, rather than only those a generator was told to emit.

\begin{table}[htbp]
  \centering
  \begin{tabular}{|| c | c ||}
    \hline
    Tensor Operator             & Description                                     \\
    \hline\hline
    $y = \tdot(x_1, x_2)$       & Matrix multiplication of $x_1$ and $x_2$        \\
    \hline
    $y = \broadcast_{dim}(x)$   & Broadcast over dimension $dim$                  \\
    \hline
    $y = \reduce_{dim}(x)$      & Reduce-sum over dimension $dim$                 \\
    \hline
    $y = \treverse_{dim}(x)$    & Reverse elements over dimension $dim$           \\
    \hline
    $y = \add(x_1, x_2)$        & Element-wise addition of $x_1$ and $x_2$        \\
    \hline
    $y = \sub(x_1, x_2)$        & Element-wise subtraction of $x_1$ and $x_2$     \\
    \hline
    $y = \negative(x)$          & Element-wise negation of $x$                    \\
    \hline
    $y = \tmin(x, c)$           & Element-wise minimum of $x$ and $c$             \\
    \hline
    $y = \tmax(x, c)$           & Element-wise maximum of $x$ and $c$             \\
    \hline
    $y = \tclamp(low, x, high)$ & Element-wise clamp of $x$ with $low$ and $high$ \\
    \hline
  \end{tabular}
  \caption{The 10 XLA-HLO operators present in the tiled kernels generated for fuzz testing using NNSmith~\cite{nnsmith}.
    Here $c$, $low$ and $high$ denote constant operands.}
  \label{app:tab-nnsmith}
  \Description{Table of the ten XLA-HLO operators appearing in the NNSmith-generated kernels, each with a one-line description.}
\end{table}

\subsection{Generation Parameters and Resulting Kernel Sizes}

Each invocation fixes a seed, a node budget, an operator include-list, a model type and a backend; Fig.~\ref{app:lst-nnsmith} shows one representative command line.

The node budget needs a word of explanation, because the size of the kernel that finally reaches \act{} is neither the budget nor the number of operators NNSmith emits.
\texttt{mgen.max\_nodes} is an upper bound on the operators generated, and the resulting graph has multiple outputs, whereas \act{} consumes single-output tiled kernels.
We therefore combine all outputs with a tree of $\add$ operators, so the node count of the final XLA-HLO graph is
\[
  (\#\text{operators}) + (\#\text{input parameters}) + (\#\text{output nodes} - 1).
\]
The input parameters and the combining $\add$ nodes are what push the final graph above the operator count: with \texttt{mgen.max\_nodes} set to 50, kernels reach up to 89 nodes.
Because the budget is only an upper bound, the number of operators NNSmith actually emits --- and hence the number of outputs that have to be combined --- varies from seed to seed.
Varying \texttt{mgen.seed} across the 100 invocations therefore yields final sizes spanning 7 to 89 nodes, which is the population reported in \S\ref{subsec:evaluation-time}.
That range brackets the 10--50 nodes typical of the TPUGraphs kernels, so the compilation-time measurements cover both smaller and substantially larger inputs than are seen in practice.

\begin{figure}[htbp]
  \begin{lstlisting}[style={shell},frame=single]
nnsmith.model_gen model.type=tensorflow backend.type=xla \
    mgen.seed=1002 \
    mgen.max_nodes=50 \
    mgen.include="[tensorflow.TFMatMul, core.Abs, core.Add, core.Max, core.Min, core.Neg, core.Sub]" \
    model.path="nnsmith_output/" \
    debug.viz=true
  \end{lstlisting}
  \caption{A representative NNSmith invocation used to generate one of the 100 synthetic kernels.
    Only \texttt{mgen.seed} varies across the population.}
  \label{app:lst-nnsmith}
  \Description{A shell command invoking nnsmith.model_gen with a TensorFlow model type, XLA backend, a fixed seed, a maximum node count of 50, and an operator include list.}
\end{figure}

\clearpage
\section{Loop Fission Factors for the Exo Kernels}
\label{app:exo}

\S\ref{subsec:evaluation-perf-gemmini} compares \act{}-Gemmini against Exo's expert-written kernel library on six GEMM shapes and reports one speedup per shape.
That single number is the outcome of a search: for each shape, \act{}-Gemmini enumerates candidate loop fission factors and its cost model selects one.
This section gives the full candidate tables behind those six numbers, so that the reported speedup can be seen in context.

\subsection{Reading the Tables}
\label{app:exo-reading}

We evaluated every candidate on Gemmini's Verilator-based RTL simulator and recorded its cycle count.
Each row is one combination of loop fission factors, together with the accumulator and scratchpad rows it occupies, its measured cycle count, and its speedup over the original hand-optimized Exo kernel.

The five loop variables split into three levels.
The outer loops $io$ and $jo$ control outer tiling.
The inner loops $i$ and $j$ are the fine-grained tile sizes that load and store into Gemmini's scratchpad.
The innermost loop $ko$ iterates over the partial sums held in the accumulator: within one $ko$ iteration the kernel loads a single $16 \times 64$ tile of $A$ and four $16 \times 64$ tiles of $B$, multiplies them as $16 \times 16$ matrix multiplications, accumulates the partial sums, and so computes one $16 \times 64$ tile of the output $C$.
Candidates whose factors exceed the accumulator or scratchpad row limits are infeasible and are not listed.

In every table the original Exo factors are \textbf{bolded} and the best factors, meaning those with the lowest cycle count, are \textcolor{myorange}{\textbf{highlighted in orange}}.
Where the two coincide, \act{}-Gemmini rediscovered Exo's hand-tuned factors and the speedup is 1.00x.
Where the highlighted row has a lower cycle count than the bolded row, \act{}-Gemmini found a better loop fission factor than the expert did.
This happened for three of the six kernels, which is the result summarized in \S\ref{subsec:evaluation-perf-gemmini}.

\subsection{Candidate Tables}
\label{app:exo-tables}

The six tables follow in the order $512{\times}512{\times}512$, $784{\times}1024{\times}256$, $12544{\times}256{\times}64$, $12544{\times}64{\times}256$, $3136{\times}512{\times}128$, $3136{\times}128{\times}512$.
Note that this is \emph{not} the left-to-right order of the bars in Fig.~\ref{fig:perf-gemmini}, which places $784{\times}1024{\times}256$ last rather than second; match tables to bars by shape rather than by position.

\begin{table}[htbp]
  \centering\footnotesize
  \begin{tabular}{||c|c|c|c|c|c|c|c|c||}
    \hline
    io                             & i                                & jo                                 & j                                   & ko                            & Accumulator Rows & Scratchpad Rows & Cycles  & Speedup \\
    \hline
    \textcolor{myorange}{\textbf{2}} & \textcolor{myorange}{\textbf{16}}  & \textcolor{myorange}{\textbf{2}}     & \textcolor{myorange}{\textbf{4}}      &
    \textcolor{myorange}{\textbf{8}} & \textcolor{myorange}{\textbf{256}} & \textcolor{myorange}{\textbf{16384}} & \textcolor{myorange}{\textbf{624170}} & \textcolor{myorange}{\textbf{1.00}}                                                          \\
    2                              & 16                               & 4                                  & 2                                   & 8                             & 128              & 12288           & 624228  & 1.00    \\
    4                              & 8                                & 2                                  & 4                                   & 8                             & 256              & 12288           & 659384  & 0.95    \\
    2                              & 16                               & 8                                  & 1                                   & 8                             & 64               & 10240           & 650725  & 0.96    \\
    8                              & 4                                & 2                                  & 4                                   & 8                             & 256              & 10240           & 855191  & 0.73    \\
    16                             & 2                                & 2                                  & 4                                   & 8                             & 256              & 9216            & 1039928 & 0.60    \\
    32                             & 1                                & 2                                  & 4                                   & 8                             & 256              & 8704            & 1351586 & 0.46    \\
    4                              & 8                                & 4                                  & 2                                   & 8                             & 128              & 8192            & 659357  & 0.95    \\
    4                              & 8                                & 8                                  & 1                                   & 8                             & 64               & 6144            & 687626  & 0.91    \\
    8                              & 4                                & 4                                  & 2                                   & 8                             & 128              & 6144            & 965931  & 0.65    \\
    16                             & 2                                & 4                                  & 2                                   & 8                             & 128              & 5120            & 1223264 & 0.51    \\
    32                             & 1                                & 4                                  & 2                                   & 8                             & 128              & 4608            & 1549222 & 0.40    \\
    8                              & 4                                & 8                                  & 1                                   & 8                             & 64               & 4096            & 1240902 & 0.50    \\
    16                             & 2                                & 8                                  & 1                                   & 8                             & 64               & 3072            & 1385353 & 0.45    \\
    32                             & 1                                & 8                                  & 1                                   & 8                             & 64               & 2560            & 1663782 & 0.38    \\
    \hline
  \end{tabular}
  \caption{Loop fission candidates for matmul $512 \times 512 \times 512$.
    \act{}-Gemmini's best factor coincides with Exo's hand-tuned factor, giving a 1.00x speedup.}
  \label{app:tab-matmul512}
  \Description{Table of loop fission candidates for a 512 by 512 by 512 matrix multiplication, listing loop factors, accumulator and scratchpad rows, cycle counts and speedups.}
\end{table}
\begin{table}[htbp]
  \centering\footnotesize
  \begin{tabular}{||c|c|c|c|c|c|c|c|c||}
    \hline
    io                             & i                               & jo                             & j                              & ko                             & Accumulator Rows                 & Scratchpad Rows                    & Cycles                              & Speedup                       \\
    \hline
    \textcolor{myorange}{\textbf{1}} & \textcolor{myorange}{\textbf{49}} & \textcolor{myorange}{\textbf{8}} & \textcolor{myorange}{\textbf{2}} & \textcolor{myorange}{\textbf{4}} & \textcolor{myorange}{\textbf{128}} & \textcolor{myorange}{\textbf{14592}} & \textcolor{myorange}{\textbf{968206}} & \textcolor{myorange}{\textbf{1.30}} \\
    1                              & 49                              & 16                             & 1                              & 4                              & 64                               & 13568                              & 1126936                             & 1.12                          \\
    \textbf{7}                     & \textbf{7}                      & \textbf{2}                     & \textbf{8}                     & \textbf{4}                     & \textbf{512}                     & \textbf{9984}                      & \textbf{1258985}                    & \textbf{1.00}                 \\
    7                              & 7                               & 4                              & 4                              & 4                              & 256                              & 5888                               & 1457055                             & 0.86                          \\
    7                              & 7                               & 8                              & 2                              & 4                              & 128                              & 3840                               & 1757805                             & 0.72                          \\
    7                              & 7                               & 16                             & 1                              & 4                              & 64                               & 2816                               & 1857018                             & 0.68                          \\
    49                             & 1                               & 2                              & 8                              & 4                              & 512                              & 8448                               & 2105258                             & 0.60                          \\
    49                             & 1                               & 4                              & 4                              & 4                              & 256                              & 4352                               & 2428470                             & 0.52                          \\
    49                             & 1                               & 8                              & 2                              & 4                              & 128                              & 2304                               & 2442627                             & 0.52                          \\
    49                             & 1                               & 16                             & 1                              & 4                              & 64                               & 1280                               & 2634956                             & 0.48                          \\
    \hline
  \end{tabular}
  \caption{Loop fission candidates for matmul $784 \times 1024 \times 256$.
    \act{}-Gemmini's best factor outperforms Exo's hand-tuned factor, giving a 1.30x speedup.}
  \label{app:tab-matmul27}
  \Description{Table of loop fission candidates for a 784 by 1024 by 256 matrix multiplication, listing loop factors, accumulator and scratchpad rows, cycle counts and speedups.}
\end{table}
\begin{table}[htbp]
  \centering\footnotesize
  \begin{tabular}{||c|c|c|c|c|c|c|c|c||}
    \hline
    io                             & i                                & jo                             & j                              & ko                             & Accumulator Rows                 & Scratchpad Rows                    & Cycles                               & Speedup                       \\
    \hline
    \textcolor{myorange}{\textbf{4}} & \textcolor{myorange}{\textbf{196}} & \textcolor{myorange}{\textbf{1}} & \textcolor{myorange}{\textbf{4}} & \textcolor{myorange}{\textbf{1}} & \textcolor{myorange}{\textbf{256}} & \textcolor{myorange}{\textbf{13568}} & \textcolor{myorange}{\textbf{1499481}} & \textcolor{myorange}{\textbf{1.00}} \\
    4                              & 196                              & 2                              & 2                              & 1                              & 128                              & 13056                              & 1513787                              & 0.99                          \\
    4                              & 196                              & 4                              & 1                              & 1                              & 64                               & 12800                              & 2027815                              & 0.74                          \\
    7                              & 112                              & 1                              & 4                              & 1                              & 256                              & 8192                               & 1569955                              & 0.96                          \\
    7                              & 112                              & 2                              & 2                              & 1                              & 128                              & 7680                               & 1602036                              & 0.94                          \\
    7                              & 112                              & 4                              & 1                              & 1                              & 64                               & 7424                               & 2059381                              & 0.73                          \\
    8                              & 98                               & 1                              & 4                              & 1                              & 256                              & 7296                               & 1579550                              & 0.95                          \\
    8                              & 98                               & 2                              & 2                              & 1                              & 128                              & 6784                               & 1604309                              & 0.93                          \\
    8                              & 98                               & 4                              & 1                              & 1                              & 64                               & 6528                               & 1842807                              & 0.81                          \\
    14                             & 56                               & 1                              & 4                              & 1                              & 256                              & 4608                               & 1702098                              & 0.88                          \\
    16                             & 49                               & 1                              & 4                              & 1                              & 256                              & 4160                               & 1781546                              & 0.84                          \\
    14                             & 56                               & 2                              & 2                              & 1                              & 128                              & 4096                               & 1540089                              & 0.97                          \\
    14                             & 56                               & 4                              & 1                              & 1                              & 64                               & 3840                               & 2003559                              & 0.75                          \\
    16                             & 49                               & 2                              & 2                              & 1                              & 128                              & 3648                               & 1696348                              & 0.88                          \\
    16                             & 49                               & 4                              & 1                              & 1                              & 64                               & 3392                               & 2103647                              & 0.71                          \\
    28                             & 28                               & 1                              & 4                              & 1                              & 256                              & 2816                               & 1737137                              & 0.86                          \\
    28                             & 28                               & 2                              & 2                              & 1                              & 128                              & 2304                               & 1720557                              & 0.87                          \\
    28                             & 28                               & 4                              & 1                              & 1                              & 64                               & 2048                               & 2119776                              & 0.71                          \\
    49                             & 16                               & 1                              & 4                              & 1                              & 256                              & 2048                               & 1738821                              & 0.86                          \\
    56                             & 14                               & 1                              & 4                              & 1                              & 256                              & 1920                               & 1733756                              & 0.86                          \\
    49                             & 16                               & 2                              & 2                              & 1                              & 128                              & 1536                               & 1742679                              & 0.86                          \\
    98                             & 8                                & 1                              & 4                              & 1                              & 256                              & 1536                               & 1738183                              & 0.86                          \\
    112                            & 7                                & 1                              & 4                              & 1                              & 256                              & 1472                               & 1738658                              & 0.86                          \\
    56                             & 14                               & 2                              & 2                              & 1                              & 128                              & 1408                               & 1743286                              & 0.86                          \\
    196                            & 4                                & 1                              & 4                              & 1                              & 256                              & 1280                               & 1716865                              & 0.87                          \\
    49                             & 16                               & 4                              & 1                              & 1                              & 64                               & 1280                               & 2183789                              & 0.69                          \\
    392                            & 2                                & 1                              & 4                              & 1                              & 256                              & 1152                               & 1720090                              & 0.87                          \\
    56                             & 14                               & 4                              & 1                              & 1                              & 64                               & 1152                               & 2184509                              & 0.69                          \\
    784                            & 1                                & 1                              & 4                              & 1                              & 256                              & 1088                               & 1793534                              & 0.84                          \\
    98                             & 8                                & 2                              & 2                              & 1                              & 128                              & 1024                               & 1775427                              & 0.84                          \\
    112                            & 7                                & 2                              & 2                              & 1                              & 128                              & 960                                & 1745616                              & 0.86                          \\
    196                            & 4                                & 2                              & 2                              & 1                              & 128                              & 768                                & 1752644                              & 0.86                          \\
    98                             & 8                                & 4                              & 1                              & 1                              & 64                               & 768                                & 2186091                              & 0.69                          \\
    112                            & 7                                & 4                              & 1                              & 1                              & 64                               & 704                                & 2105043                              & 0.71                          \\
    392                            & 2                                & 2                              & 2                              & 1                              & 128                              & 640                                & 1802506                              & 0.83                          \\
    784                            & 1                                & 2                              & 2                              & 1                              & 128                              & 576                                & 1883025                              & 0.80                          \\
    196                            & 4                                & 4                              & 1                              & 1                              & 64                               & 512                                & 2175743                              & 0.69                          \\
    392                            & 2                                & 4                              & 1                              & 1                              & 64                               & 384                                & 2114435                              & 0.71                          \\
    784                            & 1                                & 4                              & 1                              & 1                              & 64                               & 320                                & 2146024                              & 0.70                          \\
    \hline
  \end{tabular}
  \caption{Loop fission candidates for matmul $12544 \times 256 \times 64$.
    \act{}-Gemmini's best factor coincides with Exo's hand-tuned factor, giving a 1.00x speedup.}
  \label{app:tab-matmul4}
  \Description{Table of loop fission candidates for a 12544 by 256 by 64 matrix multiplication, listing loop factors, accumulator and scratchpad rows, cycle counts and speedups.}
\end{table}
\begin{table}[htbp]
  \centering\footnotesize
  \begin{tabular}{||c|c|c|c|c|c|c|c|c||}
    \hline
    io                              & i                               & jo                             & j                              & ko                             & Accumulator Rows                  & Scratchpad Rows                   & Cycles                               & Speedup                       \\
    \hline
    \textcolor{myorange}{\textbf{49}} & \textcolor{myorange}{\textbf{16}} & \textcolor{myorange}{\textbf{1}} & \textcolor{myorange}{\textbf{1}} & \textcolor{myorange}{\textbf{4}} & \textcolor{myorange}{\textbf{1024}} & \textcolor{myorange}{\textbf{5120}} & \textcolor{myorange}{\textbf{1462706}} & \textcolor{myorange}{\textbf{1.29}} \\
    56                              & 14                              & 1                              & 1                              & 4                              & 896                               & 4608                              & 1800241                              & 1.05                          \\
    \textbf{98}                     & \textbf{8}                      & \textbf{1}                     & \textbf{1}                     & \textbf{4}                     & \textbf{512}                      & \textbf{3072}                     & \textbf{1884887}                     & \textbf{1.00}                 \\
    112                             & 7                               & 1                              & 1                              & 4                              & 448                               & 2816                              & 2114836                              & 0.89                          \\
    196                             & 4                               & 1                              & 1                              & 4                              & 256                               & 2048                              & 2110773                              & 0.89                          \\
    392                             & 2                               & 1                              & 1                              & 4                              & 128                               & 1536                              & 1883721                              & 1.00                          \\
    784                             & 1                               & 1                              & 1                              & 4                              & 64                                & 1280                              & 2171151                              & 0.87                          \\
    \hline
  \end{tabular}
  \caption{Loop fission candidates for matmul $12544 \times 64 \times 256$.
    \act{}-Gemmini's best factor outperforms Exo's hand-tuned factor, giving a 1.29x speedup.}
  \label{app:tab-matmul6}
  \Description{Table of loop fission candidates for a 12544 by 64 by 256 matrix multiplication, listing loop factors, accumulator and scratchpad rows, cycle counts and speedups.}
\end{table}
\begin{table}[htbp]
  \centering\footnotesize
  \begin{tabular}{||c|c|c|c|c|c|c|c|c||}
    \hline
    io                             & i                               & jo                             & j                              & ko                             & Accumulator Rows                 & Scratchpad Rows                    & Cycles                               & Speedup                       \\
    \hline
    \textcolor{myorange}{\textbf{2}} & \textcolor{myorange}{\textbf{98}} & \textcolor{myorange}{\textbf{2}} & \textcolor{myorange}{\textbf{4}} & \textcolor{myorange}{\textbf{2}} & \textcolor{myorange}{\textbf{256}} & \textcolor{myorange}{\textbf{14592}} & \textcolor{myorange}{\textbf{1113009}} & \textcolor{myorange}{\textbf{1.06}} \\
    2                              & 98                              & 4                              & 2                              & 2                              & 128                              & 13568                              & 1205199                              & 0.98                          \\
    2                              & 98                              & 8                              & 1                              & 2                              & 64                               & 13056                              & 1433162                              & 0.82                          \\
    \textbf{4}                     & \textbf{49}                     & \textbf{1}                     & \textbf{8}                     & \textbf{2}                     & \textbf{512}                     & \textbf{10368}                     & \textbf{1179333}                     & \textbf{1.00}                 \\
    4                              & 49                              & 2                              & 4                              & 2                              & 256                              & 8320                               & 1239771                              & 0.95                          \\
    7                              & 28                              & 1                              & 8                              & 2                              & 512                              & 7680                               & 1156398                              & 1.02                          \\
    4                              & 49                              & 4                              & 2                              & 2                              & 128                              & 7296                               & 1224648                              & 0.96                          \\
    4                              & 49                              & 8                              & 1                              & 2                              & 64                               & 6784                               & 1328323                              & 0.89                          \\
    14                             & 14                              & 1                              & 8                              & 2                              & 512                              & 5888                               & 1426823                              & 0.83                          \\
    7                              & 28                              & 2                              & 4                              & 2                              & 256                              & 5632                               & 1173115                              & 1.01                          \\
    28                             & 7                               & 1                              & 8                              & 2                              & 512                              & 4992                               & 1562962                              & 0.75                          \\
    49                             & 4                               & 1                              & 8                              & 2                              & 512                              & 4608                               & 1641041                              & 0.72                          \\
    7                              & 28                              & 4                              & 2                              & 2                              & 128                              & 4608                               & 1172983                              & 1.01                          \\
    98                             & 2                               & 1                              & 8                              & 2                              & 512                              & 4352                               & 1667040                              & 0.71                          \\
    196                            & 1                               & 1                              & 8                              & 2                              & 512                              & 4224                               & 1691607                              & 0.70                          \\
    7                              & 28                              & 8                              & 1                              & 2                              & 64                               & 4096                               & 1324928                              & 0.89                          \\
    14                             & 14                              & 2                              & 4                              & 2                              & 256                              & 3840                               & 1393443                              & 0.85                          \\
    28                             & 7                               & 2                              & 4                              & 2                              & 256                              & 2944                               & 1520771                              & 0.78                          \\
    14                             & 14                              & 4                              & 2                              & 2                              & 128                              & 2816                               & 1393532                              & 0.85                          \\
    49                             & 4                               & 2                              & 4                              & 2                              & 256                              & 2560                               & 1611814                              & 0.73                          \\
    14                             & 14                              & 8                              & 1                              & 2                              & 64                               & 2304                               & 1589328                              & 0.74                          \\
    98                             & 2                               & 2                              & 4                              & 2                              & 256                              & 2304                               & 1656962                              & 0.71                          \\
    196                            & 1                               & 2                              & 4                              & 2                              & 256                              & 2176                               & 1692347                              & 0.70                          \\
    28                             & 7                               & 4                              & 2                              & 2                              & 128                              & 1920                               & 1520284                              & 0.78                          \\
    49                             & 4                               & 4                              & 2                              & 2                              & 128                              & 1536                               & 1615518                              & 0.73                          \\
    28                             & 7                               & 8                              & 1                              & 2                              & 64                               & 1408                               & 1689620                              & 0.70                          \\
    98                             & 2                               & 4                              & 2                              & 2                              & 128                              & 1280                               & 1656610                              & 0.71                          \\
    196                            & 1                               & 4                              & 2                              & 2                              & 128                              & 1152                               & 1691181                              & 0.70                          \\
    49                             & 4                               & 8                              & 1                              & 2                              & 64                               & 1024                               & 1782025                              & 0.66                          \\
    98                             & 2                               & 8                              & 1                              & 2                              & 64                               & 768                                & 1817532                              & 0.65                          \\
    196                            & 1                               & 8                              & 1                              & 2                              & 64                               & 640                                & 1836277                              & 0.64                          \\
    \hline
  \end{tabular}
  \caption{Loop fission candidates for matmul $3136 \times 512 \times 128$.
    \act{}-Gemmini's best factor outperforms Exo's hand-tuned factor, giving a 1.06x speedup.}
  \label{app:tab-matmul14}
  \Description{Table of loop fission candidates for a 3136 by 512 by 128 matrix multiplication, listing loop factors, accumulator and scratchpad rows, cycle counts and speedups.}
\end{table}
\begin{table}[htbp]
  \centering\footnotesize
  \begin{tabular}{||c|c|c|c|c|c|c|c|c||}
    \hline
    io                              & i                               & jo                             & j                              & ko                             & Accumulator Rows                 & Scratchpad Rows                    & Cycles                               & Speedup                       \\
    \hline
    \textcolor{myorange}{\textbf{14}} & \textcolor{myorange}{\textbf{14}} & \textcolor{myorange}{\textbf{1}} & \textcolor{myorange}{\textbf{2}} & \textcolor{myorange}{\textbf{8}} & \textcolor{myorange}{\textbf{128}} & \textcolor{myorange}{\textbf{11264}} & \textcolor{myorange}{\textbf{1145575}} & \textcolor{myorange}{\textbf{1.00}} \\
    14                              & 14                              & 2                              & 1                              & 8                              & 128                              & 11264                              & 1205532                              & 0.95                          \\
    28                              & 7                               & 1                              & 2                              & 8                              & 128                              & 7680                               & 1150215                              & 1.00                          \\
    28                              & 7                               & 2                              & 1                              & 8                              & 128                              & 7680                               & 1215381                              & 0.94                          \\
    49                              & 4                               & 1                              & 2                              & 8                              & 128                              & 6144                               & 1408841                              & 0.81                          \\
    49                              & 4                               & 2                              & 1                              & 8                              & 128                              & 6144                               & 1419196                              & 0.81                          \\
    98                              & 2                               & 1                              & 2                              & 8                              & 128                              & 5120                               & 1599727                              & 0.72                          \\
    98                              & 2                               & 2                              & 1                              & 8                              & 128                              & 5120                               & 1639071                              & 0.70                          \\
    196                             & 1                               & 1                              & 2                              & 8                              & 128                              & 4608                               & 1699842                              & 0.67                          \\
    196                             & 1                               & 2                              & 1                              & 8                              & 128                              & 4608                               & 1739270                              & 0.66                          \\
    \hline
  \end{tabular}
  \caption{Loop fission candidates for matmul $3136 \times 128 \times 512$.
    \act{}-Gemmini's best factor coincides with Exo's hand-tuned factor, giving a 1.00x speedup.}
  \label{app:tab-matmul16}
  \Description{Table of loop fission candidates for a 3136 by 128 by 512 matrix multiplication, listing loop factors, accumulator and scratchpad rows, cycle counts and speedups.}
\end{table}

\clearpage
\section{Detailed Analysis of \act{}-Gemmini versus the Gemmini SW Library}
\label{app:gemmini}

\S\ref{subsec:evaluation-perf-gemmini} evaluates \act{}-Gemmini against the Gemmini software library over three groups of kernels: \emph{``supported''}, where the library has a hand-written implementation; \emph{``composite''}, which compose several library calls; and \emph{``not supported''}, whose operators have no Gemmini implementation at all.
It reports the headline numbers and defers the per-kernel discussion here.
This section takes the last two groups in turn, explains where the speedups come from, and then reports two further results: an experiment with Gemmini's CISC-type instructions, and the convolution kernels.

\begin{figure}[htbp]
  \centering
  \includegraphics[width=0.98\textwidth]{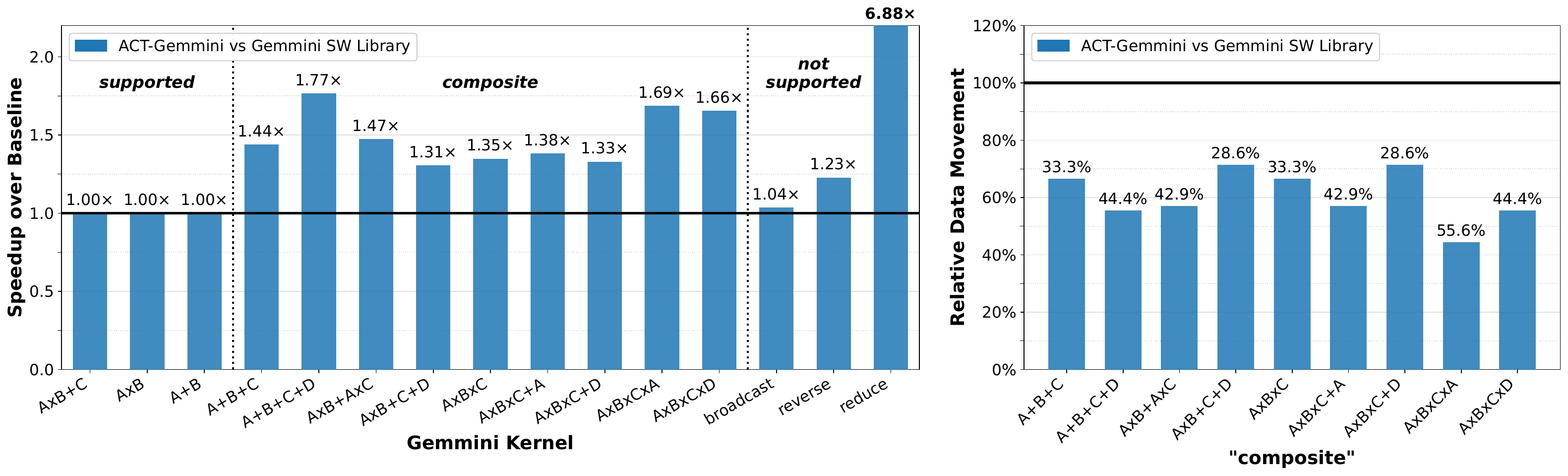}
  \caption{\act{}-Gemmini against the Gemmini SW library.
    Left: speedup of the compiled code across all three kernel groups, reproduced from Fig.~\ref{fig:perf-gemmini} so that the two panels can be read side by side.
    Right: data movement of \act{}-Gemmini relative to the library for the \emph{``composite''} kernels, where lower is better; the number printed on each bar is the \emph{reduction}, that is $100\%$ minus the plotted height.
    Note that the \emph{``not supported''} bars on the left are measured against host execution, since those operators have no Gemmini implementation in the library.}
  \label{app:fig-perf-gemmini}
  \Description{Two bar charts. The left chart shows per-kernel speedup of ACT-Gemmini over the Gemmini SW library, grouped into supported, composite and not-supported kernels. The right chart shows relative data movement for the composite kernels.}
\end{figure}

\subsection{\emph{``composite''} Kernels}
\label{app:gemmini-composite}

A \emph{``composite''} kernel is a composition of several GEMM ($\times$) and ADD ($+$) operations that the library can only execute as a sequence of separate calls.
Each library call must leave its result in HBM, because the library's calling convention cannot assume anything about scratchpad residency across calls.
The next call then loads that same tensor straight back in.
\act{}-Gemmini has no such constraint: it compiles the whole kernel at once, so an intermediate that is produced and immediately consumed never leaves the accelerator.

\begin{figure}[htbp]
  \centering
  \includegraphics[width=0.52\textwidth]{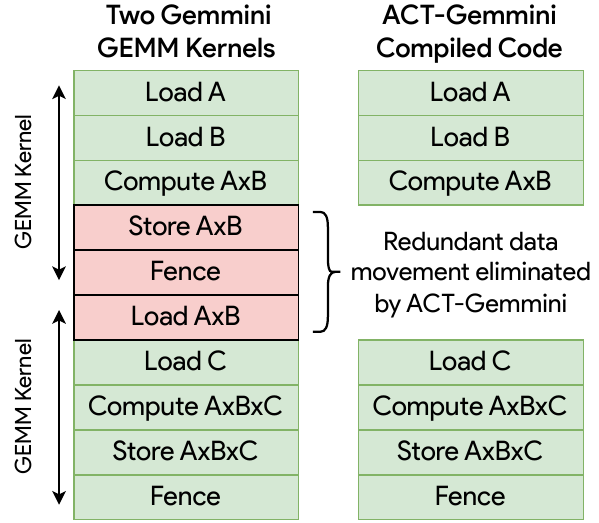}
  \caption{Breakdown of the Gemmini SW library kernels and the \act{}-Gemmini compiled code for $A \times B \times C$.
    \act{}-Gemmini eliminates the redundant load-store instructions between library calls (shaded red).}
  \label{app:fig-composite}
  \Description{Side-by-side instruction sequences. The library version issues two GEMM kernels separated by a store, a fence and a reload; the ACT-Gemmini version issues one fused sequence without them.}
\end{figure}

Fig.~\ref{app:fig-composite} makes the difference concrete for $A \times B \times C$.
The library issues two GEMM kernels, and the store of $A{\times}B$, the fence, and the reload of $A{\times}B$ between them (shaded red) are pure overhead.
\act{}-Gemmini keeps the intermediate in the scratchpad and emits neither.
For $A \times B \times C$ the library moves six tiles where \act{}-Gemmini moves four, a \textbf{33.3\%} reduction, and the measured speedup is \textbf{1.35x}.
The saving grows with scratchpad utilization, because a larger resident intermediate displaces proportionally more traffic, and it grows with the length of the chain, because each additional library call contributes another store-fence-load round trip that \act{}-Gemmini does not emit.

The same pattern holds across the other \emph{``composite''} kernels, since each additional library call in the baseline adds another store-fence-load round trip that \act{}-Gemmini removes.
The gain is not monotone in kernel size, however: how much is saved also depends on how much of the intermediate stays resident, so two of the three-call kernels in Fig.~\ref{app:fig-perf-gemmini} score slightly below the best two-call one.
For chains of GEMM, the speedup goes from 1x for $A \times B$, to 1.35x for $A \times B \times C$, to 1.66x for $A \times B \times C \times D$; for chains of ADD it goes from 1x, to 1.44x, to 1.77x.
Taken over the group, Fig.~\ref{app:fig-perf-gemmini} shows \act{}-Gemmini outperforming the library by up to \textbf{1.77x} (geomean 1.48x) while reducing data movement by up to \textbf{55.6\%} (average 39.3\%).

\subsection{\emph{``not supported''} Kernels}
\label{app:gemmini-unsupported}

The Gemmini SW library has no implementation of $\broadcast$, $\treverse$ or $\reduce$, so a conventional toolchain has no choice but to run them on the host Rocket CPU.
\act{}-Gemmini instead compiles all three onto the systolic array, and it does so without any operator-specific support.

\begin{figure}[htbp]
  \centering
  \includegraphics[width=0.95\textwidth]{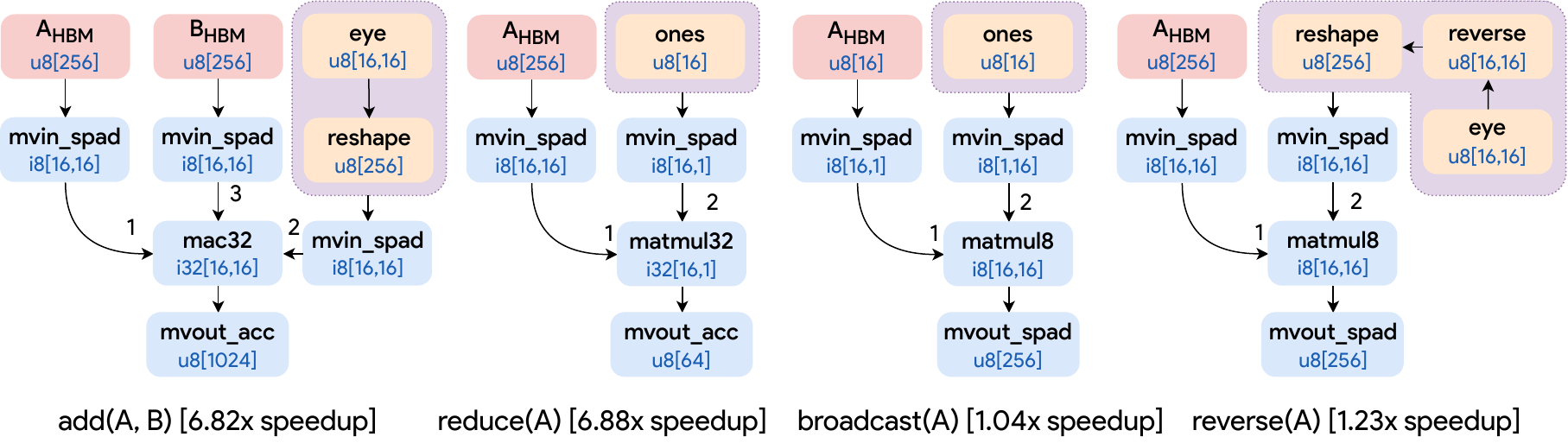}
  \caption{\pii{} graphs with constant memory (purple region) generated by \act{}-Gemmini.
    The orange nodes hold the values materialized in constant memory: \textsf{eye} is an identity matrix and \textsf{ones} is a vector of all ones.
    Speedups are with respect to the Gemmini host processor (Rocket chip).}
  \label{app:fig-gemmini-constant}
  \Description{Four pii graphs, for add, reduce, broadcast and reverse, each showing an input loaded into the scratchpad and multiplied against a constant matrix materialized in a shaded constant-memory region.}
\end{figure}

The mechanism is compile-time constant detection (\S\ref{subsec:module-3}).
Each of these three operators can be written as a matrix multiplication against a fixed matrix: $\broadcast$ against a $[1,16]$ all-ones vector, $\reduce$ against a $[16,1]$ all-ones vector, and $\treverse$ against a reversed identity matrix.
\act{}-Gemmini detects that those operands are compile-time constants, materializes them in constant memory, and emits an ordinary \textsf{matmul}.
Fig.~\ref{app:fig-gemmini-constant} shows the resulting \pii{} graphs, with the constant memory region shaded purple and the constant-valued nodes \textsf{eye} (the identity matrix) and \textsf{ones} (the all-ones vector) in orange.
Nothing in this is Gemmini-specific: the same constant detection would apply to any accelerator whose ISA description exposes a matrix-multiply instruction.

Because the baseline here is host execution rather than an accelerator kernel, the comparison is \emph{cross-architecture}, and it should be read as the practical cost of host fallback rather than as an accelerator-to-accelerator speedup.

The outcome varies sharply across the three operators, and it is worth being precise about why.
$\reduce$ gains the most, at \textbf{6.88x} (\S\ref{subsec:evaluation-perf-gemmini}): it is the only one of the three that does real arithmetic proportional to its input, so moving it off the Rocket CPU removes a genuine compute bottleneck.
$\treverse$ gains \textbf{1.23x} and $\broadcast$ only \textbf{1.04x}.
Both are pure data movement, so on the host they are already memory-bound; expressing them as a systolic-array \textsf{matmul} against a constant removes the host round trip but replaces it with an accelerator that is no faster at moving the same bytes.
The value of compiling them is therefore not raw speed but \emph{coverage} --- the kernel no longer has to leave the accelerator at all, which is what makes the surrounding \emph{``composite''} fusions of \S\ref{app:gemmini-composite} possible in the first place.
For reference on the same axis, \texttt{add(A,B)} reaches 6.82x over host execution, though it is a \emph{``supported''} kernel that the library does implement.

\subsection{CISC-type Instructions}
\label{app:gemmini-cisc}

The \idl{} description used to generate \act{}-Gemmini covers Gemmini's 11 RISC-type instructions.
Gemmini~\cite{gemmini} also provides experimental CISC-type instructions, which perform a matrix multiplication over a large tile in a single instruction and stream data directly to and from HBM.
These are implemented with a hardware FSM that double-buffers data movement against computation.
That gives them an advantage a RISC-type sequence cannot match: a compiler emitting separate load, compute and store instructions is limited by the instruction scheduler's reordering and pipelining window, whereas the FSM overlaps the two by construction.
We therefore expected CISC-type instructions to win even where a RISC-type schedule moves less data.

To test this, we added the CISC-type instructions to the \idl{} description and generated a new backend, \act{}-Gemmini+.
The outcome was as predicted.
\act{}-Gemmini+ still enumerated the RISC-only assembly among its candidates --- the RISC-type instructions remained in the description, so those candidates were never lost --- but its cost model ranked a CISC-type candidate above them, and that is what it emitted.

The result itself is unsurprising; what matters is what it took to obtain.
\act{} required \emph{no} algorithmic changes to exploit the new instruction class --- only the \idl{} description was extended, and the cost-model-driven enumeration of \S\ref{sec:cost-model} preferred the faster code automatically.
A hand-written backend would have needed a new scheduling strategy to use an FSM-driven instruction well.
This is evidence that \act{} generalizes across RISC- and CISC-style instructions of the same accelerator.

\subsection{Convolution Kernels}
\label{app:gemmini-conv}

Convolutions on Gemmini are lowered onto the same systolic array as matrix multiplications and share its compute unit, so they exercise the same instruction selection and memory allocation decisions.
We observed correspondingly similar results for convolution kernels, both against the Gemmini SW library and against Exo's expert-written kernel library.
Since they add no separate insight, \S\ref{subsec:evaluation-perf-gemmini} reports no convolution kernels.

\clearpage
\section{The oneDNN Kernels Compiled by \act{}-AMX}
\label{app:amx}

\S\ref{subsec:evaluation-perf-amx} evaluates \act{}-AMX against oneDNN, a library hand-written and heavily optimized for Intel ISA including AMX and AVX512.
The comparison is unusual among our evaluations in that the target is not a speedup but a \emph{match}: oneDNN's assembly is already expert-tuned, and the question is whether an automatically generated backend can reach it.
This section presents the five kernels behind that claim so the match can be inspected.

For each kernel we give three artifacts: its tensor computation graph, the XLA-HLO IR that \act{}-AMX takes as input, and the assembly \act{}-AMX emits.
The emitted assembly is the same hand-optimized code the oneDNN library produces --- \act{}-AMX enumerates it as one candidate among many and its cost model selects it (\S\ref{subsec:evaluation-perf-amx}) --- which is why the two match in performance.
For the fp32 and mixed kernels (K3--K5) the assembly listings show a representative excerpt, with elided lines marked \texttt{...}.

\subsection{Kernel Selection and Statistics}
\label{app:amx-stats}

The XLA compiler emits tiled kernels which are then compiled to assembly using the oneDNN library, via the \texttt{xla\_cpu\_use\_mkl\_dnn} flag.
We selected five kernels commonly observed in the oneDNN examples~\cite{onednn-examples}, listed in Table~\ref{app:tab-amx}.
They are deliberately simple as computations --- all five are matrix multiplications with element-wise post-processing --- but carry complex memory layouts on their inputs, which is what makes them a demanding test of Module~1's layout modeling (Appendix~\ref{app:solution}).

Table~\ref{app:tab-amx} separates the size of the input from the size of the output.
\#Nodes counts every instruction of the kernel's tensor computation graph, parameters and constants included, except the ROOT \textsf{tuple} that packages K5's two outputs, which is bookkeeping rather than computation.
\#AMX and \#AVX512 count the AMX tile instructions and AVX512 instructions in the compiled assembly.
The split across the five kernels is clean and follows their element types: the two int8 kernels (K1, K2) compile purely to AMX tile instructions, the two fp32 kernels (K3, K4) purely to AVX512, and the mixed kernel (K5) to a combination of both.
This much is forced rather than chosen --- the \idl{} description we used covers AMX instructions only for int8 and AVX512 instructions only for fp32, so no other assignment is type-correct.
What the cost model decides is the code \emph{within} each class: which tiling, which register blocking, which instruction order.

\begin{table}[!ht]
  \centering
  \begin{tabular}{||c|c|c|c|c||}
    \hline
    Kernel & oneDNN example        & \#Nodes & \#AMX & \#AVX512 \\
    \hline\hline
    K1     & \texttt{cnn\_inf\_amx}   & 6       & 16    & 0        \\
    K2     & \texttt{rnn\_inf\_amx}   & 6       & 9     & 0        \\
    K3     & \texttt{mem\_format\_avx} & 7       & 0     & 100      \\
    K4     & \texttt{sgemm\_avx}      & 7       & 0     & 107      \\
    K5     & \texttt{cnn\_inf\_mix}   & 23      & 16    & 37       \\
    \hline
  \end{tabular}
  \caption{The five selected oneDNN kernels and their optimized assembly code in the oneDNN library.
    \#Nodes is the number of operators in the tensor computation graph; \#AMX and \#AVX512 count instructions of each class in the compiled assembly.}
  \label{app:tab-amx}
  \Description{Table of five oneDNN kernels K1 to K5, giving for each the corresponding oneDNN example name, the number of graph nodes, and the number of AMX and AVX512 instructions in the compiled assembly.}
\end{table}

\FloatBarrier
\subsection{K1: \texttt{cnn\_inf\_amx}}
\label{app:amx-k1}

K1 is the int8 convolution-inference kernel and the simplest of the five: a single matrix multiplication of a \textsf{u8[32,64]} activation against an \textsf{i8[32,64]} weight tensor carried in the AMX-native tiled layout \textsf{\{T(16,4)\}}.
That layout is the VNNI packing the \tmul{} instruction requires, with four elements of the reduction dimension interleaved; \S\ref{subsec:evaluation-support} shows the corresponding transformation on tile register \textsf{src1}.
The compiled code is 16 AMX instructions with no AVX512 at all: four \textsf{tilezero}, four \textsf{tileloadd}, four \tmul{}, and four \textsf{tilestored}, tiling the $32 \times 32$ output into four $16 \times 16$ tile registers.

\begin{figure}[!ht]
  \centering
  \includegraphics[width=0.321\textwidth]{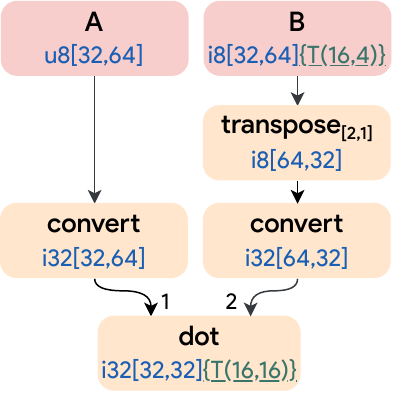}
  \caption{Tensor computation graph for oneDNN kernel K1.}
  \label{app:fig-k1}
  \Description{Tensor computation graph for kernel K1: two input tensors A and B, with B transposed, both converted to i32, feeding a dot operator with a tiled output layout.}
\end{figure}

\begin{figure}[!ht]
  \begin{lstlisting}[style={hlo},frame=single]
ENTRY k1 {
  A = u8[32,64] parameter(0)
  B = i8[32,64]{T(16,4)} parameter(1)
  convert.0 = i32[32,64] convert(A)
  transpose.1 = i8[64,32] transpose(B), dimensions=[2,1]
  convert.2 = i32[64,32] convert(transpose.1)
  ROOT dot.3 = i32[32,32]{T(16,16)} dot(convert.0, convert.2)
}
\end{lstlisting}
  \caption{XLA-HLO code for kernel K1.}
  \label{app:lst-hlo-k1}
  \Description{XLA-HLO listing for kernel K1 with two parameters, two converts, a transpose and a dot.}
\end{figure}

\begin{figure}[!ht]
  \begin{lstlisting}[style={amx-asm},frame=single]
tilezero(dst = tmm0)
tilezero(dst = tmm1)
tilezero(dst = tmm2)
tilezero(dst = tmm3)
tileloadd(dst = tmm4, mem_addr = 0x0000)
tileloadd(dst = tmm5, mem_addr = 0x0400)
tileloadd(dst = tmm6, mem_addr = 0x0800)
tileloadd(dst = tmm7, mem_addr = 0x0c00)
tdpbusd(dst = tmm0, src0 = tmm4, src1 = tmm6)
tdpbusd(dst = tmm1, src0 = tmm4, src1 = tmm7)
tdpbusd(dst = tmm2, src0 = tmm5, src1 = tmm6)
tdpbusd(dst = tmm3, src0 = tmm5, src1 = tmm7)
tilestored(src = tmm0, mem_addr = 0x1000)
tilestored(src = tmm1, mem_addr = 0x1400)
tilestored(src = tmm2, mem_addr = 0x1800)
tilestored(src = tmm3, mem_addr = 0x1c00)
\end{lstlisting}
  \caption{Compiled assembly code for kernel K1.}
  \label{app:lst-asm-k1}
  \Description{AMX assembly for kernel K1: four tilezero, four tileloadd, four tdpbusd and four tilestored instructions.}
\end{figure}

\FloatBarrier
\subsection{K2: \texttt{rnn\_inf\_amx}}
\label{app:amx-k2}

K2 is the int8 recurrent-inference kernel, structurally identical to K1 but with a half-height activation, \textsf{u8[16,64]} instead of \textsf{u8[32,64]}.
The consequence is visible in the assembly: because the output is $16 \times 32$ rather than $32 \times 32$, it occupies two tile registers instead of four, and the instruction count falls from 16 to 9.
Nothing else changes, which is the point --- the same generated backend handles both without any shape-specific path.

\begin{figure}[!ht]
  \centering
  \includegraphics[width=0.321\textwidth]{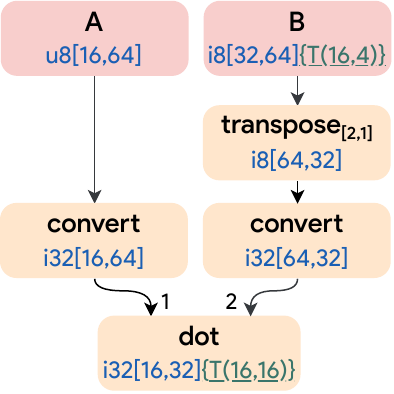}
  \caption{Tensor computation graph for oneDNN kernel K2.}
  \label{app:fig-k2}
  \Description{Tensor computation graph for kernel K2, identical in shape to K1 but with a 16 by 64 activation input.}
\end{figure}

\begin{figure}[!ht]
  \begin{lstlisting}[style={hlo},frame=single]
ENTRY k2 {
  A = u8[16,64] parameter(0)
  B = i8[32,64]{T(16,4)} parameter(1)
  convert.0 = i32[16,64] convert(A)
  transpose.1 = i8[64,32] transpose(B), dimensions=[2,1]
  convert.2 = i32[64,32] convert(transpose.1)
  ROOT dot.3 = i32[16,32]{T(16,16)} dot(convert.0, convert.2)
}
\end{lstlisting}
  \caption{XLA-HLO code for kernel K2.}
  \label{app:lst-hlo-k2}
  \Description{XLA-HLO listing for kernel K2 with a 16 by 64 activation, a transposed weight tensor, two converts and a dot.}
\end{figure}

\begin{figure}[!ht]
  \begin{lstlisting}[style={amx-asm},frame=single]
tilezero(dst = tmm0)
tilezero(dst = tmm1)
tileloadd(dst = tmm2, mem_addr = 0x0000)
tileloadd(dst = tmm3, mem_addr = 0x0400)
tileloadd(dst = tmm4, mem_addr = 0x0800)
tdpbusd(dst = tmm0, src0 = tmm2, src1 = tmm3)
tdpbusd(dst = tmm1, src0 = tmm2, src1 = tmm4)
tilestored(src = tmm0, mem_addr = 0x0c00)
tilestored(src = tmm1, mem_addr = 0x1000)
\end{lstlisting}
  \caption{Compiled assembly code for kernel K2.}
  \label{app:lst-asm-k2}
  \Description{AMX assembly for kernel K2: two tilezero, three tileloadd, two tdpbusd and two tilestored instructions.}
\end{figure}

\FloatBarrier
\subsection{K3: \texttt{mem\_format\_avx}}
\label{app:amx-k3}

K3 is an fp32 kernel exercising memory-format propagation.
It is an outer product: a \textsf{f32[32]} vector reshaped to a column, multiplied by a single sliced column of \textsf{f32[14,4]} transposed to a row, with the result transposed into the tiled layout \textsf{\{1,0:T(14,16)\}}.
Since AMX has no fp32 tile instruction, \act{}-AMX compiles it entirely to AVX512: 28 accumulator registers are zeroed, the two input rows are loaded once, and 14 iterations each broadcast one scalar and issue two fused multiply-adds, followed by 28 stores.
The excerpt below shows the first two iterations and the last.

\begin{figure}[!ht]
  \centering
  \includegraphics[width=0.321\textwidth]{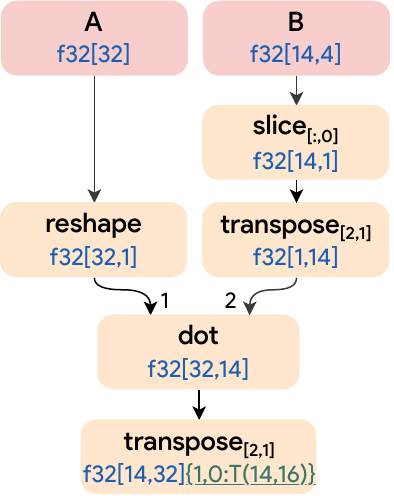}
  \caption{Tensor computation graph for oneDNN kernel K3.}
  \label{app:fig-k3}
  \Description{Tensor computation graph for kernel K3: a vector reshaped to a column and a sliced transposed row feeding a dot, whose result is transposed into a tiled layout.}
\end{figure}

\begin{figure}[!ht]
  \begin{lstlisting}[style={hlo},frame=single]
ENTRY k3 {
  A = f32[32] parameter(0)
  B = f32[14,4] parameter(1)
  reshape.0 = f32[32,1] reshape(A)
  slice.1 = f32[14,1] slice(B), dimensions=[:,0]
  transpose.2 = f32[1,14] transpose(slice.1), dimensions=[2,1]
  dot.3 = f32[32,14] dot(reshape.0, transpose.2)
  ROOT transpose.4 = f32[14,32]{1,0:T(14,16)} transpose(dot.3), dimensions=[2,1]
}
\end{lstlisting}
  \caption{XLA-HLO code for kernel K3.}
  \label{app:lst-hlo-k3}
  \Description{XLA-HLO listing for kernel K3 with a reshape, a slice, two transposes and a dot.}
\end{figure}

\begin{figure}[!ht]
  \begin{lstlisting}[style={amx-asm},frame=single]
vpxord(dst = zmm31, src0 = zmm31, src1 = zmm31)
...
vpxord(dst = zmm4, src0 = zmm4, src1 = zmm4)
vmovupsz(dst = zmm2, mem_addr=0x000)
vmovupsz(dst = zmm3, mem_addr=0x040)
vbroadcastssl(dst = zmm0, mem_addr=0x080)
vfmadd231ps(dst = zmm31, src0 = zmm3, src1 = zmm0)
vfmadd231ps(dst = zmm30, src0 = zmm2, src1 = zmm0)
vbroadcastssl(dst = zmm0, mem_addr=0x090)
vfmadd231ps(dst = zmm29, src0 = zmm3, src1 = zmm0)
vfmadd231ps(dst = zmm28, src0 = zmm2, src1 = zmm0)
...
vbroadcastssl(dst = zmm0, mem_addr=0x150)
vfmadd231ps(dst = zmm5, src0 = zmm3, src1 = zmm0)
vfmadd231ps(dst = zmm4, src0 = zmm2, src1 = zmm0)
vmovupsz(src = zmm31, mem_addr=0x160)
vmovupsz(src = zmm30, mem_addr=0x1a0)
...
vmovupsz(src = zmm4, mem_addr=0x820)
\end{lstlisting}
  \caption{Compiled assembly code for kernel K3.
    Iteration $n$ accumulates into the register pair $(\textsf{zmm}_{31-2(n-1)}, \textsf{zmm}_{30-2(n-1)})$, so the 14 iterations consume the 28 accumulators zeroed at the top.}
  \label{app:lst-asm-k3}
  \Description{AVX512 assembly excerpt for kernel K3 showing accumulator zeroing, two input loads, broadcast and fused multiply-add pairs, and stores.}
\end{figure}

\FloatBarrier
\subsection{K4: \texttt{sgemm\_avx}}
\label{app:amx-k4}

K4 is the fp32 general matrix multiplication.
Its graph has the same topology as K3's, but the two dimensions move in opposite directions: the vector grows from \textsf{f32[32]} to \textsf{f32[48]} while the sliced row shrinks from 14 elements to 8.
That is what changes the register blocking. K3 holds 28 accumulators and runs 14 iterations of two fused multiply-adds; K4 holds 24 and runs 8 iterations of three, against three loaded input registers rather than two, giving 107 AVX512 instructions.
The \textsf{vaddpsz} block near the end accumulates a further operand read from memory before the results are stored.

\begin{figure}[!ht]
  \centering
  \includegraphics[width=0.321\textwidth]{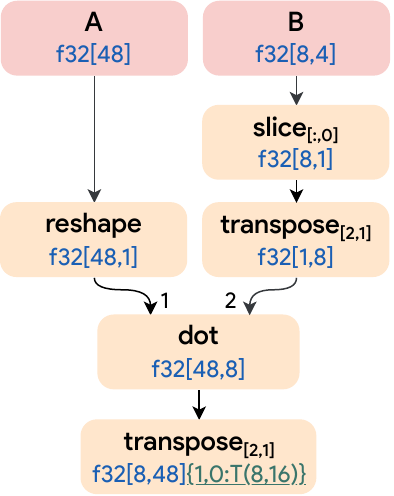}
  \caption{Tensor computation graph for oneDNN kernel K4.}
  \label{app:fig-k4}
  \Description{Tensor computation graph for kernel K4, structurally identical to K3 with a 48-element vector and an 8 by 4 matrix.}
\end{figure}

\begin{figure}[!ht]
  \begin{lstlisting}[style={hlo},frame=single]
ENTRY k4 {
  A = f32[48] parameter(0)
  B = f32[8,4] parameter(1)
  reshape.0 = f32[48,1] reshape(A)
  slice.1 = f32[8,1] slice(B), dimensions=[:,0]
  transpose.2 = f32[1,8] transpose(slice.1), dimensions=[2,1]
  dot.3 = f32[48,8] dot(reshape.0, transpose.2)
  ROOT transpose.4 = f32[8,48]{1,0:T(8,16)} transpose(dot.3), dimensions=[2,1]
}
\end{lstlisting}
  \caption{XLA-HLO code for kernel K4.}
  \label{app:lst-hlo-k4}
  \Description{XLA-HLO listing for kernel K4 with a reshape, a slice, two transposes and a dot.}
\end{figure}

\begin{figure}[!ht]
  \begin{lstlisting}[style={amx-asm},frame=single]
vpxord(dst = zmm8, src0 = zmm8, src1 = zmm8)
...
vpxord(dst = zmm31, src0 = zmm31, src1 = zmm31)
vmovupsz(dst = zmm0, mem_addr=0x000)
vmovupsz(dst = zmm1, mem_addr=0x040)
vmovupsz(dst = zmm2, mem_addr=0x080)
vbroadcastssl(dst = zmm6, mem_addr=0x0c0)
vfmadd231ps(dst = zmm8,  src0 = zmm6, src1 = zmm0)
vfmadd231ps(dst = zmm16, src0 = zmm6, src1 = zmm1)
vfmadd231ps(dst = zmm24, src0 = zmm6, src1 = zmm2)
vbroadcastssl(dst = zmm7, mem_addr=0x0d0)
vfmadd231ps(dst = zmm9,  src0 = zmm7, src1 = zmm0)
vfmadd231ps(dst = zmm17, src0 = zmm7, src1 = zmm1)
vfmadd231ps(dst = zmm25, src0 = zmm7, src1 = zmm2)
...
vbroadcastssl(dst = zmm6, mem_addr=0x120)
vfmadd231ps(dst = zmm14, src0 = zmm6, src1 = zmm0)
vfmadd231ps(dst = zmm22, src0 = zmm6, src1 = zmm1)
vfmadd231ps(dst = zmm30, src0 = zmm6, src1 = zmm2)
vbroadcastssl(dst = zmm7, mem_addr=0x130)
vfmadd231ps(dst = zmm15, src0 = zmm7, src1 = zmm0)
vfmadd231ps(dst = zmm23, src0 = zmm7, src1 = zmm1)
vfmadd231ps(dst = zmm31, src0 = zmm7, src1 = zmm2)
vaddpsz(dst = zmm8, src0 = zmm8, mem_addr=0x140)
...
vaddpsz(dst = zmm31, src0 = zmm31, mem_addr=0x700)
vmovupsz(src = zmm8, mem_addr=0x740)
...
vmovupsz(src = zmm31, mem_addr=0xd00)
\end{lstlisting}
  \caption{Compiled assembly code for kernel K4.
    The accumulators \textsf{zmm8}--\textsf{zmm31} are consumed in ascending order by both the \textsf{vaddpsz} block and the stores that follow it.}
  \label{app:lst-asm-k4}
  \Description{AVX512 assembly excerpt for kernel K4 showing accumulator zeroing, three input loads, broadcast and triple fused multiply-add groups, an accumulate-from-memory block, and stores.}
\end{figure}

\FloatBarrier
\subsection{K5: \texttt{cnn\_inf\_mix}}
\label{app:amx-k5}

K5 is the mixed-precision convolution-inference kernel and the most interesting of the five, because it is the only one whose graph spans both element types and therefore both instruction classes.
Its graph has two independent halves.
The first is exactly K1's int8 matrix multiplication, which compiles to the same 16 AMX instructions.
The second is an fp32 post-processing chain --- scale, bias, ReLU, scale again, clamp to $[0, 255]$, and convert down to \textsf{u8} --- which compiles to 37 AVX512 instructions.
Neither half is annotated as AMX or AVX512 anywhere in the input; \act{}-AMX enumerates candidates for the whole graph and emits the mixture without any per-kernel direction.
This is the concrete sense in which \act{} does not need per-kernel tuning: the same generated backend produces pure-AMX code for K1, pure-AVX512 code for K3, and the correct mixture here.

\begin{figure}[!ht]
  \centering
  \includegraphics[width=0.95\textwidth]{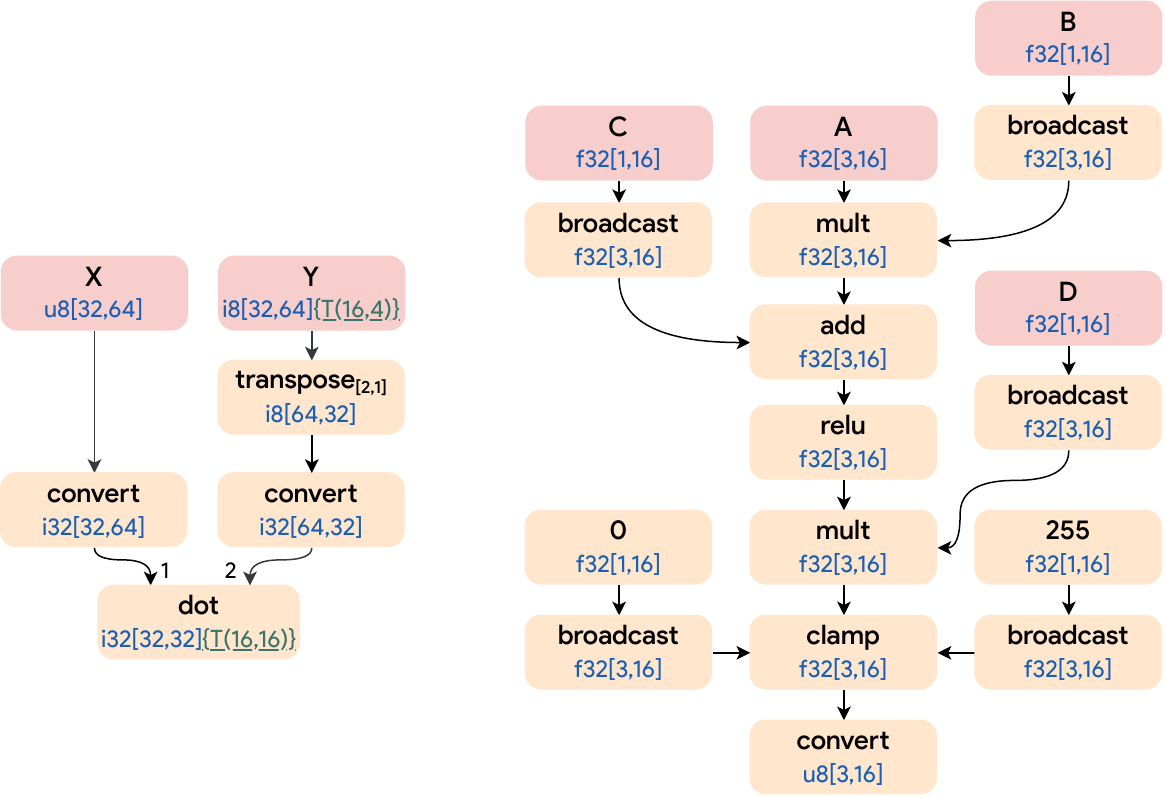}
  \caption{Tensor computation graph for oneDNN kernel K5.
    The left half is K1's int8 matrix multiplication; the right half is the fp32 post-processing chain.}
  \label{app:fig-k5}
  \Description{Tensor computation graph for kernel K5 with two disconnected halves: an int8 dot product on the left, and on the right a chain of broadcasts, multiplies, an add, a relu, a clamp and a convert.}
\end{figure}

\begin{figure}[!ht]
  \begin{lstlisting}[style={hlo},frame=single]
ENTRY k5 {
  X = u8[32,64] parameter(0)
  Y = i8[32,64]{T(16,4)} parameter(1)
  convert.0 = i32[32,64] convert(X)
  transpose.1 = i8[64,32] transpose(Y), dimensions=[2,1]
  convert.2 = i32[64,32] convert(transpose.1)
  dot.3 = i32[32,32]{T(16,16)} dot(convert.0, convert.2)
  A = f32[3,16] parameter(2)
  B = f32[1,16] parameter(3)
  C = f32[1,16] parameter(4)
  D = f32[1,16] parameter(5)
  constant.0 = f32[1,16] constant(0)
  constant.255 = f32[1,16] constant(255)
  broadcast.4 = f32[3,16] broadcast(B), dimensions=[1]
  multiply.5 = f32[3,16] multiply(A, broadcast.4)
  broadcast.6 = f32[3,16] broadcast(C), dimensions=[1]
  add.7 = f32[3,16] add(multiply.5, broadcast.6)
  broadcast.11 = f32[3,16] broadcast(constant.0), dimensions=[1]
  broadcast.12 = f32[3,16] broadcast(constant.255), dimensions=[1]
  relu.8 = f32[3,16] maximum(broadcast.11, add.7)
  broadcast.9 = f32[3,16] broadcast(D), dimensions=[1]
  multiply.10 = f32[3,16] multiply(relu.8, broadcast.9)
  clamp.13 = f32[3,16] clamp(broadcast.11, multiply.10, broadcast.12)
  convert.14 = u8[3,16] convert(clamp.13)
  ROOT tuple.15 = tuple(dot.3, convert.14)
}
\end{lstlisting}
  \caption{XLA-HLO code for kernel K5.
    The ROOT \textsf{tuple} packages the two outputs and is not counted as an operator in Table~\ref{app:tab-amx}.}
  \label{app:lst-hlo-k5}
  \Description{XLA-HLO listing for kernel K5 combining an int8 dot product with an fp32 chain of broadcasts, multiplies, an add, a maximum, a clamp and a convert, packaged in a root tuple.}
\end{figure}

\begin{figure}[!ht]
  \begin{lstlisting}[style={amx-asm},frame=single]
tilezero(dst = tmm0)
tilezero(dst = tmm1)
tilezero(dst = tmm2)
tilezero(dst = tmm3)
tileloadd(dst = tmm4, mem_addr = 0x0000)
tileloadd(dst = tmm5, mem_addr = 0x0400)
tileloadd(dst = tmm6, mem_addr = 0x0800)
tileloadd(dst = tmm7, mem_addr = 0x0c00)
tdpbusd(dst = tmm0, src0 = tmm4, src1 = tmm6)
tdpbusd(dst = tmm1, src0 = tmm4, src1 = tmm7)
tdpbusd(dst = tmm2, src0 = tmm5, src1 = tmm6)
tdpbusd(dst = tmm3, src0 = tmm5, src1 = tmm7)
tilestored(src = tmm0, mem_addr = 0x1000)
tilestored(src = tmm1, mem_addr = 0x1400)
tilestored(src = tmm2, mem_addr = 0x1800)
tilestored(src = tmm3, mem_addr = 0x1c00)
vmovupsz(dst = zmm10, mem_addr = 0x000)
vmovupsz(dst = zmm15, mem_addr = 0x040)
mov(dst = ebx, imm = 0x437f0000)
vmovq(dst = xmm9, src = rbx)
vbroadcastss(dst = zmm9, src = xmm9)
vmovupsz(dst = zmm31, mem_addr = 0x080)
vcvtdq2ps(dst = zmm31, src = zmm31)
vmovupsz(dst = zmm30, mem_addr = 0x0c0)
vcvtdq2ps(dst = zmm30, src = zmm30)
vmovupsz(dst = zmm29, mem_addr = 0x100)
vcvtdq2ps(dst = zmm29, src = zmm29)
vmulps(dst = zmm31, src0 = zmm31, src1 = zmm15)
...
vaddps(dst = zmm31, src0 = zmm31, src1 = zmm10)
...
vmaxps(dst = zmm31, src0 = zmm31, src1 = zmm8)
...
vbroadcastssl(dst = zmm0, mem_addr = 0x144)
vmulps(dst = zmm31, src0 = zmm31, src1 = zmm0)
...
vmaxps(dst = zmm31, src0 = zmm31, src1 = zmm8)
vminps(dst = zmm31, src0 = zmm31, src1 = zmm9)
vcvtps2dq(dst = zmm31, src = zmm31)
vpmovusdbx(src = zmm31, mem_addr = 0x144)
...
\end{lstlisting}
  \caption{Compiled assembly code for kernel K5.
    The AMX half is emitted first and is identical to Fig.~\ref{app:lst-asm-k1}; the AVX512 half follows.
    The immediate \texttt{0x437f0000} is the fp32 encoding of $255.0$, broadcast into \textsf{zmm9} as the clamp upper bound.}
  \label{app:lst-asm-k5}
  \Description{Assembly excerpt for kernel K5 showing the AMX tile instruction sequence followed by an AVX512 sequence of loads, converts, multiplies, adds, maximum, minimum and a narrowing store.}
\end{figure}


\end{document}